\documentclass[a4paper]{article}

\usepackage[left=2.5cm,right=2.5cm,top=3cm,bottom=3cm]{geometry}
\usepackage{authblk}
\usepackage{hyperref}
\usepackage{paralist}
\usepackage{tikz} 
\usetikzlibrary{arrows}
\usetikzlibrary{arrows,automata}
\usetikzlibrary{automata,positioning,arrows}
\usetikzlibrary{shapes}
\usetikzlibrary{shapes,backgrounds}
\usetikzlibrary{matrix}
\usetikzlibrary{chains,fit,shapes}
\usepackage{amsmath}
\usepackage{amssymb}
\usepackage{amsfonts}
\usepackage{amsthm}
\usepackage[ruled]{algorithm2e}

\usepackage{wrapfig}
\usepackage{stmaryrd}
\usepackage{tensor}

\usepackage[ruled]{algorithm2e}

\SetAlFnt{\small}
\SetAlCapFnt{\small}
\SetAlCapNameFnt{\small}
\SetAlCapHSkip{0pt}
\IncMargin{-\parindent}

\usepackage{bbold}

\usepackage{textcomp}

\usepackage{wasysym}

\newcommand{\Sort}[1]{\ensuremath{\textit{sort}(#1)}}

\newcommand{\msSort}[1]{\langle\!\langle #1 \rangle\!\rangle}

\newcommand{\compMatrix}[1]{\textsf{comp}_{#1}}

\DeclareMathOperator{\undefin}{\perp}
\DeclareMathOperator{\emptyMarker}{\text{\textestimated}}
\DeclareMathOperator{\nonEmptyMarker}{\mathbb{1}}
\DeclareMathOperator{\baseCaseMarker}{\mathbb{b}}

\DeclareMathOperator{\msleq}{\preceq}
\DeclareMathOperator{\mseq}{\prec}
\DeclareMathOperator{\seqleq}{\lesssim}

\DeclareMathOperator{\doc}{\mathbf{D}}
\DeclareMathOperator{\docSize}{\mathbf{d}}

\DeclareMathOperator{\states}{\mathnormal{q}}

\DeclareMathOperator{\bigO}{O}

\newcommand{\msprod}[1]{\otimes_{#1}}

\newcommand{\recProc}[3]{\mathsf{Comp}\mathfrak{M}(#1, #2, #3)}

\newcommand{\card}[1]{\lvert #1 \rvert}

\newcommand{\transfuncNFA}[1]{\xrightarrow{#1}}

\newcommand{\size}[1]{\mathsf{size}(#1)}

\newcommand{\doclength}[1]{|#1|_d}

\DeclareMathOperator{\VA}{\mathsf{VA}}

\DeclareMathOperator{\SLP}{\mathsf{SLP}}
\DeclareMathOperator{\NFA}{\mathsf{NFA}}
\DeclareMathOperator{\DFA}{\mathsf{DFA}}

\DeclareMathOperator{\spans}{\textsf{Spans}}

\DeclareMathOperator{\eword}{\varepsilon}
\newcommand{\lang}{L}

\DeclareMathOperator{\altop}{\vee}

\newcommand{\open}[1]{\tensor[^{#1}]{\triangleright}{_{}}}
\newcommand{\close}[1]{\tensor[_{}]{\triangleleft}{^{#1}}}

\newcommand{\spann}[2]{\ensuremath{[#1,#2\rangle}}

\newcommand{\varsx}{\ensuremath{\mathsf{x}}}
\newcommand{\varsy}{\ensuremath{\mathsf{y}}}
\newcommand{\varsz}{\ensuremath{\mathsf{z}}}

\newcommand{\varset}{\ensuremath{\mathcal{X}}}

\newcommand{\ta}{\ensuremath{\mathtt{a}}}
\newcommand{\tb}{\ensuremath{\mathtt{b}}}
\newcommand{\tc}{\ensuremath{\mathtt{c}}}

\newcommand{\getWord}[1]{\mathfrak{e}(#1)}
\newcommand{\insertmarkers}[2]{\mathfrak{m}(#1, #2)}
\newcommand{\markedpositions}[1]{\mathfrak{p}(#1)}
\newcommand{\setrep}[1]{\widehat{#1}}

\newcommand{\rightshift}[2]{\mathsf{rs}_{#2}(#1)}

\DeclareMathOperator{\derivationstepsymbol}{\mathsf{D}}
\DeclareMathOperator{\derivsymbol}{\mathfrak{D}}

\newcommand{\deriv}[1]{\derivsymbol(#1)}

\newcommand{\derivationstepsub}[2]{\derivationstepsymbol_{#1}(#2)}
\newcommand{\derivsub}[2]{\mathfrak{D}_{#1}(#2)}

\providecommand{\card}[1]{\lvert#1\rvert}

\newcommand{\powerset}[1]{\mathcal{P}(#1)}

\newcommand{\depth}[1]{\mathsf{depth}(#1)}

\newcommand{\allTrees}[4]{\mathbf{Trees}(#1, #2, #3, #4)}

\DeclareMathOperator{\markerTransMat}{\mathfrak{M}}
\DeclareMathOperator{\reachTransMat}{\mathfrak{R}}
\DeclareMathOperator{\intStatesTransMat}{\mathfrak{I}}

\newcommand{\yield}[1]{\mathsf{yield}(#1)}
\newcommand{\yieldPara}[2]{\mathsf{yield}_{#1}(#2)}

\newcommand{\mainNode}[4]{\mathfrak{K}_{#1}^{#4}[#2, #3]}

\newcommand{\treeNode}[4]{#1\langle#2 \, \text{\pointer} \, #3 \, \text{\pointer} \, #4\rangle}
\newcommand{\treeNodeTerminal}[3]{#1\langle#2 \, \text{\pointer} \, #3, \nonEmptyMarker\rangle}
\newcommand{\treeNodeEmpty}[3]{#1\langle#2 \, \text{\pointer} \, #3, \emptyMarker\rangle}

\DeclareMathOperator{\eol}{\mathsf{EOE}}

\DeclareMathOperator{\enumAll}{\textsf{\upshape EnumAll}}

\DeclareMathOperator{\enumSingleTree}{\textsf{EnumSingleTree}}
\DeclareMathOperator{\enumSingleRoot}{\textsf{EnumSingleRoot}}

\DeclareMathOperator{\enumA}{\mathcal{A}}

\DeclareMathOperator{\constAlgo}{\mathsf{ConstTree}}

\newcommand{\State}[1]{#1}

\renewcommand{\mid}{:}

\newcommand{\nc}[1]{\newcommand{#1}}

\nc{\Dom}[1]{\ensuremath{\textup{dom}(#1)}}
\nc{\emptyword}{\ensuremath{\varepsilon}}

\newtheorem{theorem}{Theorem}[section]

\newtheorem{proposition}[theorem]{Proposition}
\newtheorem{lemma}[theorem]{Lemma}
\newtheorem{definition}[theorem]{Definition}
\newtheorem{observation}[theorem]{Observation}
\newtheorem{remark}[theorem]{Remark}
\newtheorem{example}[theorem]{Example}

\begin{document}

\title{Spanner Evaluation over SLP-Compressed Documents\footnote{%
The first author has been funded by the German Research Foundation (Deutsche Forschungsgemeinschaft, DFG) -- project number 416776735 (gef\"ordert durch die Deutsche Forschungsgemeinschaft (DFG) -- Projektnummer 416776735).
The second author has been partially supported by the ANR project EQUUS ANR-19-CE48-0019; funded by the Deutsche Forschungsgemeinschaft (DFG,  German Research Foundation) -- project number 431183758 (gef\"ordert durch die Deutsche Forschungsgemeinschaft (DFG) -- Projektnummer 431183758).
}}

\author[1]{Markus L.\ Schmid}
\author[2]{Nicole Schweikardt}

\affil[1]{Humboldt-Universit\"at zu Berlin, Unter den Linden 6, D-10099, Berlin, Germany, \texttt{MLSchmid@MLSchmid.de}}
\affil[2]{Humboldt-Universit\"at zu Berlin, Unter den Linden 6, D-10099, Berlin, Germany, \texttt{schweikn@informatik.hu-berlin.de}}

\maketitle

\begin{abstract}
We consider the problem of evaluating regular spanners over compressed documents, i.\,e., we wish to solve evaluation tasks directly on the compressed data, without decompression. As compressed forms of the documents we use straight-line programs ($\SLP$s) --- a lossless compression scheme for textual data widely used in different areas of theoretical computer science and particularly well-suited for algorithmics on compressed data. \par
In data complexity, our results are as follows. For a regular spanner $M$ and an $\SLP$ $\mathcal{S}$ of size $\mathbf{s}$ that represents a document $\doc$, we can solve the tasks of model checking and of checking non-emptiness in time $\bigO(\mathbf{s})$. Computing the set $\llbracket M \rrbracket(\doc)$ of all span-tuples extracted from $\doc$ can be done in time $\bigO(\mathbf{s} {\cdot} \card{\llbracket M \rrbracket(\doc)})$, and enumeration of $\llbracket M \rrbracket(\doc)$ can be done with linear preprocessing $\bigO(\mathbf{s})$ and a delay of $\bigO(\depth{\mathcal{S}})$, where $\depth{\mathcal{S}}$ is the depth of $\mathcal{S}$'s derivation tree. \par
Note that $\mathbf{s}$ can be exponentially smaller than the document's size $\card{\doc}$; and, due to known balancing results for $\SLP$s, we can always assume that $\depth{\mathcal{S}} = \bigO(\log(\card{\doc}))$ independent of $\doc$'s compressibility. Hence, our enumeration algorithm has a delay logarithmic in the size of the non-compressed data and a preprocessing time that is at best (i.\,e., in the case of highly compressible
documents) also logarithmic, but at worst still linear. Therefore, in a big-data perspective, our enumeration algorithm for $\SLP$-compressed documents may nevertheless beat the known linear preprocessing and constant delay algorithms for non-compressed documents.
\end{abstract}

\section{Introduction}\label{sec:intro}

The information extraction framework of \emph{document spanners} has been introduced in~\cite{FaginEtAl2015} as a formalisation of the query language AQL, which is used in IBM's information extraction engine SystemT. A document spanner performs  
information extraction by mapping a \emph{document} $\doc$ (i.\,e., a string) over a finite alphabet $\Sigma$, to a relation over so-called \emph{spans} of $\doc$, which are intervals $\spann{i}{j}$ with $0 \leq i < j \leq \card{\doc}{+}1$. For example, a spanner may map documents $\doc = d_1 d_2 \ldots d_n$ over $\Sigma = \{\ta, \tb, \tc\}$ to the binary relation that contains all pairs $(\spann{i}{i{+}1}, \spann{j}{\ell})$ such that $d_i$ is the first occurrence of symbol $\ta$ and $d_{j} d_{j + 1} \ldots d_{\ell-1}$ is some factor over $\{\tc\}$. Thus, $\doc = \ta \tb \tc \tc \ta$ would be mapped to the relation 
\begin{equation*}
\{\ (\spann{1}{2},\, \spann{3}{4}), \ (\spann{1}{2},\, \spann{4}{5}), \ (\spann{1}{2},\, \spann{3}{5})\ \}\,.
\end{equation*}
It is common to let the attributes of the extracted relations be given by a set $\varset$ of \emph{variables} (i.\,e., span-tuples are mappings from $\varset$ to the set of spans) and associate a pair of parentheses $\open{\varsx}$ and $\close{\varsx}$ with each $\varsx \in \varset$. These parentheses can be used as \emph{markers} that mark subwords directly in a document (therefore they mark spans), e.\,g., the \emph{subword-marked words} 
\begin{align*}
&\open{\varsx} \ta \close{\varsx} \tb \open{\varsy} \tc \close{\varsy} \tc \ta,& &\open{\varsx} \ta \close{\varsx} \tb  \tc \open{\varsy} \tc \close{\varsy} \ta,& &\open{\varsx} \ta \close{\varsx} \tb \open{\varsy} \tc  \tc  \close{\varsy} \ta& 
\end{align*}
represent $\doc$ from above with the three mentioned span-tuples encoded by the marker symbols. In this way, spanners can be represented by sets (or languages) $L$ of subword-marked words, i.\,e., $L$ represents the spanner $\llbracket L \rrbracket$ that maps any document $\doc$ to the set $\llbracket L \rrbracket(\doc)$ of all span-tuples $t$ with the property that marking $\doc$ with $t$'s spans in the way explained above yields a word from $L$. In this sense, the  subword-marked language given by the regular expression $(\tb \altop \tc)^* \open{\varsx} \ta \close{\varsx} \Sigma^* \open{\varsy} \tc^+ \close{\varsy} \Sigma^*$ describes the spanner mentioned above. Spanners that can be expressed by \emph{regular} languages in this way are called \emph{regular spanners} and have been studied extensively since the introduction of spanners in~\cite{FaginEtAl2015}; we discuss the respective related work in detail below. 
An example of a regular spanner represented by an 
automaton can be found in Figure~\ref{fig:exampleDFASpanner}.\par
For regular spanners, typical evaluation tasks can be solved in linear time in data complexity, including the enumeration of all span-tuples of $\llbracket L \rrbracket(\doc)$ with linear preprocessing and constant delay~\cite{FlorenzanoEtAl2018,AmarilliEtAl2019}. Under the assumption that we have to fully process the document at least once, this can be considered optimal.\par
As a new angle to the evaluation of regular spanners, we consider the setting where the input documents are given in a compressed form, and we want to evaluate spanners directly on the compressed documents without decompressing them. This is especially of interest in a big-data scenario, where the documents are huge, but it is also in general reasonable to assume that textual data is managed in compressed form, simply because the state of the art in algorithms allows for it. 
Due to redundancies, textual data (especially over natural languages) is often highly compressible by practical compression schemes, and, maybe even more importantly (and in contrast to relational data), many basic algorithmic tasks can be efficiently solved directly on compressed textual data.\par
As our underlying compression scheme, we use so-called \emph{straight-line programs} ($\SLP$s), which compress a document $\doc$ by a context-free grammar that represents the singleton language $\{\doc\}$.

\subsection{Algorithmics on SLP-Compressed Strings} 
See Example~\ref{definitionsExample} for an $\SLP$ of size $16$ that represents a document of size $25$. An illustrative way to represent $\SLP$s is in form of their derivation trees (see Figure~\ref{fig:exampleDerivTreeCNF}). While the full derivation tree is an uncompressed representation, it nevertheless reveals in an intuitive way the structural redundancies exploited by the $\SLP$: for every node label (i.\,e., non-terminal) we have to store only one subtree rooted by this label. In this regard, Figure~\ref{fig:exampleDerivTreeCNF} only shows the actual $\SLP$ in bold, while the redundancies are shown in grey.\par
The task investigated in this work is to evaluate a spanner, e.\,g., the one represented by the automaton of Figure~\ref{fig:exampleDFASpanner}, on a document given as an $\SLP$, e.\,g., the one represented by the bold parts of Figure~\ref{fig:exampleDerivTreeCNF}. However, we want to avoid to completely construct the document (or the full derivation tree). \par
$\SLP$s play a prominent role in the context of string algorithms and other areas of theoretical computer science. They are mathematically easy to handle and therefore very appealing for theoretical considerations. Independent of their data-compression applications, they have been used in many different contexts as a natural tool for representing (and reasoning about) hierarchical structure in sequential data (see, e.\,g.,~\cite{NevWit97a,Nev96,Lohrey2012,Loh2014,KieYan2000,StoSzy82}).\par
$\SLP$s are also of high practical relevance, mainly because many practically applied dictionary-based compression schemes (e.\,g., run-length encoding, and --~most notably~-- the Lempel-Ziv-family LZ77, LZ78, LZW, etc. which is relevant for practical tools like the built-in Unix utility \textsf{compress} or data formats like GIF, PNG, PDF and some ZIP archive file formats) can be converted efficiently into $\SLP$s of similar size, i.\,e., with size blow-ups by only moderate constants or log-factors (see~\cite{Lohrey2012, Cording2015PhD, AbboudEtAl2017, GotoEtAl2011, Ryt2003}). Hence, algorithms for $\SLP$-compressed data carry over to these practical formats.\par
While in the early days of computer science fast compression and decompression was an important factor, it is nowadays common to also rate compression schemes according to how suitable they are for solving problems directly on the compressed data without prior decompression (also called algorithmics on compressed strings). In this regard, $\SLP$s have very good properties: many basic problems on strings like comparison, pattern matching, membership in a regular language, retrieving subwords, etc. can all be efficiently solved directly on $\SLP$s~\cite{Lohrey2012}. As demonstrated by our results, this is even true for spanner evaluation. \par
A possible drawback of $\SLP$s is that computing a minimal size $\SLP$ for a given document is intractable (even for fixed alphabets)~\cite{CaselEtAl2020}. However, this has never been an issue for the application of $\SLP$s, since many approximations and heuristics are known that efficiently (i.\,e., in (near) linear time) compute $\SLP$s that are only a log-factor larger than minimal ones (see~\cite{Chaetal2005,LehPhdThesis,CaselEtAl2020}).\par
Since we cannot discuss all relevant papers in the context of algorithmics on $\SLP$-compressed data here, we refer for further reading to the survey~\cite{Lohrey2012}, the PhD-thesis~\cite{Cording2015PhD} and the comprehensive introductions of the papers~\cite{CaselEtAl2020, AbboudEtAl2017}. 

\subsection{Regular Spanner Evaluation}
The original framework of~\cite{FaginEtAl2015} uses regular spanners to extract relations directly from documents, which can then be further manipulated by relational algebra. Since the string-compression aspect applies only to the first stage of this approach, we are only concerned with regular spanners (for non-regular aspects of spanners see~\cite{SchmidSchweikardt2021, FreydenbergerHolldack2018,Freydenberger2019,Peterfreund2021}). We note that~ \cite{Peterfreund2021} is also concerned with grammars in the context of spanners, but in a different way: while in our case the documents are represented by grammars (i.\,e., $\SLP$s), but the spanners are classical regular spanners,~\cite{Peterfreund2021} considers spanners that are represented by grammars.\par
We follow the conceptional approach of~\cite{SchmidSchweikardt2021} and consider spanners as regular languages of subword-marked words, as sketched above. In this way, we can abstract from specialised machine models and represent our spanners as classical finite automata (we discuss this aspect in some more detail in Section~\ref{sec:spanners}). In order to avoid that the same span-tuple can be represented by different markings, we represent sequences of consecutive marker symbols by sets of marker symbols (e.\,g., $\ta \open{\varsx} \tb \close{\varsx} \open{\varsy} \tc \tc \close{\varsy}$ is represented as $\ta \open{\varsx} \tb \{\close{\varsx}, \open{\varsy}\} \tc \tc \close{\varsy}$). This is a common approach and is analogous to the \emph{extended sequential VAs} introduced in~\cite{FlorenzanoEtAl2018} (also used in~\cite{AmarilliEtAl2019, AmarillietAl2020Survey}). Our spanners can be non-functional, i.\,e., we allow span-tuples with undefined variables (also called the \emph{schemaless semantics} in~\cite{MaturanaEtAl2018}). \par
Regular spanners can be evaluated very efficiently since they inherit the good algorithmic properties of regular languages (e.\,g., model checking for regular spanners is a special variant of the membership problem for regular languages); see~\cite{FaginEtAl2015, AmarilliEtAl2019, AmarillietAl2020Survey,MaturanaEtAl2018,Peterfreund2019PhD} for further details. A new aspect that has not been considered in formal language theory is that of enumerating all query results (i.\,e., span-tuples). This has been considered in~\cite{FreydenbergerEtAl2018,FlorenzanoEtAl2018,AmarilliEtAl2019} and it is a major result that constant delay enumeration is possible after linear preprocessing (even if the spanners are given by non-deterministic automata); see especially the survey~\cite{AmarillietAl2020Survey}. The 
algorithmic approach is to construct the product graph of the automaton that represents the spanner (e.\,g., the one of Figure~\ref{fig:exampleDFASpanner}) and the input document (treated as a path). This yields a directed acyclic graph that fully represents the solution set and which can be used for enumeration (Figure~$1$ of~\cite{AmarillietAl2020Survey} illustrates this construction in a single picture). \par
The main 
challenge of the present paper is that the above described construction is not possible in our setting, since it requires the input document to be decompressed. We aim to represent all runs of the automaton on the decompressed document, while respecting the document's compressed form given by the $\SLP$.

\subsection{Our Contribution}
We investigate the following tasks, for which we get as input an $\SLP$ $\mathcal{S}$ (of size $\mathbf{s}$) for a document $\doc$ (of size $\mathbf{d}$) and a spanner represented by an automaton $M$:
\begin{enumerate}[\ \ (1)]
\item[$\bullet$] \emph{non-emptiness}: check if $\llbracket M \rrbracket(\doc) \neq \emptyset$ 
\item[$\bullet$] \emph{model checking}: check if $t \in \llbracket M \rrbracket(\doc)$ for a given span-tuple $t$
\item[$\bullet$] \emph{computation}: compute the whole set $\llbracket M \rrbracket(\doc)$
\item[$\bullet$] \emph{enumeration}: enumerate the elements of $\llbracket M \rrbracket(\doc)$
\end{enumerate}
Let $\mathbf{r}$ denote the number of result tuples (i.e., span-tuples) in $\llbracket M \rrbracket(\doc)$. In terms of data complexity, our main results solve 
\begin{enumerate}[\ \ (1)]
\item\label{item:intro:modelchecking}
 \emph{non-emptiness} and \emph{model checking} in time $O(\mathbf{s})$,
\item\label{item:intro:comp}
 \emph{computation} in time $O(\mathbf{s}\cdot\mathbf{r})$,
\item\label{item:intro:enum}
 \emph{enumeration} with delay $O(\log \mathbf{d})$ after $O(\mathbf{s})$ preprocessing.
\end{enumerate}
Note that \eqref{item:intro:enum} also implies a solution for 
\emph{computation} in time $O(\mathbf{s} + \mathbf{r} {\cdot} \log \mathbf{d})$ (however, our direct algorithm for computing $\llbracket M \rrbracket(\doc)$ is much simpler and better in combined complexity).\par
These runtimes are incomparable to the known runtimes on uncompressed documents, which solve
non-emptiness and model checking in time $O(\mathbf{d})$, computation in time $O(\mathbf{d}+\mathbf{r})$, and enumeration with delay $O(1)$ after $O(\mathbf{d})$ preprocessing.
But note that, for highly compressible documents, $\mathbf{s}$ might be exponentially smaller than $\mathbf{d}$, and in these cases our algorithms will outperform the approach of first decompressing the entire document and then applying an efficient algorithm on uncompressed documents.
In the case of highly compressible documents, 
our setting can also be considered as spanner evaluation with sublinear data complexity. 

In terms of combined complexity, the O-notation in our runtime guarantees
hides some (low degree) polynomial factors in $|M|$ (the total size of the automaton), $|Q|$ (the number of $M$'s states), and $|\varset|$ (the number of span variables); 
the precise bounds in combined complexity are stated in Theorems~\ref{basicEvalTasksTheorem},~\ref{computeSetTheorem}~and~\ref{enumSetTheorem}. 
We wish to point out that the aspect of 
conciseness of different spanner representations is hidden in the factor $\card{M}$.
The automata we use are, in terms of conciseness, like (nondeterministic) \emph{extended} VAs (see~\cite{FlorenzanoEtAl2018, AmarilliEtAl2019, AmarillietAl2020Survey}); and for 
enumeration (but only for enumeration)
we additionally need the automata to be \emph{deterministic}.

\subsection{Technical approach}
Model checking and checking non-empti\-ness  can be done in a rather straightfoward way by a reduction to the problem of checking membership of an $\SLP$-compressed document to a regular language.
For computing or enumerating the solution set, we have to come up with new ideas. \par
Intuitively speaking, the compression of $\SLP$s is done by representing several occurrences of the same factor of a document by just a single non-terminal, e.\,g., the three occurrences of factor $\ta \ta$ are represented by $E$ in the $\SLP$ $\mathcal{S}$ of Figure~\ref{fig:exampleDerivTreeCNF}. However, the span-tuples to be extracted may treat different occurrences of the same factor compressed by the same non-terminal in different ways. For example, the spanner $M$ of Figure~\ref{fig:exampleDFASpanner} may extract the span-tuple that corresponds to $\ta \ta \tb \tc \tc \ta \open{\varsx} \ta \tb \ta \close{\varsx} \ta$. This messes up the compression, since the three occurrences of $\ta \ta$ have now become three different factors: $\ta \ta$, $\ta \open{\varsx} \ta$ and $\ta \close{\varsx} \ta$. So it seems that extracting a span-tuple enforces at least a partial decompression of $\mathcal{S}$ (since different occurrences of the same factor need to be treated differently).
\par
The technical challenge that we face also becomes clear by a comparison to the approach of~\cite{AmarilliEtAl2019} (for spanner evaluation in the uncompressed case), which first computes in the preprocessing \emph{one} data structure that represents the whole solution set (i.\,e., the product graph of spanner and document), and then the enumeration is done by systematically searching this data structure (with the help of
additional, pre-computed information). Since each position of the document might be the start or end position of some extracted span, it is difficult to imagine such a data structure that is not at least as large as the whole document. Therefore, this approach seems impossible in our setting.
\par
In our approach, we enumerate $\SLP$s that represent marked variants of the document. As illustrated above, these $\SLP$s must be at least partially decompressed. However, since we must only accommodate the at most $2|\varset|$ positions of the document that are start or end positions of the spans of a fixed span tuple, the required decompression is still bounded in terms of the spanner. We show that the breadth of these partially decompressed $\SLP$s is bounded by $\bigO(|\varset|)$. Their depth, however, can be as large as the depth of the input $\SLP$ representing the document. By a well-known 
balancing theorem~\cite{GanardiEtAl2019},
this depth can be assumed to be logarithmic in the size of the (uncompressed) document.

\subsection{Organisation}
Section~\ref{section:preliminaries} fixes basic notation, 
Sections~\ref{sec:spanners} and \ref{section:SLP} provide background on document spanners and
$\SLP$s, respectively.
Section~\ref{section:ModelChecking} is devoted to model checking and checking non-emptiness.
Section~\ref{sec:alogPrelim} develops a tool box that is used in Sections~\ref{sec:computation} and
\ref{sec:enumeration} for computing and for enumerating the result set. 
We conclude in Section~\ref{sec:conclusion}. 
We only provide proof sketches for some results in the main part of
this paper; full proofs for all results can be found in the appendix.

\section{Basic Definitions}\label{section:preliminaries}

Let $\mathbb{N} = \{1, 2, 3, \ldots\}$ and $[n] = \{1, 2, \ldots, n\}$ for $n \in \mathbb{N}$.
For a (partial) mapping $f : X \to Y$, we write $f(x) = \undefin$ for some $x \in X$ to denote that $f(x)$ is not defined; and we set $\Dom{f} = \{x \mid f(x) \neq \undefin\}$.
By $\mathcal{P}(A)$ we denote the power set of a set $A$, and $A^+$ denotes the set of non-empty words over $A$, and $A^* = A^+ \cup \{\eword\}$, where $\eword$ is the empty word. For a word $w \in A^*$, $|w|$ denotes its length (in particular, $\card{\eword} = 0$), and for every $b \in A$, $|w|_{b}$ denotes the number of occurrences of $b$ in $w$. A word $v \in A^+$ is a \emph{factor} of a word $w \in \Sigma^+$ if there are $u_1, u_2 \in \Sigma^*$ with $w = u_1 v u_2$. \par
For all our algorithmic considerations, we assume the RAM-model with logarithmic word-size as our computational model.\par
A \emph{nondeterministic finite automaton} ($\NFA$ for short) is a tuple $M = (Q, \Sigma, \delta, q_0, F)$ with a finite set $Q$ of states, a finite alphabet $\Sigma$, a start state $q_0 \in Q$, a set $F \subseteq Q$ of accepting states and a transition function $\delta : Q \times (\Sigma \cup \{\eword\}) \to \mathcal{P}(Q)$. We also interpret $\NFA$ as directed, edge-labelled graphs in the obvious way.\par 
We extend the transition function to $\delta : Q \times \Sigma^* \to \mathcal{P}(Q)$ in the usual way, i.\,e., for $w \in \Sigma^*$, $x \in \Sigma \cup \{\eword\}$ and $p \in Q$, we set $\delta(p, wx) = \bigcup_{q \in \delta(p, w)} \delta(q, x)$. If $M$ and its transition function $\delta$ is clear from the context, we also write $p \transfuncNFA{w} q$ to express that $q \in \delta(p, w)$. In particular, we also write $p \transfuncNFA{v} q \transfuncNFA{w} r$ instead of $p \transfuncNFA{v} q$ and $q \transfuncNFA{w} t$, and we write $p \transfuncNFA{w} F$ to denote that there is some $q \in F$ with $p \transfuncNFA{w} q$. A word $w \in \Sigma^*$ is \emph{accepted} by $M$ if $q_0 \transfuncNFA{w} F$; and $\lang(M) = \{w \mid q_0 \transfuncNFA{w} F\}$ is the \emph{language accepted by $M$}.\par
An $\NFA$ $M = (Q, \Sigma, \delta, q_0, F)$ is a \emph{deterministic finite automaton} ($\DFA$ for short) if, for every $p \in Q$ and $x \in \Sigma \cup \{\eword\}$, $\delta(p, x) = \emptyset$ if $x = \eword$, and $|\delta(p, x)| \leq 1$ if $x \in  \Sigma$. In this case we view $\delta$ as a function from $Q \times \Sigma$ to $Q$, and we extend it to $\delta : Q \times \Sigma^* \to Q$ by setting $\delta(p, wx) = \delta(\delta(p, w), x)$ and we write $p \transfuncNFA{w} q$ to denote $\delta(p, w) = q$.\par
The size $|M|$ of an $\NFA$ is the number of its transitions. As a convention for the rest of the paper, we always assume for $\NFA$ that $Q = \{\State{1}, \State{2}, \ldots, \State{\states}\}$, for some $\states \in \mathbb{N}$, and $q_0 = \State{1}$. In particular, this means that $\states = |Q|$ throughout the rest of this paper.

\section{Document Spanners}\label{sec:spanners}

Let $\Sigma$ be a terminal alphabet of constant size, and in the following we call words $\doc \in \Sigma^*$ \emph{documents}. For a document $\doc \in \Sigma^*$, we denote by $\docSize$ its \emph{length} and for every $i, j \in [\docSize {+} 1]$ with $i \leq j$, $\spann{i}{j}$ is a \emph{span of $\doc$} and its \emph{value}, denoted by $\doc\spann{i}{j}$, is the substring of $\doc$ from symbol $i$ to symbol $j{-}1$. 
The special case $\doc\spann{i}{i + 1}$ is denoted by $\doc[i]$. $\spans(\doc)$ denotes the set of spans of $\doc$, and by $\spans$ we denote the set of spans for any document, i.\,e., $\{\spann{i}{j} \mid i, j \in \mathbb{N}, i \leq j\}$ (elements from $\spans$ are simply called \emph{spans}). \par
For a finite set of variables $\varset$, an \emph{$(\varset, \doc)$-tuple} is a partial function $\varset \to \spans(\doc)$
For simplicity, we usually denote $(\varset, \doc)$-tuples in tuple-notation, for which we assume an order on $\varset$ and use the symbol $\undefin$ for undefined variables, e.\,g., $(\spann{1}{5}, \undefin, \spann{5}{7})$ describes a $(\{\varsx_1, \varsx_2, \varsx_3\}, \doc)$-tuple that maps $\varsx_1$ to $\spann{1}{5}$, $\varsx_3$ to $\spann{5}{7}$, and is undefined for $\varsx_2$. Since the dependency on the document $\doc$ is often negligible, we also use the term \emph{$\varset$-tuple} (or 
\emph{span-tuple} (\emph{over $\varset$})) to denote an $(\varset, \doc)$-tuple.\par
We also define an obvious set-representation of span-tuples that will be convenient in the context of this work. For any set $\varset$ of variables, we use a special alphabet $\Gamma_{\varset} = \{\open{\varsx}, \close{\varsx} \mid \varsx \in \varset\}$. This alphabet shall play an important role in the remainder of this work; its elements are also called \emph{markers}. For any $(\varset, \doc)$-tuple $t$, its \emph{marker set} $\setrep{t} \subseteq \varset \times [\docSize + 1]$ is defined as $\setrep{t} = \{(\open{\varsx}, i), (\close{\varsx}, j) \mid t(\varsx) = \spann{i}{j}, \varsx \in \Dom{t}\}$. It is obvious that there is a one-to-one correspondence between span-tuples and their marker sets.\par
An \emph{$(\varset, \doc)$-relation} (or \emph{$\varset$-relation} if the dependency on $\doc$ is negligible) is a set of $(\varset, \doc)$-tuples. 
As a measure of the size of a reasonable representation of an $(\varset, \doc)$-relation $R$ we use $\size{R} = \card{\varset}{\cdot} \card{R}$.\par
A \emph{spanner} (\emph{over terminal alphabet $\Sigma$ and variables $\varset$}) is a function that maps every document $\doc \in \Sigma^*$ to an $(\varset, \doc)$-relation (note that the empty relation $\emptyset$ is also a valid image of a spanner). \par
We next introduce some terminology that will be crucial for reasoning about spanners and span-tuples. We follow the common approach in the literature to represent a pair of document $\doc$ and span-tuple $t$ as a single word (which will be called \emph{subword-marked word}) by means of special marker symbols that are inserted into the document (for which we use the symbols of $\Gamma_{\varset}$). For example $\doc = \ta \tb \ta \tb$ and span-tuple $t$ with $t(\varsx) = \spann{2}{4}$ and $t(\varsy) = \spann{3}{5}$ can be represented by the subword-marked word $\ta \open{\varsx} \tb \open{\varsy} \ta \close{\varsx} \tb\close{\varsy}$.

\subsection{Subword-Marked Words}
For any set $\varset$ of variables, we shall use the set $\Gamma_{\varset} = \{\open{\varsx}, \close{\varsx} \mid \varsx \in \varset\}$ and its powerset as alphabets. The intuitive meaning of an occurrence of symbol $\open{\varsx}$ (or $\close{\varsx}$) at position $i$ is that the span of variable $\varsx$ starts at position $i$ (or ends at position $i$, respectively). If spans of several variables start or end at the same position, we encode this by using a subset of $\Gamma_{\varset}$ as a single symbol.\par

\begin{definition}\label{subwordMarkedWordDef}
A \emph{subword-marked word} (\emph{over $\Sigma$ and $\varset$}) is a word 
$w = A_1 b_1 A_2 b_2 \ldots A_n b_n A_{n + 1}$
with $b_i \in \Sigma$ for every $i \in [n]$, and $A_{i'} \in \powerset{\Gamma_{\varset}}$ for every $i' \in [n + 1]$, that satisfies the properties:
\begin{enumerate}[\ (1)]
\item[$\bullet$] for all distinct $i, j \in [n + 1]$, $A_i \cap A_j = \emptyset$,
\item[$\bullet$] if $\open{\varsx} \in A_i$ and $\close{\varsx} \in A_j$ for $\varsx \in \varset$, then $i \leq j$,
\item[$\bullet$] for all $\varsx \in \varset$, $\{\open{\varsx}, \close{\varsx}\}$ is contained in or disjoint from $\bigcup^{n + 1}_{i = 1} A_i$.
\end{enumerate}
\end{definition}

We define the \emph{document-length} of $w$ as $\doclength{w} = n$ (note that the actual length of $w$ is $|w| = 2\doclength{w} + 1$; the document-length will be the more relevant size measure for us). 
For convenience, we also omit symbols $A_i$ if they are the empty set. \par
We claimed above that subword-marked words represent a document and a span-tuple as a single word. We shall now substantiate this interpretation of subword-marked words by defining the function $\getWord{\cdot}$ that retrieves the document and the function $\markedpositions{\cdot}$ that retrieves the span-tuple (as marker set) encoded by a subword-marked word. 
To this end, let $w = A_1 b_1 A_2 b_2 \ldots A_n b_n A_{n + 1}$ be a subword-marked word over $\Sigma$ and $\varset$. By $\getWord{w}$, we denote the document over $\Sigma$ obtained by erasing all occurrences of symbols from $\powerset{\Gamma_{\varset}}$ from $w$, i.\,e., $\getWord{w} = b_1 b_2 \ldots b_n$ (note that $\doclength{w} = |\getWord{w}|$). 
Furthermore, let $\markedpositions{w}$ be the set $\{(\sigma, i) \mid \sigma \in A_i, i \in [n + 1]\}$. It can be easily seen that $\markedpositions{w}$ is the marker set $\setrep{t}$ of an $(\varset, \getWord{w})$-tuple $t$.\par
For given document $\doc$ and an $(\varset, \doc)$-tuple $t$, it is obvious how to construct a subword-marked word $w$ with $\getWord{w} = \doc$ and $\markedpositions{w} = \setrep{t}$. We will nevertheless formally define this. 
For any $(\varset, \doc)$-tuple $t$, we denote by $\insertmarkers{\doc}{\setrep{t}}$ the word $A_1 b_1 A_2 b_2 \ldots A_{\docSize} b_{\docSize} A_{\docSize + 1}$, where $b_i = \doc[i]$ for every $i \in [\docSize]$, and, for every $i' \in [\docSize + 1]$, $A_{i'} = \{\sigma \mid (\sigma, i') \in \setrep{t}\}$. It can be easily seen that $\insertmarkers{\doc}{\setrep{t}}$ is in fact a subword-marked word with $\getWord{w} = \doc$ and $\markedpositions{w} = \setrep{t}$. \par
\begin{figure}
\begin{center}
\scalebox{1}{\includegraphics{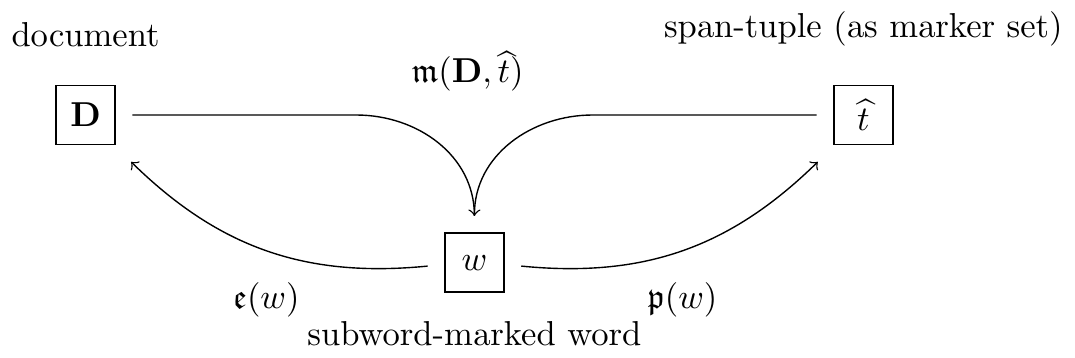}}
\end{center}
\caption{How documents, subword-marked words (marked words), and span-tuples (marker sets) relate to each other via $\getWord{\cdot}$, $\markedpositions{\cdot}$ and $\insertmarkers{\cdot}{\cdot}$.}
\label{fig:mappingsDiagram}
\end{figure}

Let us illustrate these definitions with a brief example (see also Figure~\ref{fig:mappingsDiagram} for an illustration of the mappings $\getWord{\cdot}$, $\markedpositions{\cdot}$ and $\insertmarkers{\cdot}{\cdot}$ that translate between the different representations). 

\begin{example}\label{firstSubwordMarkedExample}
Let $\Sigma = \{\ta, \tb, \tc\}$ and let $\varset = \{\varsx, \varsy, \varsz\}$. Then 
\begin{align*}
w = \{\open{\varsx}\} \ta \tb \{\open{\varsy}, \open{\varsz}, \close{\varsx}\} \tb \tc \{\close{\varsz}\} \ta \tb \{\close{\varsy}\} \ta \tc
\end{align*}
is a 
subword-marked word with $\getWord{w} = \ta \tb \tb \tc \ta \tb \ta \tc$ and
\begin{align*}
\markedpositions{w} = \{(\open{\varsx}, 1), (\close{\varsx}, 3), (\open{\varsy}, 3), (\close{\varsy}, 7), (\open{\varsz}, 3), (\close{\varsz}, 5)\}\,,
\end{align*}
where $\markedpositions{w}$ is the set representation of $(\spann{1}{3}, \spann{3}{7}, \spann{3}{5})$.\par
Moreover, for $\doc = \ta \ta \ta \tb \tc \tb \tb$ and $t = (\spann{6}{8}, \undefin, \spann{3}{8})$, we have 
$\insertmarkers{\doc}{\setrep{t}} = \ta \ta \{\open{\varsz}\} \ta \tb \tc \{\open{\varsx}\} \tb \tb \{\close{\varsx}, \close{\varsz}\}$.
\end{example}

In the following, if this causes no confusion, we shall also use span-tuples and their marker sets interchangeably.

\subsection{Regular Spanners} 
A set $L$ of subword-marked words (over $\Sigma$ and $\varset$) is a \emph{subword-marked language} (\emph{over $\Sigma$ and $\varset$}).
Since every subword-marked word $w$ over $\Sigma$ and $\varset$ describes the $(\varset, \getWord{w})$-tuple $\markedpositions{w}$, a subword-marked language $L$ can be interpreted as a spanner $\llbracket L \rrbracket$ (over $\Sigma$ and $\varset$) as follows: for every $\doc \in \Sigma^*$, $\llbracket L \rrbracket(\doc) = \{\markedpositions{w} \mid w \in L, \getWord{w} = \doc\}$. 

\begin{proposition}\label{basicSubwordMarkedLanguageProposition}
Let $L$ be a subword-marked language over $\Sigma$ and $\varset$, let $\doc \in \Sigma^*$ and let $t$ be an $(\varset, \doc)$-tuple. Then $t \in \llbracket L \rrbracket(\doc)$ if and only if $\insertmarkers{\doc}{t} \in L$.
\end{proposition}

\begin{figure}
\begin{center}
\scalebox{1.5}{\includegraphics{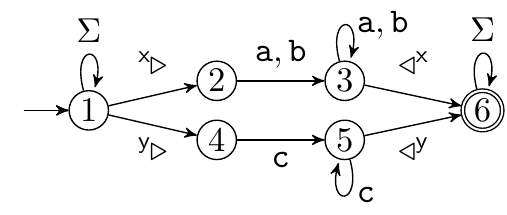}}
\end{center}
\caption{A $\DFA$ that represents a $(\{\ta, \tb, \tc\}, \{\varsx, \varsy\})$ spanner ($1$~is the initial and $6$ is the only accepting state).}
\label{fig:exampleDFASpanner}
\end{figure}

A spanner $S$ over $\Sigma$ and $\varset$ is called a \emph{regular $(\Sigma, \varset)$-spanner} (or simply \emph{$(\Sigma, \varset)$-spanner}) if $S = \llbracket L \rrbracket$ for some regular subword-marked language $L$ over $\Sigma$ and $\varset$. We will represent $(\Sigma, \varset)$-spanners as $\NFA$s or $\DFA$s accepting subword-marked languages (see Figure~\ref{fig:exampleDFASpanner} for an example). For the sake of conciseness, we do not explicitly mention the alphabet $\powerset{\Gamma_{\varset}}$ for such automata
over $\Sigma$ and $\varset$, 
i.\,e., we denote them by $M = (Q, \Sigma, \State{1}, \delta, F)$ but have in mind that $\Sigma$ has to be replaced by $\Sigma\cup\powerset{\Gamma_{\varset}}$. 
We will write $\llbracket M \rrbracket$ instead of $\llbracket L(M) \rrbracket$.\par

\begin{remark}
For $\NFA$ $M$ that accept subword-marked languages over $\Sigma$ and $\varset$, we assume that for given $i, j \in [\states]$ and $y \in \Sigma \cup \powerset{\Gamma_{\varset}}$, we can check whether $j \in \delta(i, y)$ in constant time. Moreover, we also assume that we can iterate through $M$'s set of arcs in time $\bigO(|M|)$.
\end{remark}

\subsection{Representations of Regular Spanners}
In the initial paper~\cite{FaginEtAl2015}, regular spanners were represented by so-called \emph{variable-set automata} ($\VA$, for short). In our terminology, $\VA$s are $\NFA$s that accept subword-marked languages with the difference that consecutive marker symbols are explicitly represented as sequences and not merged into sets. As a result, a document and a span-tuple do not describe a subword-marked word in a unique way (i.\,e., the function $\insertmarkers{\cdot}{\cdot}$ is not well-defined), which means that for solving model checking according to Proposition~\ref{basicSubwordMarkedLanguageProposition}, we potentially need to consider an exponential number of subword-marked words. This is a well-known problem and can be dealt with by restricting spanners to be functional (i.\,e., span-tuples are total functions)~\cite{FreydenbergerEtAl2018, FlorenzanoEtAl2018}, by imposing a fixed order on sequences of marker symbols in the subword-marked words~\cite{SchmidSchweikardt2021, DoleschalEtAl2019}, or by using sets of marker symbols as symbols, as done for \emph{extended} $\VA$s~\cite{FlorenzanoEtAl2018, AmarilliEtAl2019} and also in this paper.\par
It is well-known
that the $\VA$s of~\cite{FaginEtAl2015} can be transformed into extended $\VA$s, or into $\VA$s with an order on the marker symbols, or into $\NFA$s for subword-marked languages (in the way defined here); see, e.\,g.,~\cite{FlorenzanoEtAl2018, AmarilliEtAl2019}. However, these translations cause an exponential size blow-up in the worst-case (this is formally proven in~\cite{FlorenzanoEtAl2018}), except for functional $\VA$s (on the other hand, functionality is a proper restriction compared to non-functional regular spanners). \par
We present our results in a way that abstracts from these well-documented issues of conversions between different representations of regular spanners, since they would distract from the actual story of this paper, which is spanner evaluation on compressed documents. In order to extend our results to other spanner formalisms, one has to keep in mind the overhead of translations between formalisms (which 
affects the combined complexity, but not the data complexity).

\section{SLP-Compressed Documents}\label{section:SLP}

We now formally describe the concept of straight-line programs ($\SLP$s, for short), that has already been discussed in the introduction. 

\subsection{Straight-Line Programs}
A \emph{context-free grammar} is a tuple $G = (N, \Sigma, R, S_0)$, where $N$ is the set of \emph{non-terminals}, $\Sigma$ is the \emph{terminal alphabet}, $S_0 \in N$ is the \emph{start symbol} and $R \subseteq N \times (N \cup \Sigma)^+$ is the set of \emph{rules} (as a convention, we write rules $(A, w) \in R$ also in the form $A \to w$). A context-free grammar $\mathcal{S} = (N, \Sigma, R, S_0)$ is a \emph{straight-line program} ($\SLP$) if $R$ is a total function $N \to (N \cup \Sigma)^+$ and the relation $\{(A, B) \mid (A, w) \in R, |w|_{B} \geq 1\}$ is acyclic. In this case, for every $A \in N$, let $\derivationstepsub{\mathcal{S}}{A}$ be the unique $w \in (N \cup \Sigma)^+$ such that $(A, w) \in R$, and let $\derivationstepsub{\mathcal{S}}{a} = a$ for every $a \in \Sigma$; we also call $A \to \derivationstepsub{\mathcal{S}}{A}$ \emph{the rule for $A$}.
For an $\SLP$ $\mathcal{S} = (N, \Sigma, R, S_0)$, we extend $\derivationstepsymbol_{\mathcal{S}}$ to a morphism $(N \cup \Sigma)^+ \to (N \cup \Sigma)^+$ by setting $\derivationstepsub{\mathcal{S}}{\alpha_1 \ldots \alpha_n} = \derivationstepsub{\mathcal{S}}{\alpha_1} \ldots \derivationstepsub{\mathcal{S}}{\alpha_n}$, for $\alpha_i \in (N \cup \Sigma)$, $1 \leq i \leq n$. 
Furthermore, for every $\alpha \in (N \cup \Sigma)^+$, we set $\derivationstepsymbol^1_{\mathcal{S}}(\alpha) = \derivationstepsub{\mathcal{S}}{\alpha}$, $\derivationstepsymbol^k_{\mathcal{S}}(\alpha) = \derivationstepsub{\mathcal{S}}{\derivationstepsymbol^{k-1}_{\mathcal{S}}(\alpha)}$, for every $k \geq 2$; and $\derivsub{\mathcal{S}}{\alpha} = \derivationstepsymbol^{|N|}_{\mathcal{S}}(\alpha)$ is the \emph{derivative} of $\alpha$. By definition, $\derivsub{\mathcal{S}}{\alpha} \in \Sigma^+$ for every $\alpha \in (N \cup \Sigma)^+$.\par
The \emph{depth} of a non-terminal $A \in N$ is defined by $\depth{A} = \min\{k \mid \derivationstepsymbol^k_{\mathcal{S}}(A) = \derivsub{\mathcal{S}}{A}\}$, and the \emph{depth} of $\mathcal{S}$ is $\depth{\mathcal{S}} = \depth{S_0}$. The \emph{size} of $\mathcal{S}$ is defined by $\size{\mathcal{S}} = |N| + \sum_{A \in N} \card{\derivationstepsub{\mathcal{S}}{A}}$. If the $\SLP$ under consideration is clear from the context, we also drop the subscript $\mathcal{S}$. Moreover, we set $\deriv{\mathcal{S}} = \deriv{S_0}$ and say that \emph{$\mathcal{S}$ is an $\SLP$ for (the word or document) $\deriv{\mathcal{S}}$}. We view $\mathcal{S}$ as a compressed representation of the document $\deriv{\mathcal{S}}$. \par
\begin{figure}
\begin{center}
\scalebox{1}{\includegraphics{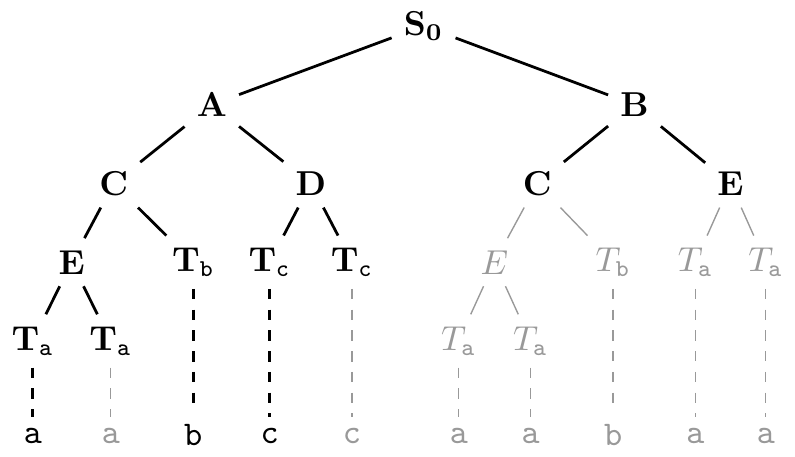}}
\end{center}
\caption{The derivation tree of the $\SLP$ from Example~\ref{normalFormSLPExample}; the actual rules of the $\SLP$ are shown in bold.}
\label{fig:exampleDerivTreeCNF}
\end{figure}

The \emph{derivation tree} of an $\SLP$ $\mathcal{S} = (N, \Sigma, R, S_0)$ is a ranked ordered tree with node-labels from $\Sigma \cup N$, inductively defined as follows. The root is labelled by $S_0$ and every
 node labelled by $A \in N$ with 
$\derivationstepsub{\mathcal{S}}{A}
= \alpha_1 \alpha_2 \ldots \alpha_n$ has $n$ children 
labelled by $\alpha_1, \alpha_2, \ldots, \alpha_n$ in exactly this order. We note that all leaves of the derivation tree are from $\Sigma$, and spelling them out from left to right yields exactly $\deriv{S_0}$; moreover, the depth of the derivation tree is exactly $\depth{\mathcal{S}}$. See Figure~\ref{fig:exampleDerivTreeCNF} for an example of a derivation tree. We stress the fact that the derivation tree of an $\SLP$ $\mathcal{S}$ is a \emph{non-compressed} representation of $\deriv{\mathcal{S}}$. In particular, algorithms on $\SLP$-compressed strings cannot afford to explicitly build the full derivation tree.

\begin{example}\label{definitionsExample}
Let $\mathcal{S} = (N, \Sigma, R, S_0)$ be an $\SLP$ with $N = \{S_0, A, B\}$, $\Sigma = \{\ta, \tb\}$, and $R = \{S_0 \to A \tb \ta A B \tb, A \to B \ta B, B \to \tb \ta \ta \tb\}$.
By definition, $\deriv{B} = \tb \ta \ta \tb$, $\deriv{A} = \deriv{B} \ta \deriv{B} = \tb \ta \ta \tb \ta \tb \ta \ta \tb$ and 
$\deriv{\mathcal{S}} = \deriv{S_0} = \tb \ta \ta \tb \ta \tb \ta \ta \tb \tb \ta \tb \ta \ta \tb \ta \tb \ta \ta \tb\tb \ta \ta \tb \tb$.
Thus, $\mathcal{S}$ is an $\SLP$ for 
\begin{equation*}
\tb \ta \ta \tb \ta \tb \ta \ta \tb \tb \ta \tb \ta \ta \tb \ta \tb \ta \ta \tb \tb \ta \ta \tb \tb\,. 
\end{equation*}
In particular, we note that $\size{\mathcal{S}} = 16 < 25 = \card{\deriv{\mathcal{S}}}$.
\end{example}

From now on, we shall always denote the document compressed by the $\SLP$ by $\doc$ (i.\,e., $\deriv{\mathcal{S}} = \doc$ for the $\SLP$s $\mathcal{S}$ that we consider). Recall that we denote by $\docSize$ the size of $\doc$.\par
An $\SLP$ $\mathcal{S} = (N, \Sigma, R, S_0)$ is in \emph{Chomsky normal form} if, for every $A \in N$, $\derivationstepsub{\mathcal{S}}{A} \in (\Sigma \cup N^2)$, and $\mathcal{S}$ is \emph{$c$-balanced} for some $c \in \mathbb{N}$ if $\depth{\mathcal{S}} \leq c \log(\docSize)$. We note that if $\mathcal{S}$ is in Chomsky normal form, then $\size{\mathcal{S}} = 3|N|$.
We say that an $\SLP$ is in \emph{normal form} if it is in Chomsky normal form and, for every $x \in \Sigma$, $T_x$ is the unique non-terminal with rule $T_x \to x$. We call the $T_x$ \emph{leaf non-terminals} and all other $A \in N \setminus \{T_x \mid x \in \Sigma\}$ \emph{inner non-terminals}. For $\SLP$s in normal form, we let the leaf non-terminals be the leaves of derivation trees. 
From now on,
we assume that all $\SLP$s are in normal form.

\begin{example}\label{normalFormSLPExample}
Let $\mathcal{S} = (N, \Sigma, R, S_0)$ be a normal form $\SLP$ with $N = \{S_0, A, B, C, D, E, T_{\ta}, T_{\tb}, T_{\tc}\}$, $\Sigma = \{\ta, \tb, \tc\}$, and $R = \{S_0 \to A B, A \to C D, B \to C E, C \to E T_{\tb}, D \to T_{\tc} T_{\tc}, E \to T_{\ta} T_{\ta}\} \cup \{T_{x} \to x \mid x \in \Sigma\}$. Figure~\ref{fig:exampleDerivTreeCNF} shows the derivation tree of $\mathcal{S}$. It can be easily verified that $\deriv{\mathcal{S}} = \ta \ta \tb \tc \tc \ta \ta \tb \ta \ta$.
\end{example}

\subsection{Further Properties of SLPs} 
The size of an $\SLP$ can be logarithmic in the size of the document, e.\,g., strings $\ta^{2^n}$ can be represented by $n + 1$ rules of the form $S \to A_1 A_1, A_1 \to A_2 A_2, \ldots, A_n \to \ta$. On the other hand, it can be shown that $\log\mathbf{d}$ is also an asymptotic lower bound for $\size{\mathcal{S}}$ (see~\cite[Lemma~1]{Chaetal2005}).
Another important parameter is $\depth{\mathcal{S}}$. E.\,g., finding in an $\SLP$ the $i^{\text{th}}$ symbol $\doc[i]$ of the document represented by $\mathcal{S}$ can be achieved by a top-down traversal of the derivation tree, which depends on $\depth{\mathcal{S}}$. For $\SLP$s with a constant branching factor (like $\SLP$s in normal form), $\depth{\mathcal{S}}$ is also lower bounded by $\log\docSize$. This optimum is achieved by balanced $\SLP$s and the following theorem shows that it is in fact without loss of generality to assume $\SLP$s to be balanced:

\begin{theorem}[$\SLP$ Balancing Theorem, Ganardi, Jez and Lohrey~\cite{GanardiEtAl2019}]\label{balancingTheorem}
There is a $c \in \mathbb{N}$ such that any given $\SLP$ $\mathcal{S}$ for document $\doc$ can be transformed in time $\bigO(\size{\mathcal{S}})$ into a $c$-balanced $\SLP$ $\mathcal{S}'$ for $\doc$ in Chomsky normal form with $\size{\mathcal{S}'} = \bigO(\size{\mathcal{S}})$.
\end{theorem}

Theorem~\ref{balancingTheorem} means that whenever a factor $\depth{\mathcal{S}}$ occurs in the running time, which, in the general case, can only be upper bounded by $\size{\mathcal{S}}$, it can be replaced by $\log\docSize$, which corresponds to $\size{\mathcal{S}}$ in the best-case compression scenario. For clarity, we nevertheless mention any dependency on $\depth{\mathcal{S}}$ in our results.\par
In the field of algorithmics on ($\SLP$-)compressed strings, it is common to assume the word-size of the underlying RAM-model to be logarithmic in $\docSize$, where $\docSize$ is the size of the \emph{non-compressed} input. This means that we can perform arithmetic operations on the positions of $\doc$ in constant time. In particular, we state the following fact, which is easy to show and well-known in the context of $\SLP$s.

\begin{lemma}\label{computeDerivationLengthsLemma}
Given an $\SLP$ $\mathcal{S}$, we can compute all the numbers $\card{\deriv{A}}$ for all non-terminals $A$ within time $\bigO(\size{\mathcal{S}})$.
\end{lemma}

\subsection{SLPs and Finite Automata}

A classical 
task in the context of algorithmics on $\SLP$-compressed strings is to check membership of an $\SLP$-compressed document $\doc$ to a given regular language $L$. It is intuitively clear that algorithms for our spanner evaluation tasks (see Section~\ref{sec:intro}) will necessarily also implicitly solve this task in some way. For example, given an $\SLP$ for $\doc$ and an $\NFA$ $M$ over $\Sigma$, checking if $\doc \in \lang(M)$ reduces to the model checking task $\emptyset \in \llbracket M \rrbracket(\doc)$. Hence, 
we discuss checking membership of $\SLP$s to regular languages in a bit more detail.\par
Let $\mathcal{S}$ be an $\SLP$ for $\doc$ and let $M$ be an $\NFA$ with $\states$ states.
The general idea is to compute, for each $A \in N$, a Boolean $(\states \times \states)$ matrix $M_A$ whose entries indicate from which state we can reach which state by reading $\deriv{A}$. This can be done recursively along the structure of $\mathcal{S}$: the matrices $M_{T_x}$ for the leaf non-terminals are directly given by $M$'s transition function, and for every inner non-terminal $A \in N$ with a rule $A \to BC$, we have $M_A = M_B \cdot M_C$ (where $\cdot$ denotes the usual Boolean matrix multiplication). This yields the following well-known result, that has been formally stated at several places in the literature (see, e.\,g.,~\cite{PlandowskiRytter99,MarkeySchnoebelen2004,Lohrey2012}):

\begin{lemma}\label{SLPNFAMembershipLemma}
Let $\mathcal{S}$ be an $\SLP$ for $\doc$ and let $M$ be an $\NFA$ with $\states$ states. Then we can check whether $\doc \in \lang(M)$ in time $\bigO(\size{\mathcal{S}}\states^3)$.
\end{lemma}

With a fast Boolean matrix multiplication algorithm that runs in time $\bigO(\states^{\omega})$, Lemma~\ref{SLPNFAMembershipLemma} can be improved to $\bigO(\size{\mathcal{S}}\states^{\omega})$. In fact, the best known upper bound 
is $\bigO(\min\{\size{\mathcal{S}}\states^{\omega}, |\doc|\states^{2}\})$ (the latter running-time is achieved by explicitly constructing $\doc$). However, for ``combinatorial algorithms'', this bound simplifies to $\bigO(\min\{\size{\mathcal{S}}\states^{3}$, $|\doc|\states^{2}\})$ and it is shown in~\cite{AbboudEtAl2017} 
that, conditional to the so-called combinatorial $k$-Clique conjecture, this is optimal in the sense that there is no ``combinatorial algorithm'' with running-time 
$\bigO(\min\{\size{\mathcal{S}}\states^{3}, |\doc|\states^{2}\}^{1 - \epsilon})$ for any $\epsilon > 0$.\par

\section{Non-Emptiness and Model Checking}\label{section:ModelChecking}

In this section, we consider the non-emptiness and the model checking problem (see Section~\ref{sec:intro}), which can be reduced to the problem of checking membership of an $\SLP$-compressed document to a regular language. Here, we provide a sketch of how this can be done.\par
For checking if $\llbracket M \rrbracket(\doc) \neq \emptyset$, it suffices to check whether $M$ can accept a subword-marked word $w$ with $\getWord{w} = \doc$. This can be easily done by treating all $\powerset{\Gamma_{\varset}}$-transitions of $M$ as $\eword$-transitions and then simply check membership of $\doc$ by using Lemma~\ref{SLPNFAMembershipLemma}. \par
For checking if $t \in \llbracket M \rrbracket(\doc)$ for a given span-tuple $t$, we proceed as follows. We transform the $\SLP$ $\mathcal{S}$ for $\doc$ into an $\SLP$ $\mathcal{S}'$ for the subword-marked word $w = \insertmarkers{\doc}{t}$ (recall from Section~\ref{sec:spanners} that $\getWord{w} = \doc$ and $\markedpositions{w} = t$). Since $t \in \llbracket M \rrbracket(\doc)$ if and only if $w \in L(M)$ (see Proposition~\ref{basicSubwordMarkedLanguageProposition}), it suffices to check whether $\deriv{\mathcal{S}'} \in L(M)$ (for which we can rely again on Lemma~\ref{SLPNFAMembershipLemma}). The only question left is how to construct $\mathcal{S}'$, and this can be done as follows. For every $i \in [\docSize]$ such that there is at least one $(\sigma, i) \in \setrep{t}$, we compute the set $\Lambda_i = \{\sigma \mid (\sigma, i) \in \setrep{t}\}$. Note that there are at most $2|\varset|$ such sets, and these can be easily obtained from $\setrep{t}$ in time $\bigO(|\varset|)$. Then, for each such set $\Lambda_i$, we traverse the derivation tree of $\mathcal{S}$ top-down in order to find the leaf corresponding to position $i$ (for this, the numbers $\card{\deriv{A}}$ are essential, which we can compute according to Lemma~\ref{computeDerivationLengthsLemma}). Then we add the symbol $\Lambda_i$ at this position, but, since this changes the meaning of all the non-terminals of this root-to-leaf path, we have to introduce $\depth{\mathcal{S}}$ new non-terminals.
Overall, we only add $\bigO(|\varset|\depth{\mathcal{S}})$ new non-terminals to $\mathcal{S}$; in particular, we never have to construct the whole derivation tree, but at most $2|\varset|$ paths of length $\depth{\mathcal{S}}$.
This leads to:

\begin{theorem}\label{basicEvalTasksTheorem}
Let $\mathcal{S}$ be an $\SLP$ for $\doc$, let $M$ be an $\NFA$ that represents a $(\Sigma, \varset)$-spanner, and let $t$ be an $(\varset, \doc)$-tuple. Checking whether
\begin{enumerate}
\item $\llbracket M \rrbracket(\doc) \neq \emptyset$ can be done in time $\bigO(|M| + \size{\mathcal{S}} \states^3)$.
\item $t \in \llbracket M \rrbracket(\doc)$ can be done in time $\bigO((\size{\mathcal{S}} + |\varset| \depth{\mathcal{S}})  \states^3)$.
\end{enumerate}
\end{theorem}

\section{Algorithmic Preliminaries}\label{sec:alogPrelim}

In this section, we develop a tool box for spanner evaluation over $\SLP$s.
On the conceptional side, we first extend 
our definitions from Section~\ref{sec:spanners} to the case of 
incomplete (or partial) span-tuples (which is necessary to reason about the subwords of the document compressed by single non-terminals of the $\SLP$). Then, we present a sequence of lemmas that allow us to regard the solution set $\llbracket M \rrbracket(\doc)$ as being decomposed according to the recursive structure of the $\SLP$. This point of view will be crucial both for the task of computing (Section~\ref{sec:computation}) and of enumerating (Section~\ref{sec:enumeration}) the set $\llbracket M \rrbracket(\doc)$.

\subsection{Representations of Partial Span-Tuples}

Recall Example~\ref{firstSubwordMarkedExample} for document $\doc = \ta \tb \tb \tc \ta \tb \ta \tc$: 
\begin{align*}
w =\:&\{\open{\varsx}\} \ta \tb \{\open{\varsy}, \open{\varsz}, \close{\varsx}\} \tb \tc \{\close{\varsz}\} \ta \tb \{\close{\varsy}\} \ta \tc\,,\\
\markedpositions{w} =\:&\{(\open{\varsx}, 1), (\close{\varsx}, 3), (\open{\varsy}, 3),(\close{\varsy}, 7), (\open{\varsz}, 3), (\close{\varsz}, 5)\}\,.
\end{align*}
If we consider the factorisation $\doc = \doc_1 \doc_2$ with $\doc_1 = \ta \tb \tb$ and $\doc_2 = \tc \ta \tb \ta \tc$, then this corresponds to the factorisation $w = w_1 w_2$ with $w_1 = \{\open{\varsx}\} \ta \tb \{\open{\varsy}, \open{\varsz}, \close{\varsx}\} \tb$ and $w_2 = \tc \{\close{\varsz}\} \ta \tb \{\close{\varsy}\} \ta \tc$.
Technically, neither $w_1$ nor $w_2$ are subword-marked words. However, it can be easily seen that  the functions $\getWord{\cdot}$ and $\markedpositions{\cdot}$ are still well-defined and $\getWord{w_1} = \doc_1$, $\getWord{w_2} = \doc_2$,
$\markedpositions{w_1} = \{(\open{\varsx}, 1), (\open{\varsy}, 3), (\open{\varsz}, 3), (\close{\varsx}, 3)\}$, $\markedpositions{w_2} = \{(\close{\varsz}, 2), (\close{\varsy}, 4)\}$. The sets $\markedpositions{w_1}$ and $\markedpositions{w_2}$ are not valid marker sets that describe valid span-tuples, but we can interpret them as representing \emph{partial} span-tuples. Moreover, we can also combine $\markedpositions{w_1}$ and $\markedpositions{w_2}$ in order to obtain the marker set of the whole span-tuple, but we have to keep in mind that $\markedpositions{w_2}$ corresponds to a factor of $\doc$ that is not a prefix and therefore the elements from $\markedpositions{w_2}$ have to be \emph{shifted} to the right by $\card{\doc_1} = 3$ positions. We now formalise these observations.\par
Any factor of a subword-marked word is called a \emph{marked word}.
Since marked words are words $w = A_1 b_1 \ldots A_n b_n A_{n + 1}$ with $b_i \in \Sigma$ and $A_{i'} \in \powerset{\Gamma_{\varset}}$ (except for the possibility that $A_1$ or $A_{n + 1}$ are missing, which we can simply interpret as $A_1 = \emptyset$ or $A_{n+1} = \emptyset$, respectively), the functions $\getWord{\cdot}$ and $\markedpositions{\cdot}$ can be defined in the same way as for subword-marked words, i\,e., $\getWord{w} = b_1 b_2 \ldots b_n$ and $\markedpositions{w} = \{(\sigma, i) \mid \sigma \in A_i, i \in [n + 1]\}$. \par
For any marked word $w$, we call the set $\markedpositions{w}$ a \emph{partial marker set}, and we shall denote partial marker sets by $\Lambda$ in order to distinguish them from span-tuples and from (non-partial) marker sets. \par
As long as a partial marker set $\Lambda$ is \emph{compatible} with a document $\doc$, i.\,e., $\max\{\ell \mid (\sigma, \ell) \in \Lambda\} \leq \docSize + 1$, we can also define $\insertmarkers{\doc}{\Lambda}$ analogously as for non-partial marker sets, i.\,e., $\insertmarkers{\doc}{\Lambda} = A_1 b_1 \ldots A_{\docSize} b_{\docSize} A_{\docSize + 1}$, where $b_i = \doc[i]$ for every $i \in [\docSize]$, and, for every $i' \in [\docSize + 1]$, $A_{i'} = \{\sigma \mid (\sigma, i') \in \Lambda\}$. Note that the diagram of Figure~\ref{fig:mappingsDiagram} still serves as an illustration (we just have to keep in mind that $\setrep{t}$ is now a partial marker set). \par
For any partial marker set $\Lambda$ and any $\ell \in \mathbb{N}$, 
the \emph{$\ell$-rightshift of $\Lambda$}, denoted by $\rightshift{\Lambda}{\ell}$, is the partial marker set $\{(\sigma, k + \ell) \mid (\sigma, k) \in \Lambda\}$.\par

\begin{example}
Let $\Sigma = \{\ta, \tb, \tc\}$, $\varset = \{\varsx, \varsy, \varsz\}$. The partial marker sets $\Lambda_1 = \{(2, \open{\varsy}), (4, \open{\varsz}), (4, \open{\varsx}), (6, \close{\varsz})\}$ and $\Lambda_2 = \{(2, \close{\varsx}), (4, \close{\varsy})\}$, which are compatible with $\doc_1 = \ta \tb \ta \tb \tc \tc$ and $\doc_2 = \tc \ta \tb \ta$, respectively, but are both not marker sets of some span-tuple. Moreover,  
$\insertmarkers{\doc_1}{\Lambda_1} = \ta \{\open{\varsy}\} \tb \ta \{\open{\varsz}, \open{\varsx}\} \tb \tc \{\close{\varsz}\} \tc$, $\insertmarkers{\doc_2}{\Lambda_2} = \tc \{\close{\varsx}\} \ta \tb \{\close{\varsy}\} \ta$. We observe that 
\begin{equation*}
\Lambda = \Lambda_1 \cup \rightshift{\Lambda_2}{\card{\doc_1}} =  \{(2, \open{\varsy}), (4, \open{\varsz}), (4, \open{\varsx}), (6, \close{\varsz}), (8, \close{\varsx}), (10, \close{\varsy})\} 
\end{equation*}
is a marker set for $\doc = \doc_1 \doc_2$, and $\insertmarkers{\doc}{\Lambda} = \insertmarkers{\doc_1}{\Lambda_1}  \insertmarkers{\doc_2}{\Lambda_2}$.
\end{example}

For any subword-marked word $w$ with $\getWord{w} = \doc$ and any factorisation $\doc = \doc_1 \doc_2$, there might be two ways of factorising $w = w_1 w_2$ such that $\getWord{w_1} = \doc_1$ and $\getWord{w_2} = \doc_2$ (i.\,e., depending on whether the symbol from $\powerset{\Gamma_{\varset}}$ at the cut point belongs to $w_1$ or to $w_2$). In order to deal with this issue, we will only consider marked words that end on a symbol from $\Sigma$. This is only possible, if all our subword-marked words are \emph{non tail-spanning}, which means that the final symbol $A_{\doclength{w}+1}$ from $\powerset{\Gamma_{\varset}}$ is empty (and therefore, can be ignored). We say that a subword-marked language $L$ (i.\,e., a spanner) is \emph{non tail-spanning} if every $w \in L$ is non tail-spanning. \par
We assume all regular spanners to be non-tail spanning in the remainder of this paper. Note that this is a very minor restriction: any $\NFA$ $M$ that represents a $(\Sigma, \varset)$-spanner can be easily transformed into an $\NFA$ $M'$ with $\lang(M') = \{w\# \mid w \in \lang(M)\}$ for some $\# \notin \Sigma$. In particular, this means that $\llbracket M' \rrbracket$ is non-tail spanning and, for every document $\doc$, we have $\llbracket M \rrbracket(\doc) = \llbracket M' \rrbracket(\doc \#)$.

\subsection{Technical Lemmas}
In the following, let $\mathcal{S} = (N, \Sigma, R, S_0)$ be an $\SLP$ for $\doc$, and let $M = (Q, \Sigma, \State{1}, \delta, F)$ be an $\NFA$ with $Q = \{\State{1}, \State{2}, \ldots, \State{\states}\}$ that represents a $(\Sigma, \varset)$-spanner. \par 
The following definition is central for our evaluation algorithms (recall that for $i, j \in [\states]$ we denote by $i \transfuncNFA{w} j$ that $w$ takes $M$ from state $i$ to state $j$, i.\,e., $j \in \delta(i, w)$). 

\begin{definition}\label{mainMatricesDefinition}
For any non-terminal $A \in N$, we define a $(\states \times \states)$-matrix $\markerTransMat_A$ as follows.
For every $i, j \in [\states]$, $\markerTransMat_A[i, j]$ is a set that contains exactly the partial marker sets $\Lambda$ such that 
\begin{itemize}
\item $\Lambda$ is compatible with $\deriv{A}$, 
\item $\insertmarkers{\deriv{A}}{\Lambda}$ is non tail-spanning, and 
\item $\State{i} \transfuncNFA{\insertmarkers{\deriv{A}}{\Lambda}} \State{j}$. 
\end{itemize}
\end{definition}

Intuitively speaking, $\markerTransMat_A$ contains all the information of how the spanner represented by $M$ operates on the word $\deriv{A}$; thus, $\markerTransMat_{S_0}$ can be interpreted as a representation of $\llbracket M \rrbracket(\doc)$. This is formalised by the next lemma. Recall that $F$ denotes $M$'s set of accepting states and $1$ is $M$'s start state.

\begin{lemma}\label{markerTransMatSolutionSetComputationLemma}
$\llbracket M \rrbracket(\doc) = \bigcup_{\State{j} \in F} \markerTransMat_{S_0}[1, j]$.
\end{lemma}

This means that computing or enumerating the set $\llbracket M \rrbracket(\doc)$ reduces to the computation or enumeration of the sets $\markerTransMat_{S_0}[1, j]$ with $j \in F$. The purpose of the remaining notions and lemmas of this section is to show how we can recursively construct the entries of the matrices $\markerTransMat_A$ along the structure of the $\SLP$. \par
Note that for each $A \in N$ and $i, j \in [\states]$, there are three possible (mutually exclusive) cases of how the set $\markerTransMat_A[i, j]$ looks like:\label{reachMatrixExplanation}
\begin{enumerate}[\ \ (1)]
\item[$\bullet$] There is no marked word $w$ with $\State{i} \transfuncNFA{w} \State{j}$ and $\getWord{w} = \deriv{A}$.
\\ This means that $\markerTransMat_A[i, j] = \emptyset$. 
\item[$\bullet$] The only possible marked word $w$ with $\State{i} \transfuncNFA{w} \State{j}$ and $\getWord{w} = \deriv{A}$ is the word $w = \deriv{A}$ (i.\,e., $\markedpositions{w} = \emptyset$). 
\\ This means that $\markerTransMat_A[i, j] = \{\emptyset\}$. 
\item[$\bullet$] There is at least one marked word $w$ with $\State{i} \transfuncNFA{w} \State{j}$ and $\getWord{w} = \deriv{A}$ that actually contains markers (i.\,e., $\markedpositions{w} \neq \emptyset$). 
\\ This means that $\markerTransMat_A[i, j]$ is neither $\{\emptyset\}$ nor $\emptyset$. 
\end{enumerate}
For the computation (and enumeration) of the sets $\markerTransMat_{S_0}[1, j]$ with $j \in F$ (and therefore the set $\llbracket M \rrbracket(\doc)$) it will be a crucial preprocessing step to compute for every $A \in N$ and $i, j \in [\states]$, which of the three cases mentioned above apply. \par
Moreover, for any rule $A \to BC$ of $\mathcal{S}$, for every marked word $w$ with $\State{i} \transfuncNFA{w} \State{j}$ and $\getWord{w} = \deriv{A}$, there must be some state $\State{k}$ that we enter after having read exactly the (non-tail spanning) portion of $w$ that corresponds to $\deriv{B}$, i.\,e., $w = w_B w_C$, where $\getWord{w_B} = \deriv{B}$, $\getWord{w_C} = \deriv{C}$ and $\State{i} \transfuncNFA{w_B} \State{k} \transfuncNFA{w_C} \State{j}$. We also want to compute all these \emph{intermediate} states for every inner non-terminal $A \in N$ and $i, j \in [\states]$. We now formally define these data structures and then show how to compute them efficiently.

\begin{definition}\label{mainMatricesTwoDefinition}
For any non-terminal $A \in N$, we define a $(\states \times \states)$-matrix $\reachTransMat_A$ as follows. For every $i, j \in [\states]$, let $\reachTransMat_A[i, j] = \undefin$ if \,$\:\markerTransMat_A[i, j] = \emptyset$, let $\reachTransMat_A[i, j] = \emptyMarker$ if $\:\markerTransMat_A[i, j] = \{\emptyset\}$, and let $\reachTransMat_A[i, j] = \nonEmptyMarker$ otherwise. For any inner non-terminal $A \in N$ with rule $A \to BC$, we define a $(\states \times \states)$-matrix $\intStatesTransMat_A$ as follows. For every $i, j \in [\states]$, $\intStatesTransMat_A[i, j] \:= \:\{k \mid \reachTransMat_B[i, k] \neq \undefin \text{ and } \reachTransMat_C[k, j] \neq \undefin\}$.
\end{definition}

The next lemma will be crucial for the precomputation phase of our algorithms for computing and enumerating $\llbracket M \rrbracket(\doc)$.

\begin{lemma}\label{inductionBaseLemma}
All the matrices $\reachTransMat_A$ for every $A \in N$, $\intStatesTransMat_{A'}$ for every inner non-terminal $A' \in N$, and\,$\markerTransMat_{T_x}$ for every $x \in \Sigma$ can be computed in total time $\bigO(|M| + \,\size{\mathcal{S}} {\cdot} \states^3)$.
\end{lemma}

\renewcommand{\proofname}{Proof Sketch}
\begin{proof}
For computing all $\markerTransMat_{T_x}$ with $x \in \Sigma$, it is helpful to observe
the following:
\begin{itemize}
\item For every $x \in \Sigma$ and every $i, j \in [\states]$, we have $\markerTransMat_{T_x}[i, j]\: = \:\{\markedpositions{A_1 x} \mid A_1 \in \powerset{\Gamma_{\varset}},\:\State{i} \transfuncNFA{A_1 x} \State{j}\}$.
\item By iterating through $M$'s arcs, we can compute the set $P_i = \{(\ell, Y) \mid Y \in \powerset{\Gamma_{\varset}}, \State{\ell} \transfuncNFA{Y} \State{i}\}$ for all $i \in [\states]$.
\item Afterwards, we initialise $\markerTransMat_{T_x}[i, j]$ to $\emptyset$ for all $x \in \Sigma$ and $i, j \in [\states]$ and then iterate through the arcs of $M$ and use the precomputed $P_i$ to simultaneously construct all the $\markerTransMat_{T_x}$
\end{itemize}
All this can be achieved in time $\bigO(\states^2 + \,|M|)$.\par
We now have all $\markerTransMat_{T_x}$ with $x \in \Sigma$, and we can directly obtain $\reachTransMat_{T_{x}}$ from $\markerTransMat_{T_x}$ in time $\bigO(|\Sigma|\states^2)$. Finally, the matrices $\reachTransMat_{A}$ and $\intStatesTransMat_A$ for inner non-terminals with $A \to BC$ can be computed recursively in a bottom-up fashion using time $\bigO(|N|\states^3)$.
\end{proof}
\renewcommand{\proofname}{Proof}

The next lemma states how for inner non-terminals $A$ with rule $A \to BC$, and $i, j \in [\states]$, the set $\markerTransMat_A[i, j]$ is composed from sets $\markerTransMat_B[i, k]$ and $\markerTransMat_C[k, j]$ with $k \in \intStatesTransMat_A[i, j]$. For formulating the lemma, we need the following notation. For partial marker sets $\Lambda, \Lambda'$
and some $s \in \mathbb{N}$, let $\Lambda \msprod{s} \Lambda' = \Lambda \cup  \rightshift{\Lambda'}{s}$.

\begin{lemma}\label{mainConstructionLemma}
Let $A \to BC$ be a rule of $\mathcal{S}$, let $i, j \in [\states]$ and let $\Lambda_A$ be a partial marker set. Then following are equivalent:
\begin{enumerate}
\item\label{mainConstructionLemmaPointOne} $\Lambda_A \in \markerTransMat_A[i, j]$.
\item\label{mainConstructionLemmaPointTwo} There are a $k \in \intStatesTransMat_A[i, j]$ and partial marker sets $\Lambda_B \in \markerTransMat_B[i, k]$ and $\Lambda_C \in \markerTransMat_C[k, j]$, such that $\Lambda_A = \Lambda_B \msprod{\card{\deriv{B}}} \Lambda_C$.
\end{enumerate}
\end{lemma}

We extend the operator $\msprod{s}$ to \emph{sets} $\Delta, \Delta'$ of partial marker sets by $\Delta \msprod{s} \Delta' \:=\: \{\Lambda \msprod{s} \Lambda' \mid \Lambda \in \Delta, \Lambda' \in \Delta'\}$.

\begin{definition}\label{kSetsDefinition}
For every inner non-terminal $A \in N$ with rule $A \to BC$, for every $i, j \in [\states]$ and $k \in  \intStatesTransMat_A[i, j]$, we define 
$\mainNode{A}{i}{j}{k} = \markerTransMat_B[i, k] \msprod{\card{\deriv{B}}} \markerTransMat_C[k, j]$.
\end{definition}

With this terminology, we can now conclude from Lemma~\ref{mainConstructionLemma} that $\markerTransMat_A[i, j]$ actually decomposes into the $\card{\intStatesTransMat_A[i, j]}$ (not necessarily disjoint) sets $\mainNode{A}{i}{j}{k}$ with $k \in \intStatesTransMat_A[i, j]$.

\begin{lemma}\label{mainInductiveLemma}
Let $A \in N$ be an inner non-terminal and let $i, j \in [\states]$. Then $\markerTransMat_A[i, j] = \bigcup_{k \in \intStatesTransMat_A[i, j]} \mainNode{A}{i}{j}{k}$.
\end{lemma}

For $k, k' \in \intStatesTransMat_A[i, j]$ with $k \neq k'$, $\mainNode{A}{i}{j}{k} \cap \mainNode{A}{i}{j}{k'} \neq \emptyset$ is possible. But for every fixed $k$, every element from $\mainNode{A}{i}{j}{k}$ can only be obtained from elements of $\markerTransMat_B[i, k]$ and $\markerTransMat_C[k, j]$ in a unique way:

\begin{lemma}\label{simpleDisjointnessLemma}
Let $A \in N$ with rule $A \to BC$, let $i, j, k \in [\states]$, let $\Lambda_B, \Lambda'_B \in \markerTransMat_B[i, k]$ and $\Lambda_C, \Lambda'_C \in \markerTransMat_C[k, j]$. Then 
\begin{equation*}
\Lambda_B \msprod{\card{\deriv{B}}} \Lambda_C = \Lambda'_B \msprod{\card{\deriv{B}}} \Lambda'_C \iff \Lambda_B = \Lambda'_B \text{ and } \Lambda_C = \Lambda'_C\,.
\end{equation*}
\end{lemma}

\section{Computation of the Solution Set}\label{sec:computation}

We now consider the problem of computing the full set $\llbracket M \rrbracket(\doc)$.
In contrast to non-emptiness and model-checking, 
this task, as well as enumerating $\llbracket M \rrbracket(\doc)$, are not decision problems anymore and, to the best of our knowledge, they do not reduce to any existing algorithm on $\SLP$-compressed documents. 

By utilising the technical machinery of
Section~\ref{sec:alogPrelim}
we obtain this section's main result.
We write $\Sort{n}$ for the time it takes to sort a set of size $O(n)$; depending on the underlying
machine model this might be interpreted as $O(n)$ or as $O(n\log n)$.

\begin{theorem}\label{computeSetTheorem}
Let $\mathcal{S}$ be an $\SLP$ for $\doc$ and let $M$ be an $\NFA$ that represents a $(\Sigma, \varset)$-spanner. The set $\llbracket M \rrbracket(\doc)$ can be computed in time $\bigO(\Sort{|M|} {\cdot} \states^2 \ + \ \, \size{\mathcal{S}} {\cdot} \states^4 {\cdot} \size{\llbracket M \rrbracket(\doc)})$. 
\end{theorem}

\renewcommand{\proofname}{Proof Sketch}
\begin{proof}
We first perform the preprocessing described by Lemma~\ref{inductionBaseLemma}.
For any given $A \in N$ and $i, j \in [\states]$, we can inductively compute $\markerTransMat_{A}[i, j]$ as follows. If $A = T_x$ is a leaf non-terminal, then we already have computed $\markerTransMat_{A}[i, j]$; this serves as the basis of the induction. If $A \to BC$ is a rule, then, according to Lemma~\ref{mainInductiveLemma}, the set $\markerTransMat_A[i, j]$ is given by $\bigcup_{k \in \intStatesTransMat_A[i, j]} \mainNode{A}{i}{j}{k}$. Therefore, for every $k \in \intStatesTransMat_A[i, j]$, we compute the set $\mainNode{A}{i}{j}{k}$. By Definition~\ref{kSetsDefinition}, $\mainNode{A}{i}{j}{k}=\markerTransMat_B[i, k] \msprod{\card{\deriv{B}}} \markerTransMat_C[k, j]$. By induction, we can assume that the sets $\markerTransMat_B[i, k]$ and $\markerTransMat_C[k, j]$ have already been computed for every $k \in \intStatesTransMat_A[i, j]$. Finally, according to Lemma~\ref{markerTransMatSolutionSetComputationLemma}, $\llbracket M \rrbracket(\doc) = \bigcup_{j \in F'}\markerTransMat_{S_0}[1,j]$, where $F' = \{\State{j} \in F \mid \reachTransMat_{S_0}[1,j] \neq \undefin\}$, so it is sufficient to recursively compute all $\markerTransMat_{S_0}[1,j]$ with $j \in F'$. There are, however, two difficulties to be dealt with. \par
In order to avoid duplicates when constructing unions of sets of marker sets, we define an order on marker sets and handle all sets of marker sets as sorted lists according to this order. More precisely, we initially construct sorted lists of the sets $\markerTransMat_{T_x}[i, j]$ for every $x \in \Sigma$ and $i, j \in [\states]$ (which is responsible for the additive term $\Sort{|M|} {\cdot} \states^2$ in the running time). Then, we can  create sorted lists of unions of sets of marker sets by merging sorted lists and directly discarding the duplicates. \par
To obtain the claimed running time, we have to show that the computed intermediate sets $\markerTransMat_{A}[i, j]$ cannot get larger than the final set $\llbracket M \rrbracket(\doc)$. In fact, this is not necessarily the case for \emph{every} $A \in N$ and $i, j \in [\states]$. However, 
if in the recursion we need to compute some set $\markerTransMat_{A}[i, j]$,
 then for every $\Lambda \in \markerTransMat_{A}[i, j]$ there is a subword-marked word $v \in \lang(M)$ with $\getWord{v} = \doc$ and $v = v_1 \insertmarkers{\deriv{A}}{\Lambda} v_3$ such that $\State{1} \transfuncNFA{v_1} \State{i} \transfuncNFA{\insertmarkers{\deriv{A}}{\Lambda}} \State{j} \transfuncNFA{v_3} F$. This directly implies that, if $\markerTransMat_{A}[i, j]$ is computed in the recursion, then for each $\Lambda \in \markerTransMat_{A}[i, j]$ there is a unique element in $\llbracket M \rrbracket(\doc)$. Thus $|\markerTransMat_{A}[i, j]| \leq |\llbracket M \rrbracket(\doc)|$. 
\end{proof}
\renewcommand{\proofname}{Proof}

\section{Enumeration of the Solution Set}\label{sec:enumeration}

In this section, we consider the problem of enumerating the set $\llbracket M \rrbracket(\doc)$. 
In the following, let $\mathcal{S} = (N, \Sigma, R, S_0)$ be an $\SLP$ for $\doc$, and let $M = (Q, \Sigma, \State{1}, \delta, F)$ be an $\NFA$ with $Q = \{\State{1}, \State{2}, \ldots, \State{\states}\}$ that represents a $(\Sigma, \varset)$-spanner. \par
The matrices $\reachTransMat_A$ (Definition~\ref{mainMatricesTwoDefinition}) shall play an important role in the following. In particular, recall the meaning of the three possible entries ``$\undefin$'' ($\markerTransMat_A[i, j] = \emptyset$), ``$\emptyMarker$'' ($\markerTransMat_A[i, j] = \{\emptyset\}$) and ``$\nonEmptyMarker$'' ($\markerTransMat_A[i, j]$ is neither $\{\emptyset\}$ nor $\emptyset$); see also the explanations on page~\pageref{reachMatrixExplanation}.

\textbf{$(M,\mathcal{S})$-Trees}:
We define certain ordered binary trees with node- and arc-labels. 
All arc-labels will be non-negative integers, namely numbers 0 or $\card{\deriv{A}}$ for $A\in N$.
The available node-labels are given as follows. For every $A\in N$ and all $i,j\in [\states]$, 
\begin{enumerate}[\ (1)]
\item[$\bullet$] if $\reachTransMat_{A}[i, j] = \emptyMarker$, then there is a node-label  $\treeNodeEmpty{A}{i}{j}$. 
\item[$\bullet$] if $\reachTransMat_{A}[i, j] = \nonEmptyMarker$, then
  \begin{itemize}
    \item[$-$] if $A$ is a leaf non-terminal, then there is a node-label $\treeNodeTerminal{A}{i}{j}$,
    \item[$-$] if $A$ is an inner non-terminal, then for every 
          $k \in \intStatesTransMat_A[i, j]$ there is
          a node-label $\treeNode{A}{i}{k}{j}$.
  \end{itemize}
\end{enumerate}
For $(A,i,j)$ with $\reachTransMat_{A}[i, j] = \undefin$, we do not define any node-label(s).

In an \emph{$(M,\mathcal{S})$-tree}, nodes labelled with $\treeNodeEmpty{A}{i}{j}$ or $\treeNodeTerminal{A}{i}{j}$ are leaves. 
Each node $v$ labelled with $\treeNode{A}{i}{k}{j}$
has a left child $v_\ell$ and a right child $v_r$.
Let $A\to BC$ be the rule for $A$. Then the arc from $v$ to $v_\ell$ is labelled $0$ and the arc from $v$ to $v_r$ is labelled $\card{\deriv{B}}$.
The node $v_\ell$ is labelled as follows:
\begin{enumerate}[\ (1)]
 \item[$\bullet$] 
  If $\reachTransMat_{B}[i, k]= \emptyMarker$, then $v_\ell$ is labelled 
  $\treeNodeEmpty{B}{i}{k}$. 
 \item[$\bullet$]
  If $\reachTransMat_{B}[i, k] = \nonEmptyMarker$, then 
  \begin{itemize}
   \item[$-$] if $B$ is a leaf non-terminal, then $v_\ell$ is labelled 
        $\treeNodeTerminal{B}{i}{k}$,
   \item[$-$] if $B$ is an inner non-terminal, then $v_\ell$ is labelled with 
        \\ $\treeNode{B}{i}{k'}{k}$
        for a $k'\in \intStatesTransMat_B[i, k]$.
  \end{itemize}
 \item[$\bullet$] $\reachTransMat_{B}[i, k] = \undefin$ cannot occur because we know that $k\in \intStatesTransMat_A[i, j]$. 
\end{enumerate}
The node $v_r$ is labelled analogously:
\begin{enumerate}[\ (1)]
 \item[$\bullet$] 
  If $\reachTransMat_{C}[k,j]= \emptyMarker$, then $v_r$ is labelled 
  $\treeNodeEmpty{C}{k}{j}$. 
 \item[$\bullet$]
  If $\reachTransMat_{C}[k, j] = \nonEmptyMarker$, then 
  \begin{itemize}
   \item[$-$] if $C$ is a leaf non-terminal, then $v_r$ is labelled 
        $\treeNodeTerminal{C}{k}{j}$,
   \item[$-$] if $C$ is an inner non-terminal, then $v_r$ is labelled with 
        \\ $\treeNode{C}{k}{k'}{j}$
        for a $k'\in \intStatesTransMat_C[k,j]$.
  \end{itemize}
 \item[$\bullet$] $\reachTransMat_{C}[k, j] = \undefin$ cannot occur because we know that $k\in \intStatesTransMat_A[i, j]$. 
\end{enumerate}

\noindent
The idea underlying this notion is that a subtree
rooted by $\treeNode{A}{i}{k}{j}$ represents \emph{some} partial
marker sets $\Lambda \in \markerTransMat_A[i, j]$ that correspond to
marked words that can be read via intermediate state $\State{k}$,
i.\,e., the subset $\Lambda_B \subseteq \Lambda$ corresponding to
$\deriv{B}$ is from $\markerTransMat_B[i, k]$ and the subset
$\Lambda_C \subseteq \Lambda$ corresponding to $\deriv{C}$ is from
$\markerTransMat_C[k, j]$. Hence, $\treeNode{A}{i}{k}{j}$ can be
interpreted as representing \emph{some} elements of
$\mainNode{A}{i}{j}{k} \subseteq \markerTransMat_A[i, j]$. Moreover,
\emph{all} possible subtrees rooted by $\treeNode{A}{i}{k}{j}$ will
represent the full set $\mainNode{A}{i}{j}{k}$. 
Then, 
by Lemma~\ref{mainInductiveLemma}, 
the set of all subtrees rooted by
$\treeNode{A}{i}{k_A}{j}$ for a $k_A \in \intStatesTransMat_A[i,
j]$ represents the complete set $\markerTransMat_A[i, j]$. \par
In the case that $\reachTransMat_{A}[i, j] = \emptyMarker$, we know that $\markerTransMat_A[i, j] = \{\emptyset\}$, i.\,e., the 
empty set is the only partial marker set in $\markerTransMat_A[i,
j]$. If $A$ is a leaf non-terminal $T_x$ with
$\reachTransMat_{T_x}[i, j] = \nonEmptyMarker$, then the set $\markerTransMat_{A}[i, j]$ can be easily computed in a preprocessing step (see Lemma~\ref{inductionBaseLemma}). Therefore, we treat these cases as leaves in our trees (i.\,e., as the base cases where the recursive branches represented by these trees terminate). \par
In this way, such a tree rooted by $\treeNode{A}{i}{k}{j}$ for some $k \in \intStatesTransMat_A[i, j]$ is a concise representation of some runs of the recursive procedure implicitly given by Lemma~\ref{mainInductiveLemma}, i.\,e., 
\begin{align*}
\markerTransMat_A[i, j] &\supseteq \mainNode{A}{i}{j}{k} = \markerTransMat_B[i, k] \msprod{\card{\deriv{B}}} \markerTransMat_C[k, j]\\
&= \{\Lambda_B \cup \rightshift{\Lambda_C}{\card{\deriv{B}}} \mid \Lambda_B \in \markerTransMat_B[i, k], \Lambda_C \in \markerTransMat_C[k, j]\}\,.
\end{align*}
This also explains why we store the shift $\card{\deriv{B}}$, which is necessary for the operation $\msprod{\card{\deriv{B}}}$, on the arc from a node $v$ labelled $\treeNode{A}{i}{k}{j}$ to its right child $v_r$ labelled $\treeNode{C}{k}{k_C}{j}$. \par

For any $(M, \mathcal{S})$-tree $\mathcal{T}$, we denote its leaves labelled by
$\treeNodeTerminal{T_x}{i}{j}$ (for $x \in \Sigma$) as
\emph{terminal-leaves} and all the other leaves, i.\,e., leaves
labelled by $\treeNodeEmpty{A}{i}{j}$, as \emph{empty-leaves}. Note
that leaves $\treeNodeEmpty{A}{i}{j}$ with $A = T_x$ are considered
empty-leaves. Obviously, different nodes of $(M, \mathcal{S})$-trees can have
the same label. 
As indicated before, the purpose of $(M, \mathcal{S})$-trees is to represent
sets of partial marker sets. We shall now define this formally by
first defining the \emph{yield} of single $(M, \mathcal{S})$-trees.
From here on, the following notation will be convenient. For trees
$\mathcal{T}_1,\mathcal{T}_2$, arc-labels $s_1,s_2$, and a node-label
$P$ we write $P((\mathcal{T}_1,s_1),(\mathcal{T}_2,s_2))$ to denote the tree
whose root is labelled $P$ and has the roots of $\mathcal{T}_1$ and
$\mathcal{T}_2$ as its left and right child, respectively, with arcs
labelled by $s_1$ and $s_2$, respectively.

\begin{definition}\label{yieldDefinition}
The \emph{yield} of an $(M, \mathcal{S})$-tree $\mathcal{T}$ is
inductively defined as follows. 
If \,$\mathcal{T}$ is a single node labelled $\treeNodeEmpty{A}{i}{j}$, then $\yield{\mathcal{T}} = \{\emptyset\}$.
If  \,$\mathcal{T}$ is a single node labelled
  $\treeNodeTerminal{T_x}{i}{j}$, then  $\yield{\mathcal{T}} = \markerTransMat_{T_x}[i, j]$.
If \, $\mathcal{T}{=}\, P((\mathcal{T}_1, 0), (\mathcal{T}_2, s))$, then
$\yield{\mathcal{T}}  =  \yield{\mathcal{T}_1} \msprod{s} \yield{\mathcal{T}_2}.$

\end{definition}

For every node $u$ of a fixed $(M, \mathcal{S})$-tree $\mathcal{T}$, we shall
denote by $\yieldPara{\mathcal{T}}{u}$ the yield of the subtree of
$\mathcal{T}$ rooted by $u$. 
An $(M,\mathcal{S})$-tree whose root node has a label including the
non-terminal $A$ will sometimes be called \emph{$(M,A)$-tree}.

\begin{figure}
\begin{center}
\scalebox{1}{\includegraphics{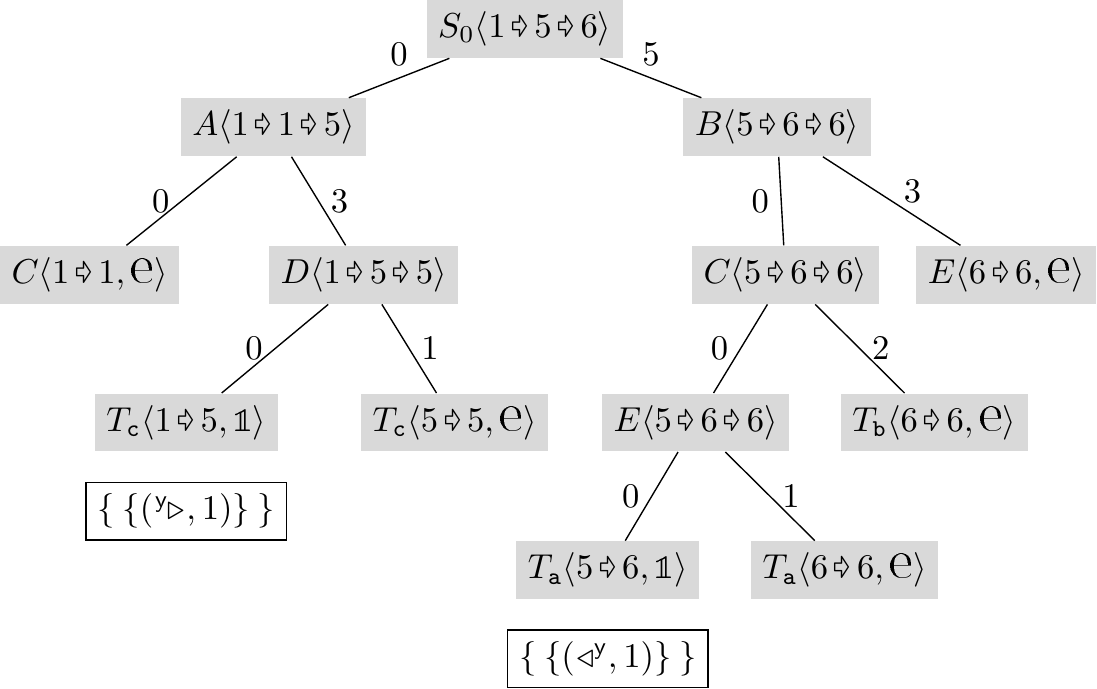}}
\end{center}
\caption{The $(M, S_0)$-tree discussed in Example~\ref{SMTreeExample}.
}
\label{fig:MATree}
\end{figure}

\begin{example}\label{SMTreeExample}
We recall the $\SLP$ $\mathcal{S}$ from Example~\ref{normalFormSLPExample} for $\doc = \ta \ta \tb \tc \tc \ta \ta \tb \ta \ta$ and the $\DFA$ $M$ from Figure~\ref{fig:exampleDFASpanner}. It can be verified that the tree $\mathcal{T}$ depicted in Figure~\ref{fig:MATree} is an $(M, S_0)$-tree.
As an example, note that according to the definition of $(M, \mathcal{S})$-trees the root can have a left child labelled by $\treeNode{A}{1}{1}{5}$, since $M$ can go from state $1$ to state $5$ by reading the marked word $\ta \ta \tb \open{\varsy} \tc \tc$ (corresponding to $\deriv{A}$), while reading the prefix $\ta \ta \tb$ (corresponding to $\deriv{C}$) between state $1$ and state $1$, and reading the suffix $\open{\varsy} \tc \tc$  (corresponding to $\deriv{D}$) between state $1$ and state $5$. Then, the node labelled by $\treeNodeEmpty{C}{1}{1}$ is an empty-leaf, since $w = \deriv{C} = \ta \ta \tb$ is the only marked word with $\getWord{w} = \deriv{C}$ that can be read going from state $1$ to state $1$. \par
The yield of all leaves of the $(M, S_0)$-tree depicted in Figure~\ref{fig:MATree} is $\{\emptyset\}$, except for the terminal-leaves labelled by $\treeNodeTerminal{T_{\tc}}{1}{5}$ and by $\treeNodeTerminal{T_{\ta}}{5}{6}$, whose yields are $\yield{\treeNodeTerminal{T_{\tc}}{1}{5}} = \{\{(\open{\varsy}, 1)\}\}$ and $\yield{\treeNodeTerminal{T_{\ta}}{5}{6}} = \{\{(\close{\varsy}, 1)\}\}$. These yields are shown in Figure~\ref{fig:MATree} below the corresponding leaves. By the recursive definition of $\yield{\cdot}$, we get $\yield{\treeNode{A}{1}{1}{5}} = \{\{(\open{\varsy}, 4)\}\}$ and $\yield{\treeNode{B}{5}{6}{6}} = \{\{(\close{\varsy}, 1)\}\}$. Since the arc from the root to the node labelled by $\treeNode{B}{5}{6}{6}$ is labelled by $5$, we get $\yield{\mathcal{T}} = \{\{(\open{\varsy}, 4), (\close{\varsy}, 6)\}\}$. \par
Note that $\Lambda = \{(\open{\varsy}, 4), (\close{\varsy}, 6)\}$ corresponds to the $(\{\varsx, \varsy\}, \doc)$-tuple $t$ with $t(\varsx) = \undefin$ and $t(\varsy) = \spann{4}{6}$, and $\insertmarkers{\doc}{\Lambda} = \ta \ta \tb \open{\varsy} \tc \tc \close{\varsy} \ta \ta \tb \ta \ta$. 
\end{example}

As an immediate consequence of Definition~\ref{yieldDefinition} we obtain:

\begin{lemma}\label{simpleYieldLemma}
Let $\mathcal{T}$ be an $(M, \mathcal{S})$-tree and let $u$ be a node of $\mathcal{T}$ labelled $\treeNode{B}{i}{k}{j}$, $\treeNodeEmpty{B}{i}{j}$ or $\treeNodeTerminal{B}{i}{j}$ for some $B \in N$, $i, j, k \in [\states]$. Then every element from $\yieldPara{\mathcal{T}}{u}$ is a partial marker set over $\varset$ compatible with $\deriv{B}$.
\end{lemma}

We measure the size of $|\mathcal{T}|$ of a tree $\mathcal{T}$ as the number of its nodes. Next, we estimate the size of $(M, A)$-trees. Recall that the depth of non-terminals has been defined in Section~\ref{section:SLP}.

\begin{lemma}\label{treeSizeLemma}
Let $A\in N$ and let $\mathcal{T}$ be an $(M,A)$-tree.\\ 
Then $|\mathcal{T}| \leq 4|\varset| {\cdot} \depth{A}$, and \,$\mathcal{T}$ has at most $2|\varset|$ terminal-leaves.
\end{lemma}

\renewcommand{\proofname}{Proof Sketch}
\begin{proof}
The following can be shown by induction. If the subtree rooted by an
inner node $u$ contains $\ell$ terminal-leaves, then, since the
yield of each terminal-leaf contains at least one non-empty partial
marker set, there must be partial marker sets in
$\yieldPara{\mathcal{T}}{u}$
 with a size of at least $\ell$ (i.\,e.,
$\yieldPara{\mathcal{T}}{u}$ must contain a partial marker set that is
constructed from $\ell$ many non-empty marker sets from the terminal-leaves). Since partial marker sets have size at most $2|\varset|$,
this means that $\mathcal{T}$ has at most $2|\varset|$
terminal-leaves.\par 
Furthermore, all inner nodes and all terminal-leaves lie on
paths (of length $\leq \depth{A}$) from some terminal-leaf to the root. Thus, there are at most $2|\varset| {\cdot} \depth{A}$ inner nodes and terminal-leaves. Moreover, each of these nodes can be adjacent to at most one empty-leaf, thus, the total number of nodes is at most $4|\varset| {\cdot} \depth{A}$.
\end{proof}
\renewcommand{\proofname}{Proof}

We next consider the algorithmic problem of enumerating the yield of a given $(M, A)$-tree. An \emph{$(M, A)$-tree with leaf-pointers} is an $(M, A)$-tree where, additionally, every terminal-leaf labelled by $\treeNodeTerminal{T_x}{i}{j}$ stores a pointer to the first element of a list that contains the elements of $\markerTransMat_{T_x}[i, j]$ (for all $x \in \Sigma$, $i, j \in [\states]$). 
This enables us to obtain the following.

\begin{lemma}\label{singleTreeEnumLemma}
Given an $(M, A)$-tree $\mathcal{T}$ with leaf-pointers, 
the set 
$\yield{\mathcal{T}}$ can be enumerated with preprocessing $\bigO(\depth{A}|\varset|)$ and delay $\bigO(|\varset|)$.
\end{lemma}

So far, we have established that $(M, A)$-trees represent partial
marker sets, that they have moderate size and that their yield can be
easily enumerated. However, we still need to show that the yields of
all $(M, A)$-trees rooted by $\treeNode{A}{i}{k}{j}$ for some $k
\in \intStatesTransMat_A[i, j]$, represent the complete set
$\markerTransMat_A[i, j]$. Moreover, in order to reduce the problem of
enumerating elements from $\markerTransMat_A[i, j]$ to enumerating
$(M, A)$-trees, we have to establish some kind of one-to-one
correspondence between $(M, A)$-trees and partial marker sets from
$\markerTransMat_A[i, j]$. These issues will be settled
next.

\subsection{A Unique Representation by $(M,\mathcal{S})$-Trees}
For $A \in N$, $i, j \in [q]$ and $k \in \intStatesTransMat_A[i, j] \cup \{\baseCaseMarker\}$, we define the set $\allTrees{A}{i}{k}{j}$ as follows. The set $\allTrees{A}{i}{\baseCaseMarker}{j}$ contains a single tree with a single node labelled $\treeNodeEmpty{A}{i}{j}$ if $\reachTransMat_{A}[i, j] = \emptyMarker$, and it contains a single tree with a single node labelled $\treeNodeTerminal{A}{i}{j}$ if $\reachTransMat_{A}[i, j] \neq \emptyMarker$ (note that in the following, we consider $\allTrees{A}{i}{\baseCaseMarker}{j}$ only in the case where $\reachTransMat_{A}[i, j] = \emptyMarker$ or $A$ is a leaf non-terminal). For non-terminals $A \in N$, $i, j \in [q]$ and $k \in \intStatesTransMat_A[i, j]$, $\allTrees{A}{i}{k}{j}$ contains all $(M, A)$-trees whose root is labelled $\treeNode{A}{i}{k}{j}$ (we shall consider $\allTrees{A}{i}{k}{j}$ only in the case where $\reachTransMat_{A}[i, j] = \nonEmptyMarker$).

We extend the yield from single $(M, A)$-trees to sets of
$(M, A)$-trees in the obvious way:
\begin{equation*}
\yield{\allTrees{A}{i}{k}{j}} \ = \ \bigcup_{\mathcal{T} \in \allTrees{A}{i}{k}{j}} \yield{\mathcal{T}}\,.
\end{equation*}

In particular, $\yield{\allTrees{A}{i}{\baseCaseMarker}{j}} =
\{\emptyset\}$ if  $\reachTransMat_{A}[i, j] = \emptyMarker$, and
$\yield{\allTrees{A}{i}{\baseCaseMarker}{j}} = \markerTransMat_A[i,
j]$ for leaf non-terminals $A$ with $\reachTransMat_{A}[i, j] \neq
\undefin$.
By Lemma~\ref{simpleYieldLemma}, the yield of any set of $(M, A)$-trees is a set of partial marker sets. 
The next lemma can
be concluded in a straightforward way from
Definition~\ref{kSetsDefinition} and  Lemma~\ref{mainInductiveLemma}.

\begin{lemma}\label{mainCorrectnessLemmaEnumPartOne}
$\yield{\allTrees{A}{i}{k}{j}}=\mainNode{A}{i}{j}{k}$, for all
inner non-terminals $A$, all $i, j \in [\states]$ with
$\reachTransMat_A[i, j] \neq \undefin$, and all $k \in
\intStatesTransMat_A[i, j]$. 
\end{lemma}

By Lemma~\ref{mainCorrectnessLemmaEnumPartOne}, we can consider the
$(M, A)$-trees of $\allTrees{A}{i}{k}{j}$ as a representation of
$\mainNode{A}{i}{j}{k}$. Hence,  $\bigcup_{k \in
  \intStatesTransMat_A[i, j]} \allTrees{A}{i}{k}{j}$ serves as a
representation of $\markerTransMat_A[i, j]$. We could thus enumerate
the trees of $\bigcup_{k \in \intStatesTransMat_A[i, j]}
\allTrees{A}{i}{k}{j}$ and, for each individual $(M, A)$-tree, use
Lemma~\ref{singleTreeEnumLemma} to enumerate its yield. However, in
addition to the question of how to enumerate all these trees (which
shall be taken care of later on), we also have to deal with the
possibility that the yields of different $(M, A)$-trees are not
disjoint, which would lead to duplicates in the enumeration. With
respect to this latter issue, we have already observed in
Section~\ref{sec:alogPrelim}, that $\mainNode{A}{i}{j}{k} \cap
\mainNode{A}{i}{j}{k'} \neq \emptyset$, for some $k, k' \in
\intStatesTransMat_A[i, j]$ with $k \neq k'$, is possible. However, 
if $M$ is a $\DFA$, the sets
$\mainNode{A}{i}{j}{k}$ with $k \in \intStatesTransMat_A[i, j]$ are in
fact pairwise disjoint:

\begin{lemma}\label{mainInductiveLemmaEnumVersion}
Let $A$ be a non-terminal, let $A'$ be an inner non-terminal, let $i, j, j' \in [\states]$ with $j \neq j'$, and let $k, k' \in \intStatesTransMat_{A'}[i, j]$ with $k' \neq k$. If $M$ is a $\DFA$, then
$\markerTransMat_A[i, j] \cap \markerTransMat_A[i, j'] = \emptyset$ and $\mainNode{A'}{i}{j}{k} \cap \mainNode{A'}{i}{j}{k'} = \emptyset$.
\end{lemma}

Using this lemma we can show that, as long as $M$ is deterministic, the yields of different $(M, A)$-trees are necessarily disjoint. We define \emph{equality} of $(M, A)$-trees $\mathcal{T}_1$ and $\mathcal{T}_2$, denoted by $\mathcal{T}_1 = \mathcal{T}_2$, as follows. The roots are called \emph{corresponding} if they have the same label; and any other node $P_1$ of $\mathcal{T}_1$ corresponds to a node $P_2$ of $\mathcal{T}_2$ if they have the same label and are both the left (or both the right) child of corresponding parent nodes. Now $\mathcal{T}_1 = \mathcal{T}_2$ if and only if this correspondence is a bijection between the nodes of $\mathcal{T}_1$ and $\mathcal{T}_2$.\par 
This means that non-equal $(M, A)$-trees have either differently labelled roots or they are extensions of the same tree (i.\,e., the tree of all the corresponding nodes) and differ in the way that a leaf of this common tree (possibly the root) has differently labelled left children or differently labelled right children in $\mathcal{T}_1$ and $\mathcal{T}_2$, respectively. Note, however, that for corresponding nodes $P_1$ and $P_2$ of non-equal $\mathcal{T}_1$ and $\mathcal{T}_2$ it is nevertheless possible that $\yieldPara{\mathcal{T}_1}{P} \neq \yieldPara{\mathcal{T}_2}{P}$.

\begin{lemma}\label{mainCorrectnessLemmaEnumPartTwo}
Let $M$ be a $\DFA$. Let $A \in N$ be an inner non-terminal, let $i, j_1, j_2 \in [\states]$ with $\reachTransMat_A[i, j_1] = \reachTransMat_A[i, j_2] = \nonEmptyMarker$, let $k_1 \in \intStatesTransMat_A[i, j_1]$ and $k_2 \in \intStatesTransMat_A[i, j_2]$. Let $\mathcal{T}_1, \mathcal{T}_2$ be non-equal $(M, A)$-trees with roots labelled by $\treeNode{A}{i}{k_1}{j_1}$ and $\treeNode{A}{i}{k_2}{j_2}$. Then $\yield{\mathcal{T}_1} \cap \yield{\mathcal{T}_2} = \emptyset$.
\end{lemma}

\renewcommand{\proofname}{Proof Sketch}
\begin{proof}
For contradiction, assume $\yield{\mathcal{T}_1}\cap
\yield{\mathcal{T}_2}\neq\emptyset$.
By Lemma~\ref{mainCorrectnessLemmaEnumPartOne}, $\yield{\mathcal{T}_1}
\subseteq \mainNode{A}{i}{j_1}{k_1} \subseteq \markerTransMat_A[i,
j_1]$ and $\yield{\mathcal{T}_2} \subseteq \mainNode{A}{i}{j_2}{k_2}
\subseteq \markerTransMat_A[i, j_2]$. Thus, 
Lemma~\ref{mainInductiveLemmaEnumVersion} implies that $j_1 = j_2$ and 
$k_1 = k_2$. This means that $\mathcal{T}_1$ and $\mathcal{T}_2$ have
corresponding roots labelled by $\treeNode{A}{i}{k}{j}$, for $j =
j_1 = j_2$ and $k = k_1 = k_2$.\par 
Let $\widehat{\mathcal{T}}$ be the tree of the nodes of
$\mathcal{T}_1$ and $\mathcal{T}_2$ that are corresponding. For any
node $P$ of $\widehat{\mathcal{T}}$ with a left child $L_1$ and a
right child $R_1$ in $\mathcal{T}_1$, and a left child $L_2$ and a
right child $R_2$ in $\mathcal{T}_2$, we can show that if
$\yieldPara{\mathcal{T}_1}{P} \cap \yieldPara{\mathcal{T}_2}{P} \neq
\emptyset$, then $\yieldPara{\mathcal{T}_1}{L_1} \cap
\yieldPara{\mathcal{T}_2}{L_2} \neq \emptyset$ and
$\yieldPara{\mathcal{T}_1}{R_1} \cap \yieldPara{\mathcal{T}_2}{R_2}
\neq \emptyset$ (for this we use
Lemmas~\ref{mainCorrectnessLemmaEnumPartOne}~and~\ref{simpleDisjointnessLemma}).

Hence, since 
$\mathcal{T}_1 \neq \mathcal{T}_2$,
there must be some node $P$ of $\widehat{\mathcal{T}}$ with
$\yieldPara{\mathcal{T}_1}{P} \cap \yieldPara{\mathcal{T}_2}{P} \neq
\emptyset$, such that $P$'s left children $L_1$ and $L_2$ in
$\mathcal{T}_1$ and $\mathcal{T}_2$, respectively, are not
corresponding (or this is the case with respect to $P$'s right
children, which can be handled analogously). 
By our above observation, $\yieldPara{\mathcal{T}_1}{L_1} \cap \yieldPara{\mathcal{T}_2}{L_2} \neq \emptyset$, but $L_1$ is labelled by $\treeNode{B}{i'}{k_{B, 1}}{k'}$, $L_2$ is labelled by $\treeNode{B}{i'}{k_{B, 2}}{k'}$ with $k_{B, 1} \neq k_{B, 2}$. Since $\yieldPara{\mathcal{T}_1}{L_1} \subseteq \mainNode{B}{i'}{k'}{k_{B, 1}}$ and $\yieldPara{\mathcal{T}_2}{L_2} \subseteq \mainNode{B}{i'}{k'}{k_{B, 2}}$, this means that $\mainNode{B}{i'}{k'}{k_{B, 1}} \cap \mainNode{B}{i'}{k'}{k_{B, 2}} \neq \emptyset$, which is a contradiction to Lemma~\ref{mainInductiveLemmaEnumVersion}.
\end{proof}
\renewcommand{\proofname}{Proof}

\subsection{The Enumeration Algorithm} 

An enumeration algorithm $\enumA$ produces, on some input $I$, an \emph{output sequence} $(s_1, s_2, \ldots, s_n, \eol)$, where $\eol$ is the \emph{end-of-enumeration} marker. We say that $\enumA$ on input $I$ \emph{enumerates} a set $S$ if and only if the output sequence is $(s_1, s_2, \ldots, s_n, \eol)$, $|S| = n$ and $S = \{s_1, s_2, \ldots, s_n\}$. The \emph{preprocessing time} (of $\enumA$ on input $I$) is the time that elapses between starting $\enumA(I)$ and the output of the first element, 
and the \emph{delay} is the time that elapses between any two elements of the output sequence. The preprocessing time and the delay of $\enumA$ is the maximum preprocessing time and maximum delay, respectively, over all possible inputs (measured as function of the input size).\par
We present an enumeration algorithm $\enumAll$ (given in
Algorithm~\ref{enumAllAlgo}), that receives as input some $A \in N$,
$i, j \in [\states]$ and 
$k \in \intStatesTransMat_A[i, j]$.
We treat recursive calls to $\enumAll$ as sets of the elements of the
output sequence, which allows to use \textbf{for}-loops to iterate
through the output sequences returned by the recursive calls (see
Lines~\ref{enumBLine}~and~\ref{enumCLine}). For this, we assume that
any recursive call of $\enumAll$ writes its output element in a
buffer and then produces the next element only when it is requested by
the \textbf{for}-loop. Consequently, the 
time used for starting the next iteration of the \textbf{for}-loop is bounded by the preprocessing time (if it is the first iteration) or the delay (for all other iterations) of the recursive call of $\enumAll$ (this includes checking that there is no iteration left, since we can only check this by receiving $\eol$ from the recursive call).\par
The algorithm requires the data-structures $\reachTransMat_A$ and
$\intStatesTransMat_A$, which, for now, we assume to be at our
disposal. We further assume that, for all $A \in N$ and all $i, j \in
[\states]$, we have the sets $\mathcal{I}_A[i, j]$ at our disposal,
which are defined as follows. If $A = T_x$ or $\reachTransMat_{A}[i,
j] = \emptyMarker$ then $\mathcal{I}_A[i, j] = \{\baseCaseMarker\}$
(here, the symbol $\baseCaseMarker$ serves as a marker for the ``base
case''), and $\mathcal{I}_A[i, j] = \intStatesTransMat_A[i, j]$
otherwise.  

\SetKw{Return}{output $\gets$}

\begin{algorithm}
\SetAlgoNoEnd
\LinesNumbered
\SetSideCommentRight
\SetFillComment
\SetKwInOut{Input}{Input}
\SetKwInOut{Output}{Output}
\Input{Non-terminal $A \in N$, $i, j \in [\states]$, $k \in \mathcal{I}_A[i, j] \cup \{\baseCaseMarker\}$.}
\Output{A sequence of the trees in $\allTrees{A}{i}{k}{j}$, followed
  by $\eol$}

\If{$k = \baseCaseMarker$}
{
	\If{$\reachTransMat_{A}[i, j] = \emptyMarker$}
	{
	\Return single node with label $\treeNodeEmpty{A}{i}{j}$, \Return $\eol$\;\label{produceEmptyTreeLine} 	
	}
	\Else
	{
	\Return single node with label $\treeNodeTerminal{A}{i}{j}$, 	\Return $\eol$\;\label{produceTerminalTreeLine}
	}
}
\ElseIf{$A$ is an inner non-terminal with $A \to BC$}
{

\For{$(k_B, k_C) \in (\mathcal{I}_B[i, k] \times \mathcal{I}_C[k, j])$\label{enumStatesLine}}
{
  \For{$\mathcal{T}_B \in \enumAll(B, i, k_B, k)$\label{enumBLine}}
    {
      \If{$\mathcal{T}_B\neq \eol$}
      {
      	 \For{$\mathcal{T}_C \in \enumAll(C, k, k_C, j)$\label{enumCLine}}
         {
	   \If{$\mathcal{T}_C\neq \eol$}
           {
	     \Return $\treeNode{A}{i}{k}{j}(\mathcal{T}_B, \mathcal{T}_C)$\;\label{recursiveOutputLine}
      	   }
         }
      }
    }	
}

}	      		
\Return $\eol$\;\label{theEndLine}
\caption{$\enumAll(A, i, k, j)$}\label{enumAllAlgo}
\end{algorithm}

\newcommand{\Max}[1]{\ensuremath{\textup{max}}(#1)}

For $A \in N$,  $i, j \in [\states]$, and  $k \in \mathcal{I}_A[i, j] \cup \{\baseCaseMarker\}$ we let
$\Max{A,i,k,j}$ be the maximum number of nodes of a tree in
$\allTrees{A}{i}{k}{j}$.

\begin{lemma}\label{enumAllLemma}
Whenever it receives as input an inner non-terminal $A$, states $i, j \in [\states]$ such that $\reachTransMat_{A}[i, j] = \nonEmptyMarker$, and a $k \in \intStatesTransMat_A[i, j]$,
the algorithm $\enumAll(A, i, k, j)$ enumerates the elements of the set $\allTrees{A}{i}{k}{j}$ with preprocessing and delay $\bigO(\Max{A,i,k,j})$.
\end{lemma}

\renewcommand{\proofname}{Proof Sketch}
\begin{proof}
We first observe that if $\reachTransMat_{A}[i, j] = \emptyMarker$ or if $A$ is a leaf non-terminal, then $\enumAll(A, i, \baseCaseMarker, j)$ enumerates the set $\allTrees{A}{i}{\baseCaseMarker}{j}$ with constant preprocessing and constant delay. This can be used as the base of an induction to show that there is a constant $c$ such that for all inputs $A \in N$, $i, j \in [\states]$, $k \in \mathcal{I}_A[i, j] \cup \{\baseCaseMarker\}$, such that $k=\baseCaseMarker$ or $\reachTransMat_{A}[i, j] = \nonEmptyMarker$, the algorithm $\enumAll(A, i, k, j)$ enumerates (without duplicates) the set $\allTrees{A}{i}{k}{j}$ such that it
 takes time at most
 \begin{enumerate}[\ \ (1)]
  \item[$\bullet$] $c\cdot \Max{A,i,k,j}$ before the first output is created,
  \item[$\bullet$] $2c\cdot \Max{A,i,k,j}$ between any two consecutive output trees,
  \item[$\bullet$] $c\cdot\Max{A,i,k,j}$ between outputting the last tree and 
  $\eol$.
 \end{enumerate}
Let $A \to BC$ be a rule. To see that 
$\enumAll(A, i, k, j)$ does in fact enumerate 
$\allTrees{A}{i}{k}{j}$, we observe that, for all $(k_B, k_C) \in
(\mathcal{I}_B[i,k] \times \mathcal{I}_C[k,j])$, all $\mathcal{T}_B \in
\allTrees{B}{i}{k_B}{k}$ and all $\mathcal{T}_C \in
\allTrees{C}{k}{k_C}{j}$, the algorithm will produce the tree with a
root labelled by $\treeNode{A}{i}{k}{j}$, and with the roots of
$\mathcal{T}_B$ and $\mathcal{T}_C$ as left and right child,
respectively. By definition of $(M, A)$-trees, the algorithm produces
exactly all $(M, A)$-trees with a root labelled by
$\treeNode{A}{i}{k}{j}$.
Note that duplicate output trees can neither be produced during 
the same iteration of the loop of Line~\ref{enumStatesLine} nor during
the executions of different iterations of the loop of Line~\ref{enumStatesLine}.\par
In order to prove the claimed runtime bounds, we assume as
induction hypothesis that these bounds hold with respect to every $k_B
\in \mathcal{I}_B[i,k]$ and every $k_C\in\mathcal{I}_C[k,j]$ 
(with $\Max{A,i,k,j}$ replaced by $\Max{B,i,k_B,k}$ and by $\Max{C,k,k_C,j}$, respectively). 
Then we can show that the first element of $\treeNode{A}{i}{k}{j}$ is produced in time $c{\cdot} \Max{A,i,k,j}$.
We also have to show that after having produced some (but not the
last) element of $\treeNode{A}{i}{k}{j}$, we only need time $2c\cdot
\Max{A,i,k,j}$ to produce the next element, and that after having
produced the last element of $\treeNode{A}{i}{k}{j}$, we need at most
time $c{\cdot} \Max{A,i,k,j}$ to produce $\eol$. There are $4$
individual cases to consider (for convenience, we call the loops of
Lines~\ref{enumStatesLine}, \ref{enumBLine} and \ref{enumCLine} by
\emph{states-loop}, \emph{$B$-loop} and \emph{$C$-loop}, respectively):
(1)~we are not in the last iteration of the $C$-loop,
(2)~we are in the last iteration of the $C$-loop (but not the $B$-loop),
(3)~we are in the last iterations of the $C$-loop and the $B$-loop (but not the states-loop),
(4)~we are in the last iterations of the $C$-loop, the $B$-loop and the states-loop.
By using our induction hypothesis, we can show that the first three cases yield in fact a delay of at most $2c{\cdot} \Max{A,i,k,j}$, while the fourth case yields a delay of $c{\cdot} \Max{A,i,k,j}$. We emphasise that for obtaining these bounds, it is absolutely vital that the delay for getting the first element and the element $\eol$ is better than the delay between two consecutive elements.
\end{proof}
\renewcommand{\proofname}{Proof}

\begin{theorem}\label{enumSetTheorem}
Let $\mathcal{S}$ be an $\SLP$ for $\doc$ and let $M$ be a $\DFA$ that represents a $(\Sigma, \varset)$-spanner. The set $\llbracket M \rrbracket(\doc)$ can be enumerated with preprocessing time 
$\bigO(|M| + \,\size{\mathcal{S}} {\cdot} \states^3)$ and delay
$\bigO(\depth{\mathcal{S}} {\cdot} |\varset|)$.
\end{theorem}

\renewcommand{\proofname}{Proof Sketch}
\begin{proof}
In the preprocessing phase, we compute all the matrices
$\reachTransMat_A$ for every $A \in N$, $\intStatesTransMat_{A'}$ for
every inner non-terminal $A' \in N$, and\,$\markerTransMat_{T_x}$ for
every $x \in \Sigma$. We also compute the set $F' = \{j \in F \mid
\reachTransMat_{S_0}[1, j] \neq \undefin\}$ and, for every $A \in N$
and for every $i, j \in [\states]$, the sets $\mathcal{I}_A[i, j]$. 
According to Lemma~\ref{inductionBaseLemma}, all this can be done in time is $\bigO(|M| + \size{\mathcal{S}} {\cdot} \states^3)$. 
Next, we present an enumeration procedure that receives an $(M, A)$-tree $\mathcal{T}$ as input.\smallskip\\
$\enumSingleTree(\mathcal{T})$:
\begin{enumerate}
\item Add the correct leaf-pointers to $\mathcal{T}$
\item Enumerate $\yield{\mathcal{T}}$ according to Lemma~\ref{singleTreeEnumLemma}.
\end{enumerate}
The following can be concluded from Lemmas~\ref{treeSizeLemma}~and~\ref{singleTreeEnumLemma}.\smallskip\\
\noindent \emph{Claim $1$}: The procedure $\enumSingleTree(\mathcal{T})$ enumerates $\yield{\mathcal{T}}$ with preprocessing time $\bigO(\depth{A}|\varset|)$ and delay $\bigO(|\varset|)$. \smallskip\\
For all $j \in F'$ and $k \in \intStatesTransMat_{S_0}[1, j]$, we use the enumeration procedure\smallskip\\
$\enumSingleRoot(j, k)$:
\begin{enumerate}
\item By calling $\enumAll(S_0, 1, k, j)$, we produce a sequence \\ $(\mathcal{T}_1, \mathcal{T}_2, \ldots, \mathcal{T}_{n_{j, k}})$ of $(M, S_0)$-trees followed by $\eol$.
\item In this enumeration, whenever we receive $\mathcal{T}_\ell$ for
  some $\ell \in [n_{j, k}]$, we carry out
  $\enumSingleTree(\mathcal{T}_{\ell})$ and produce its output
  sequence 
  as output.
\end{enumerate}
\noindent \emph{Claim $2$}: The procedure $\enumSingleRoot(j, k)$ enumerates $\mainNode{S_0}{1}{j}{k}$ with preprocessing time and delay $\bigO(\depth{S_0}|\varset|)$. \smallskip\\
This claim is mainly a consequence of Lemma~\ref{enumAllLemma} and
Claim~$1$; but we also need Lemma~\ref{treeSizeLemma} to bound the preprocessing time and delay, Lemma~\ref{mainCorrectnessLemmaEnumPartOne} to argue that exactly the set $\mainNode{S_0}{1}{j}{k}$ is enumerated, and Lemma~\ref{mainCorrectnessLemmaEnumPartTwo} to show that the enumeration is without duplicates.\par
The complete enumeration phase now consists of performing
$\enumSingleRoot(j, k)$ for 
every $j \in F'$ and  every $k \in \intStatesTransMat_{S_0}[1, j]$.
By Claims~$1$ and~$2$, and Lemmas~\ref{mainInductiveLemma},
\ref{markerTransMatSolutionSetComputationLemma},
\ref{mainInductiveLemmaEnumVersion}, this enumeration produces the
correct output within the claimed time bounds.
\end{proof}
\renewcommand{\proofname}{Proof}

Note that we need $M$ to be a $\DFA$ to apply
Lemma~\ref{mainCorrectnessLemmaEnumPartTwo}, i.\,e., to argue that the
yields of different $(M, S_0)$-trees are disjoint. Observe that running
the algorithm of Theorem~\ref{enumSetTheorem} directly on an
$\NFA$ yields a correct enumeration with the same
complexity bounds, but with possible duplicates. 
But since we can transform $\NFA$s into $\DFA$s (at the cost of an exponential blow-up
in automata size), Theorem~\ref{enumSetTheorem}, without producing duplicates, holds
also for $\NFA$s, but $|M|$ and $q$ in the preprocessing become
$2^{|M|}$ and $2^q$.
However, this affects only the preprocessing time, and it
does not change the data complexity. 

\section{Conclusion}\label{sec:conclusion}

We showed that regular spanners can be efficiently evaluated directly on $\SLP$-compressed documents.
In the best-case scenario where the $\SLP$s have a size logarithmic in the size $\mathbf{d}$ of the uncompressed document, our approach solves all the considered evaluation tasks with only a logarithmic dependency on $\mathbf{d}$. Our enumeration algorithm's delay is $O(\log \mathbf{d})$; and
the most important question left open is whether this can be improved to a constant delay --- we believe this to be difficult. 

In terms of combined complexity, it might be interesting to know whether fast Boolean matrix multiplication can lower the degree of the polynomial with respect to the number of states, as it is the case for checking membership of an $\SLP$-compressed document in a regular language (see Section~\ref{section:SLP}). Another intriguing question is whether spanner evaluation on compressed documents can handle updates of the document.

\newpage

\appendix

\section*{APPENDIX}

\bigskip

\section{Proof omitted in Section~\ref{section:SLP}}
\subsection*{Proof of Lemma~\ref{computeDerivationLengthsLemma}}

\begin{proof}
For every $A \in N$ with $A \to a$, we set $|\deriv{A}| = 1$, for every $A \in N$ with $A \to BC$, we set $|\deriv{A}| = |\deriv{B}| + |\deriv{C}|$. Thus, we can recursively compute $|\deriv{A}|$ for every $A \in N$ in time $\bigO(|N|)$.
\end{proof}

\section{Proof  omitted in Section~\ref{section:ModelChecking}}

\subsection*{Proof of Theorem~\ref{basicEvalTasksTheorem}}

\begin{proof}
We prove the two statements separately.
\begin{enumerate}
\item We first obtain an $\NFA$ $M'$ from $M$ by replacing all $\powerset{\Gamma_{\varset}}$-transitions by $\eword$-transitions. This can be done in time $\bigO(|M|)$. Note that $M'$ is an $\NFA$ over the alphabet $\Sigma$. We can now observe the following:
\begin{align*}
&\llbracket M \rrbracket(\doc) \neq \emptyset &\iff\\
&\text{there is an $(\varset, \doc)$-tuple $t$ with } t \in \llbracket M \rrbracket(\doc) &\iff\\
&\text{there is an $(\varset, \doc)$-tuple $t$ with } \insertmarkers{\doc}{t} \in \lang(M) &\iff\\
&\doc \in \lang(M')
\end{align*}

The second equivalence is a consequence of Proposition~\ref{basicSubwordMarkedLanguageProposition}, the third equivalence follows by construction of $M'$. Finally, according to Lemma~\ref{SLPNFAMembershipLemma}, $\doc \in \lang(M')$ can be checked in time $\bigO(|N| |Q'|^3) = \bigO(|N| \states^3)$. Therefore, the total running time is $\bigO(|M| + |N| \states^3)$.
\item According to Proposition~\ref{basicSubwordMarkedLanguageProposition}, $t \in \llbracket M \rrbracket(\doc)$ if and only if $\insertmarkers{\doc}{t} \in L(M)$. Thus, it is sufficient to construct an $\SLP$ $\mathcal{S}' = (N', \Sigma \cup \powerset{\Gamma_{\varset}}, R', S'_0)$ with $\deriv{\mathcal{S}'} = \insertmarkers{\doc}{t}$ and then check whether $\deriv{\mathcal{S}'} \in L(M)$. According to Lemma~\ref{SLPNFAMembershipLemma}, the latter be done in time $\bigO(\size{\mathcal{S}'}\states^3)$.\par 
We conclude the proof by explaining how $\mathcal{S}'$ can be obtained from $\mathcal{S}$. We first compute all numbers $\card{\deriv{A}}$ with $A \in N$, which, according to Lemma~\ref{computeDerivationLengthsLemma}, can be done in time $\bigO(\size{\mathcal{S}})$. Let $j_1, j_2, \ldots, j_{\ell} \in \mathbb{N}$ be such that, for every $i \in \{j_1, j_2, \ldots, j_{\ell}\}$, there is some $(\sigma, i) \in \setrep{t}$. Moreover, for every $i \in \{j_1, j_2, \ldots, j_{\ell}\}$, let $\Lambda_{i} = \{\sigma \mid (\sigma, i) \in \setrep{t}\}$.
Note that $\ell \leq 2|\varset|$ and that the sets $\Lambda_{i}$ with $i \in \{j_1, j_2, \ldots, j_{\ell}\}$ can can be obtained from $\setrep{t}$ in time $\bigO(|\varset|)$. 
For every $i \in \{j_1, j_2, \ldots, j_{\ell}\}$ we proceed as follows. 
We search the derivation tree of $\mathcal{S}$ for the node $T_x$ that corresponds to position $i$ of $\doc$. This can be done by starting in the root $S_0$ and for every encountered internal node $A$ with rule $A \to BC$, we use the numbers $\card{\deriv{B}}$ and $\card{\deriv{C}}$ to decide where to descend. More precisely, we initialise a counter $c = 1$ and for every internal node $A$ (this includes the initial case $A = S_0$) that we encounter, we descend to the left child $B$ if $i \leq c + \card{\deriv{B}} - 1$, and we descend to the right child $C$ if $i > c + \card{\deriv{B}} - 1$; moreover, if we descend to the right child, we update $c$ by adding $\card{\deriv{B}}$ to it. We interrupt this procedure if we encounter $c = i$ and the current node is $T_x$ (note that $c = i$ can also happen at some internal node, in which case we have to descend further, but only to left children). In each step of this procedure we have to do a constant number of arithmetic operations with respect to numbers of size $\log(\docSize)$, and there are at most $\bigO(\depth{\mathcal{S}})$ steps to perform. Thus, we can find this leaf in $\mathcal{S}$'s derivation tree in time $\bigO(\depth{\mathcal{S}})$. \par
Now we replace the leaf $T_{x}$ by a new non-terminal $T_{\Lambda_i x}$ with rule $T_{\Lambda_i x} \to T_{\Lambda_i} T_x$, where $T_{\Lambda_i}$ is a new non-terminal with rule $T_{\Lambda_i} \to \Lambda_i$. Moreover, every non-terminal $A$ with rule $A \to CD$ encountered in the path from the root $S_0$ to the leaf $T_x$ has to be replaced by $A_i \to C_i D$ or $A_i \to C D_i$ depending on whether the path proceeds in $C$ or in $D$ (where $A_i, C_i$ and $D_i$ are new non-terminals). We note that the thus modified $\SLP$ is in Chomsky normal form.\par
For each $i \in \{j_1, j_2, \ldots, j_{\ell}\}$, this construction can be carried out in time $\bigO(\depth{\mathcal{S}})$ and adds $\bigO(\depth{\mathcal{S}})$ new non-terminals to the $\SLP$. Thus, $\mathcal{S}'$ is constructed in time $\bigO(|\varset|\depth{\mathcal{S}})$; and $|\mathcal{S}'| = \bigO(|\mathcal{S}| + |\varset| \depth{\mathcal{S}})$.
Finally, it can be easily verified that $\deriv{\mathcal{S}'} = \insertmarkers{\doc}{t}$, i.\,e., it has the desired property. Hence, 
the total time required for checking $t \in \llbracket M \rrbracket(w)$ is 
\begin{equation*}
\bigO(\size{\mathcal{S}'}|Q|^3) = \bigO((|\mathcal{S}| + |\varset| \depth{\mathcal{S}})|Q|^3)\,.
\end{equation*}
\end{enumerate}
\end{proof}

\section{Proofs omitted in Section~\ref{sec:alogPrelim}}

\subsection*{Proof of Lemma~\ref{markerTransMatSolutionSetComputationLemma}}

\begin{proof}
Every $\Lambda \in \llbracket M \rrbracket(\doc)$ is a partial marker set compatible with $\doc$ and $\insertmarkers{\doc}{\Lambda} \in \lang(M)$. Thus, for some $\State{j} \in F$, we have $\State{1} \transfuncNFA{\insertmarkers{\doc}{\Lambda}} \State{j}$. Moreover, since $\lang(M)$ is non tail-spanning, also $\insertmarkers{\doc}{\Lambda}$ is non tail-spanning, which means that $\Lambda \in \markerTransMat_{S_0}[1, j]$.\par
On the other hand, let $\Lambda \in \markerTransMat_{S_0}[1, j]$ for some $\State{j} \in F$. This means that $\Lambda$ is compatible with $\doc$, $\insertmarkers{\deriv{A}}{\Lambda}$ is non tail-spanning, and $\State{1} \transfuncNFA{\insertmarkers{\doc}{\Lambda}} \State{j}$.
Consequently, $\insertmarkers{\doc}{\Lambda} \in \lang(M)$ with $\getWord{\insertmarkers{\doc}{\Lambda}} = \doc$. Thus, $\markedpositions{\insertmarkers{\doc}{\Lambda}} = \Lambda \in \llbracket M \rrbracket(\doc)$.
\end{proof}

\subsection*{Proof of Lemma~\ref{inductionBaseLemma}}

\begin{proof}
We first show how to compute all $\markerTransMat_{T_x}$ with $x \in \Sigma$, from which we immediately get all $\reachTransMat_{T_x}$. Then, using $\reachTransMat_{T_x}$ as the base of an induction, we can compute the matrices $\reachTransMat_A$ and $\intStatesTransMat_A$ for all inner non-terminals $A \in N$.\par
We first prove the following claim.
\smallskip\\
\noindent \emph{Claim $1$}: For every $x \in \Sigma$ and every $i, j \in [\states]$, we have
\begin{equation*}
\markerTransMat_{T_x}[i, j]\: = \:\{\markedpositions{A_1 x} \mid A_1 \in \powerset{\Gamma_{\varset}},\:\State{i} \transfuncNFA{A_1 x} \State{j}\}\,.
\end{equation*}
\noindent \emph{Proof of Claim $1$}: Let $\Lambda \in \markerTransMat_{T_x}[i, j]$ be arbitrarily chosen. By definition, $\Lambda$ is compatible with $\deriv{T_x} = x$ and $\insertmarkers{x}{\Lambda}$ is non tail-spanning. This directly implies that, for some $A_1 \in \powerset{\Gamma_{\varset}}$, we have that $\Lambda = \{(\sigma, 1) \mid \sigma \in A_1\}$ and therefore $\Lambda = \markedpositions{A_1 x}$. Moreover, since $\State{i} \transfuncNFA{\insertmarkers{\deriv{T_x}}{\Lambda}} \State{j}$, we have that $\State{i} \transfuncNFA{A_1 x} \State{j}$. \par
On the other hand, if for some $A_1 \in \powerset{\Gamma_{\varset}}$ we have that $\State{i} \transfuncNFA{A_1x} \State{j}$, then $\markedpositions{A_1 x}$ is compatible with $x$ and $A_1 x$ is non-tail-spanning. Moreover, since $\insertmarkers{\deriv{T_x}}{\markedpositions{A_1 x}} = A_1 x$, this means that $\markedpositions{A_1 x} \in \markerTransMat_{T_x}[i, j]$. \hfill \qed (\emph{Claim $1$})\smallskip\\
This means that we can compute all $\markerTransMat_{T_x}$ with $x \in \Sigma$ as follows. For every $i \in [\states]$, we compute the set $P_i = \{(\ell, Y) \mid Y \in \powerset{\Gamma_{\varset}}, \State{\ell} \transfuncNFA{Y} \State{i}\}$. This can be done in time $\bigO(|M|)$ by iterating through each arc $(\State{i}, K, \State{j})$ of $M$ and adding $(i, K)$ to $P_j$ if and only if $K \in \powerset{\Gamma_{\varset}}$. Then, for every $x \in \Sigma$, and for every $i, j \in [\states]$, we initialise $\markerTransMat_{T_x}[i, j] = \emptyset$, which can be done in time $\bigO(|\Sigma|\states^2)$. Then, we iterate through each arc $(\State{i}, x, \State{j})$ of $M$ with $x \in \Sigma$ and do the following:
\begin{itemize}
\item We add $\emptyset$ to $\markerTransMat_{T_x}[i, j]$.
\item For every $(\ell, Y) \in P_i$, we add $\{(\sigma, 1) \mid \sigma \in Y\}$ to $\markerTransMat_{T_x}[\ell, j]$.
\end{itemize}
Since $|\bigcup_{i \in [\states]} P_i| = \bigO(|M|)$, the above procedure can be carried out in time $\bigO(|M| + \states^2)$.\par
As observed above, for every $x \in \Sigma$, we can directly obtain $\reachTransMat_{T_{x}}$ from $\markerTransMat_{T_x}$ in time $\bigO(|\Sigma|\states^2)$. \par
Next, we recursively compute all $\reachTransMat_{A}$ and $\intStatesTransMat_A$ with $A \to BC$ in a bottom-up fashion, i.\,e., we show how $\reachTransMat_{A}$ and $\intStatesTransMat_A$ can be computed under the assumption that $\reachTransMat_{B}$, $\reachTransMat_{C}$, $\intStatesTransMat_B$ and $\intStatesTransMat_C$ are already computed.\par
Let $A \in N$ with $A \to BC$ and $i, j \in [\states]$ be fixed. We first set $\reachTransMat_{A}[i, j] = \undefin$ and then we iterate through all $k \in [\states]$ and if $\reachTransMat_{B}[i, k] \neq \undefin$ and $\reachTransMat_{C}[k, j] \neq \undefin$, then we set $\reachTransMat_{A}[i, j] = \emptyMarker$ and add $k$ to $\intStatesTransMat_A[i, j]$. Now, the set $\intStatesTransMat_A[i, j]$ is computed correctly, but the entry $\reachTransMat_{A}[i, j]$ is only correct if $\reachTransMat_{A}[i, j] = \undefin$ or $\reachTransMat_{A}[i, j] = \emptyMarker$. Therefore, we iterate again through all $k \in [\states]$ and if $\reachTransMat_{B}[i, k] = \nonEmptyMarker$ or $\reachTransMat_{C}[k, j] = \nonEmptyMarker$, then we set $\reachTransMat_{A}[i, j] = \nonEmptyMarker$.\par
Since we have to do this for each $A \in N$ and $i, j \in [\states]$, we can do this in time $\bigO(|N|\states^3)$. Consequently, the total time required is $\bigO(|M| + |N|\states^3)$.
\end{proof}

\subsection*{Proof of Lemma~\ref{mainConstructionLemma}}

\begin{proof}
For convenience, we set $w_A = \deriv{A}$, $w_B = \deriv{B}$ and $w_C = \deriv{C}$. \par
``(\ref{mainConstructionLemmaPointOne}) $\Rightarrow$ (\ref{mainConstructionLemmaPointTwo})'': We assume that $\Lambda_A \in \markerTransMat_A[i, j]$. Let $v_A = \insertmarkers{w_A}{\Lambda_A}$. Since $w_A = w_B w_C$, there must be marked words $v_B$ and $v_C$ such that $v_A = v_B v_C$, $\getWord{v_B} = w_B$, $\getWord{v_C} = w_C$ and both $v_B$ and $v_C$ are non tail-spanning (note that, by assumption, $v_A$ is non-tail-spanning). Next, we set $\Lambda_B = \markedpositions{v_B}$ and $\Lambda_C = \markedpositions{v_C}$, which also means that $v_B = \insertmarkers{w_B}{\Lambda_B}$ and $v_C = \insertmarkers{w_C}{\Lambda_C}$. In particular, $\Lambda_B$ is compatible with $w_B$, $\Lambda_C$ is compatible $w_C$, and both $\insertmarkers{w_B}{\Lambda_B}$ and $\insertmarkers{w_C}{\Lambda_C}$ are non-tail-spanning. Furthermore, $\Lambda_A = \Lambda_B \msprod{|w_{B}|} \Lambda_C$. From $\State{i} \transfuncNFA{v_A} \State{j}$ and $v_A = v_B v_C$, we can directly conclude that there is some $k \in [\states]$ with $\State{i} \transfuncNFA{v_B} \State{k} \transfuncNFA{v_C} \State{j}$. Moreover, since $\Lambda_B = \markedpositions{v_B}$ and $\Lambda_C = \markedpositions{v_C}$, and $\getWord{v_B} = w_B = \deriv{B}$ and $\getWord{v_C} = w_C = \deriv{C}$, we have $\Lambda_B \in \markerTransMat_B[i, k]$ and $\Lambda_C \in \markerTransMat_C[k, j]$. In particular, this means that $k \in \intStatesTransMat_A[i, j]$.\par
``(\ref{mainConstructionLemmaPointTwo}) $\Rightarrow$ (\ref{mainConstructionLemmaPointOne})'': We assume that there are a $k \in \intStatesTransMat_A[i, j]$ and partial marker sets $\Lambda_B \in \markerTransMat_B[i, k]$ and $\Lambda_C \in \markerTransMat_C[k, j]$, such that $\Lambda_A = \Lambda_B \msprod{\card{\deriv{B}}} \Lambda_C$. Since $\Lambda_B$ and $\Lambda_C$ are compatible with $w_{B}$ and $w_{C}$, respectively, and since $w_A = w_B w_C$, we can conclude that $\Lambda_A$ is compatible with $w_A$. In particular, this means that $\insertmarkers{w_{A}}{\Lambda_A}$ is defined. Since $\insertmarkers{w_{B}}{\Lambda_B}$ is non-tail-spanning, we also know that $\insertmarkers{w_{B}}{\Lambda_B} \insertmarkers{w_{C}}{\Lambda_C} = \insertmarkers{w_{A}}{\Lambda_A}$. Since $\insertmarkers{w_{C}}{\Lambda_C}$ is non-tail-spanning, we can conclude that $\insertmarkers{w_{A}}{\Lambda_A}$ is non-tail-spanning. Finally, since 
\begin{equation*}
\State{i} \transfuncNFA{\insertmarkers{w_{B}}{\Lambda_B}} \State{k} \transfuncNFA{\insertmarkers{w_{C}}{\Lambda_C}} \State{j}\,, 
\end{equation*}
we obtain that $\State{i} \transfuncNFA{\insertmarkers{w_{B}}{\Lambda_B} \insertmarkers{w_{C}}{\Lambda_C}} \State{j}$, which means that  $\State{i} \transfuncNFA{\insertmarkers{w_{A}}{\Lambda_A}} \State{j}$.
Thus, $\Lambda_A \in \markerTransMat_A[i, j]$.
\end{proof}

\subsection*{Proof of Lemma~\ref{mainInductiveLemma}}

\begin{proof}
``$\subseteq$'': Let $A \to BC$ be the rule of $A$. Let $\Lambda \in \markerTransMat_A[i, j]$. By Lemma~\ref{mainConstructionLemma}, there is a $k \in \intStatesTransMat_A[i, j]$ and partial marker sets $\Lambda_B \in \markerTransMat_B[i, k]$ and $\Lambda_C \in \markerTransMat_C[k, j]$, such that $\Lambda = \Lambda_B \msprod{\card{\deriv{B}}} \Lambda_C$. Thus, $\Lambda \in \mainNode{A}{i}{j}{k} \subseteq \bigcup_{\ell \in \intStatesTransMat_A[i, j]} \mainNode{A}{i}{j}{\ell}(A)$.\par
``$\supseteq$'': Let $\Lambda \in \mainNode{A}{i}{j}{k}$ for some $k \in \intStatesTransMat_A[i, j]$. By definition, this means that $\Lambda = \Lambda_B \msprod{\card{\deriv{B}}} \Lambda_C$ for some $\Lambda_B \in \markerTransMat_B[i, k]$ and $\Lambda_C \in \markerTransMat_C[k, j]$. By Lemma~\ref{mainConstructionLemma}, we obtain that $\Lambda \in \markerTransMat_A[i, j]$.
\end{proof}

\subsection*{Proof of Lemma~\ref{simpleDisjointnessLemma}}

\begin{proof}
The direction ``$\Leftarrow$'' is trivial. For the direction ``$\Rightarrow$'', we assume that $\Lambda_B \msprod{\card{\deriv{B}}} \Lambda_C = \Lambda'_B \msprod{\card{\deriv{B}}} \Lambda'_C$ for $\Lambda_B, \Lambda'_B \in \markerTransMat_B[i, k]$ and $\Lambda_C, \Lambda'_C \in \markerTransMat_C[k, j]$. Our goal is to show that $\Lambda_B = \Lambda'_B$ and $\Lambda_C = \Lambda'_C$. Since every $(\sigma, p) \in \Lambda_B \cup \Lambda'_B$ satisfies $p \leq \card{\deriv{B}}$, and every $(\sigma, p) \in \rightshift{\Lambda_C}{\card{\deriv{B}}} \cup \rightshift{\Lambda'_C}{\card{\deriv{B}}}$ satisfies $p > \card{\deriv{B}}$, equality between $\Lambda_B \msprod{\card{\deriv{B}}}\Lambda_C$ and $\Lambda'_B \msprod{\card{\deriv{B}}} \Lambda'_C$ is only possible if $\Lambda_B = \Lambda'_B$ and $\rightshift{\Lambda_C}{\card{\deriv{B}}} = \rightshift{\Lambda'_C}{\card{\deriv{B}}}$. Since the mapping $\rightshift{\cdot}{\card{\deriv{B}}}$ is injective on the set of partial marker sets, this yields that $\Lambda_C = \Lambda'_C$.
\end{proof}

\section{Details omitted in Section~\ref{sec:computation}}
\subsection*{Proof of Theorem~\ref{computeSetTheorem}}

\begin{proof}
We first give a high-level description of the algorithm. In general, for given $A \in N$ and $i, j \in [\states]$, we can recursively compute $\markerTransMat_{A}[i, j]$ as follows. If $A = T_x$ is a leaf non-terminal, then we can assume that we have computed $\markerTransMat_{A}[i, j]$ already in a preprocessing phase according to Lemma~\ref{inductionBaseLemma}. If $A \to BC$ is a rule, then, for every $k \in \intStatesTransMat_A[i, j]$, we recursively compute $\markerTransMat_B[i, k]$ and $\markerTransMat_C[k, j]$, and then set $\markerTransMat_A[i, j] = \bigcup_{k \in \intStatesTransMat_A[i, j]} \mainNode{A}{i}{j}{k}$ (see Lemma~\ref{mainInductiveLemma}), where $\mainNode{A}{i}{j}{k} \: = \: \markerTransMat_B[i, k] \msprod{\card{\deriv{B}}} \markerTransMat_C[k, j]$ (Definition~\ref{kSetsDefinition}). Then, according to Lemma~\ref{markerTransMatSolutionSetComputationLemma}, $\llbracket M \rrbracket(\doc) = \bigcup_{j \in F'}\markerTransMat_{S_0}[1,j]$, where $F' = \{\State{j} \in F \mid \reachTransMat_{S_0}[1,j] \neq \undefin\}$.\medskip\\
\textbf{Sets of Marker Sets as Sorted Lists}: In order to handle the problem of duplicates in unions of sets of marker sets, we use an order on marker sets as follows. First, we define a way to extend a total order $\seqleq$ on some alphabet $A$ to words from $A^*$. Let $u, v \in A^*$, then we set $u \seqleq v$ if, for some $i$ with $i \leq |u|$ and $i \leq |v|$, 
\begin{equation*}
u\spann{1}{i+1} = v\spann{1}{i+1} 
\end{equation*}
and either $i = |v|$ or $u[i+1] \seqleq v[i+1]$. This means that words are ordered according to the leftmost position where they differ, and if one is a prefix of the other, then the prefix is larger according to $\seqleq$ (and not the smaller one as it is the case for the normal lexicographic order). \par
Now let $\msleq$ be any order on $\Gamma_{\varset}$. We extend $\msleq$ to an order on $\Gamma_{\varset} \times \mathbb{N}$ and then to an order on marker sets as follows. For $(\sigma_1, i_1), (\sigma_2, i_2) \in (\Gamma_{\varset} \times \mathbb{N})$, we set $(\sigma_1, i_1) \msleq (\sigma_2, i_2)$ if either $i_1 < i_2$ or $i_1 = i_2$ and $\sigma_1 \msleq \sigma_2$. Now for any marker set $\Lambda$, let $\msSort{\Lambda}$ be the word over alphabet $\Gamma_{\varset} \times \mathbb{N}$ obtained by appending $\Lambda$'s elements in ascending order with respect to $\msleq$. Finally, we extend $\msleq$ to words over $\Gamma_{\varset} \times \mathbb{N}$ in the way described above, and, for marker sets $\Lambda_1, \Lambda_2$, we set $\Lambda_1 \msleq \Lambda_2$ if $\msSort{\Lambda_1} \msleq \msSort{\Lambda_2}$.\par
In the following, when we talk about \emph{sorted lists} of some sets $\Delta$ of marker sets, we always mean a list that contains $\Delta$'s elements (without duplicates) as words over $\Gamma_{\varset} \times \mathbb{N}$ as described above, and sorted in increasing order according to $\msleq$.\par
We observe the following important property of the order $\msleq$. Let $A \to BC$ be a rule, let $\Lambda_{B}, \Lambda_C$ be marker sets compatible with $\deriv{B}$ and $\deriv{C}$, respectively. We let $\Lambda_{BC} = \Lambda_{B} \msprod{\card{\deriv{B}}} \Lambda_{C}$ and recall that $\Lambda_{BC} = \Lambda_{B} \cup \rightshift{\Lambda_C}{\card{\deriv{B}}}$. Now let $\Lambda'_{B}, \Lambda'_C$ be other marker sets compatible with $\deriv{B}$ and $\deriv{C}$, respectively, and let $\Lambda'_{BC} = \Lambda'_{B} \msprod{\card{\deriv{B}}} \Lambda'_{C}$. By our choice of the order $\msleq$, we can directly conclude that $\Lambda_B \mseq \Lambda'_{B}$ implies $\Lambda_{BC} \mseq \Lambda'_{BC}$. On the other hand, if $\Lambda_B = \Lambda'_{B}$, then $\Lambda_{BC} \mseq \Lambda'_{BC}$ if and only if $\Lambda_C \mseq \Lambda'_{C}$. \par
This also means that if we have sets $\markerTransMat_B[i, k]$ and $\markerTransMat_C[k, j]$ given as sorted lists, then we can obtain a sorted list of $\markerTransMat_B[i, k] \msprod{\card{\deriv{B}}} \markerTransMat_C[k, j]$ in time $\bigO(|\varset| \cdot |\markerTransMat_B[i, k]| \cdot |\markerTransMat_C[k, j]|)$ by iterating through all elements $\Lambda_B \in \markerTransMat_B[i, k]$ and for each such element iterating through all elements $\Lambda_C \in \markerTransMat_C[k, j]$ and spending time $\bigO(|\varset|)$ for constructing $\Lambda_B \msprod{\card{\deriv{B}}} \Lambda_C$.\par
Analogously, if we have sets $\Delta_1, \Delta_2, \ldots, \Delta_n$ of marker sets given as sorted lists, then we can construct a sorted list of $\bigcup_{i \in [n]} \Delta_i$ (without duplicates) in time $\bigO(n \cdot |\varset| \cdot \sum_{i \in [n]} |\Delta_i|)$.\medskip\\
\textbf{Algorithm}: We now describe the algorithm in detail. For convenience, we state the algorithm and the proof of correctness in terms of sets of marker sets. However, when estimating the time complexity, then we assume that the actual implementation will represent sets of marker sets as sorted lists as defined above.
This aspect will be made precise in the running time estimation of the algorithm, but for our argument of correctness, these issues do not matter.\par
We use Boolean matrices $\compMatrix{A}$ in order to store which entries of matrices $\markerTransMat_{A}$ have already been computed. 
\begin{enumerate}
\item\label{CompAlgoStepOne} We initially compute the following matrices.
\begin{itemize}
\item For every $A \in N$ and $i, j \in [\states]$, set $\compMatrix{A}[i, j] = 0$.
\item Compute all the matrices $\reachTransMat_A$ for every $A \in N$, $\intStatesTransMat_{A'}$ for every inner non-terminal $A' \in N$, and\,$\markerTransMat_{T_x}$ for every $x \in \Sigma$ according to Lemma~\ref{inductionBaseLemma}.
For every $x \in \Sigma$ and every $i, j \in [\states]$, set $\compMatrix{T_x}[i, j] = 1$.
\item Compute $F' = \{\State{j} \in F \mid \reachTransMat_{S_0}[1,j] \neq \undefin\}$.
\end{itemize}
\item\label{CompAlgoStepTwo} 
For every $\State{j} \in F'$,
we compute $\markerTransMat_{S_0}[1, j]$ by calling the recursive procedure $\recProc{S_0}{1}{j}$, which is defined as follows:\smallskip\\
$\recProc{A}{i}{j}$:
\begin{itemize}
\item If $\compMatrix{A}[i, j] = 1$, then return $\markerTransMat_{A}[i, j]$. 
\item If $\compMatrix{A}[i, j] = 0$ and $A \to BC$ is a rule, then, 
for every $k \in \intStatesTransMat_A[i, j]$, compute 
\begin{equation*}
M^k_A = \recProc{B}{i}{k} \msprod{\card{\deriv{B}}} \recProc{C}{k}{j}\,,
\end{equation*}
\item Return $M_A = \bigcup_{k \in \intStatesTransMat_A[i, j]} M^k_A$.
\end{itemize}
\item\label{CompAlgoStepThree} 
Produce $\bigcup_{\State{j} \in F'} \markerTransMat_{S_0}[1, j]$ as output.
\end{enumerate}
\textbf{Correctness}: Lemma~\ref{markerTransMatSolutionSetComputationLemma} implies that the output $\bigcup_{\State{j} \in F'} \markerTransMat_{S_0}[1, j]$ equals $\llbracket M \rrbracket(\doc)$. Thus, in order to conclude the proof of correctness, we only have to show that all the entries $\markerTransMat_{A}[i, j]$ are correctly computed by the call of $\recProc{A}{i}{j}$. For $A = T_x$ this follows from Lemma~\ref{inductionBaseLemma}. Now assume that $A$ is an inner non-terminal with a rule $A \to BC$, and assume that, for every $k \in \intStatesTransMat_A[i, j]$, the sets $\markerTransMat_{B}[i, k]$ and $\markerTransMat_{C}[k, j]$ are already computed (by our recursive approach, we know that we can assume this). Then $\recProc{A}{i}{j}$ computes $\bigcup_{k \in \intStatesTransMat_A[i, j]} (\markerTransMat_{B}[i, k] \msprod{\card{\deriv{B}}} \markerTransMat_{C}[k, j]) = \bigcup_{k \in \intStatesTransMat_A[i, j]} \mainNode{A}{i}{j}{k}$. By Lemma~\ref{mainInductiveLemma}, $\bigcup_{k \in \intStatesTransMat_A[i, j]} \mainNode{A}{i}{j}{k} = \markerTransMat_A[i, j]$, so we correctly compute $\markerTransMat_A[i, j]$.
\medskip\\
\textbf{Complexity}: According to Lemma~\ref{inductionBaseLemma}, all the required matrices of Step~(\ref{CompAlgoStepOne}) can be computed in total time $\bigO(|M| + |N|\states^3)$. However, we also require a sorted list of each set $\markerTransMat_{T_{x}}[i, j]$ with $x \in \Sigma$ and $i, j \in [\states]$, which can be achieved as follows.\par
Each $\Lambda \in \markerTransMat_{T_{x}}[i, j]$ is a subset of $\{(1, \sigma) \mid \sigma \in \Gamma_{\varset}\}$, and hence $\msSort{\Lambda}$ can be constructed in time $\bigO(|\varset|)$ for each such $\Lambda$. Furthermore, since $|\markerTransMat_{T_{x}}[i, j]| \leq |M|$, we can obtained $\{\msSort{\Lambda} \mid \Lambda \in \markerTransMat_{T_{x}}[i, j]\}$ in time $\bigO(|M| \cdot |\varset|)$. By sorting $\{\msSort{\Lambda} \mid \Lambda \in \markerTransMat_{T_{x}}[i, j]\}$, we obtain a sorted list of $\markerTransMat_{T_{x}}[i, j]$ in time $\bigO(\Sort{|\markerTransMat_{T_{x}}[i, j]|}) = \bigO(\Sort{|M|})$.
Thus, Step~(\ref{CompAlgoStepOne}) is accomplished in total time $\bigO(\Sort{|M|} \cdot \states^2 +\;|N|\states^3)$. \par
In Step~(\ref{CompAlgoStepThree}), we have to compute $\bigcup_{\State{j} \in F'} \markerTransMat_{S_0}[1, j]$. Under the assumption that all $\markerTransMat_{S_0}[1, j]$ are provided as sorted lists (we shall in the discussion of Step~(\ref{CompAlgoStepTwo}) that we can achieve this), then this can be done in time 
\begin{align*}
\bigO(|F'| \cdot |\varset| \cdot \sum_{j \in F'} \card{\markerTransMat_{S_0}[1, j]})\,. 
\end{align*}
Since every $\card{\markerTransMat_{S_0}[1, j]}$ is bounded by $\card{\llbracket M \rrbracket(\doc)}$ we can therefore compute $\bigcup_{\State{j} \in F'} \markerTransMat_{S_0}[1, j]$ in time
\begin{align*}
\bigO(\states \cdot \size{\llbracket M \rrbracket(\doc)})\,.
\end{align*}\par
Estimating the complexity of Step~(\ref{CompAlgoStepTwo}) is more complicated. We first note that for a fixed $A \in N$ with rule $A \to BC$ and $i, j \in [\states]$
computing $\markerTransMat_A[i, j]$ is done by computing $\bigcup_{k \in \intStatesTransMat_A[i, j]} \mainNode{A}{i}{j}{k}$, where $\mainNode{A}{i}{j}{k} = \markerTransMat_{B}[i, k] \msprod{\card{\deriv{B}}} \markerTransMat_{C}[k, j]$. For every $k \in \intStatesTransMat_A[i, j]$, assuming that we have $\markerTransMat_B[i, k]$ and $\markerTransMat_C[k, j]$ at our disposal as sorted lists, then a sorted list of $\mainNode{A}{i}{j}{k}$ can be computed in time 
\begin{align*}
\bigO\left(|\varset| \cdot |\markerTransMat_B[i, k]| \cdot |\markerTransMat_C[k, j]|\right)\,. 
\end{align*}
Next, we will show, for every $k \in \intStatesTransMat_A[i, j]$, that $|\markerTransMat_B[i, k]| \cdot |\markerTransMat_C[k, j]|$ is in fact upper bounded by $|\llbracket M \rrbracket(\doc)|$.\par
First, we introduce some helpful notation. For $A \in N$ and $i, j \in [\states]$, we say that the triple $(A, i, j)$ satisfies \emph{condition $(\dagger)$} if the following holds:
\begin{quote}
There is some subword-marked word $v \in \lang(M)$ with $\getWord{v} = \doc$ and $v = v_1 v_2 v_3$ with $\getWord{v_2} = \deriv{A}$, $\State{1} \transfuncNFA{v_1} \State{i} \transfuncNFA{v_2} \State{j} \transfuncNFA{v_3} F$. 
\end{quote}
I.\,e. if $(A, i, j)$ satisfies condition $(\dagger)$, then there is some $\Lambda \in \markerTransMat_{A}[i, j]$, and there is a subword-marked word $v$ with $\getWord{v} = \doc$ that is accepted by $M$ in such a way that between state $\State{i}$ and state $\State{j}$ the marked word $\insertmarkers{\deriv{A}}{\Lambda}$ is read. \smallskip\\
\noindent \emph{Claim $1$}: For every $A \in N$ and $i, j \in [\states]$ such that $\markerTransMat_{A}[i, j]$ is computed in Step~(\ref{CompAlgoStepTwo}), the triple $(A, i, j)$ satisfies property $(\dagger)$. \smallskip\\
\noindent \emph{Proof of Claim $1$}: We proceed by induction. For all $j \in [\states]$ with $\State{j} \in F$, $\markerTransMat_{S_0}[1, j]$ is only computed if $\reachTransMat_{S_0}[1, j] \neq \undefin$, which means that there is a partial marker set $\Lambda$ compatible with $\deriv{S_0}$ such that $\State{1} \transfuncNFA{\insertmarkers{\deriv{S_0}}{\Lambda}} \State{j}$. This means that property $(\dagger)$ is satisfied with $v = \insertmarkers{\deriv{S_0}}{\Lambda}$ and with respect to the factorisation $v = v_1 v_2 v_3$ with $v_1 = v_3 = \emptyword$. \par
Now assume that in the recursion we compute $\markerTransMat_{A}[i, j]$ for some $A \in N$ and $i, j \in [\states]$, and that $(A, i, j)$ satisfies property $(\dagger)$. We know that there is a rule $A \to BC$, since if $A = T_x$, then $\markerTransMat_{A}$ has already been computed in Step~(\ref{CompAlgoStepOne}), which is a contradiction to the assumption that $\markerTransMat_{A}[i, j]$ is computed in Step~(\ref{CompAlgoStepTwo}). Moreover, since $(A, i, j)$ satisfies property $(\dagger)$, there is some subword-marked word $v \in \lang(M)$ with $\getWord{v} = \doc$ and $v = v_1 v_2 v_3$ with $\getWord{v_2} = \deriv{A}$, $\State{1} \transfuncNFA{v_1} \State{i} \transfuncNFA{v_2} \State{j} \transfuncNFA{v_3} F$. \par
Now assume that for some $k \in \intStatesTransMat_A[i, j]$ the sets $\markerTransMat_B[i, k]$ and $\markerTransMat_C[k, j]$ are computed in Step~(\ref{CompAlgoStepTwo}). This means that there are partial marker sets $\Lambda_B \in \markerTransMat_{B}[i, k]$ and $\Lambda_C \in \markerTransMat_{C}[k, j]$. Let $v_B = \insertmarkers{\deriv{B}}{\Lambda_B}$ and let $v_C = \insertmarkers{\deriv{C}}{\Lambda_C}$ (this is well-defined since $\Lambda_B$ is compatible with $\deriv{B}$ and $\Lambda_C$ is compatible with $\deriv{C}$). In particular, this also means that $\State{i} \transfuncNFA{v_B} \State{k} \transfuncNFA{v_C} \State{j}$. In summary, we know the following facts:
\begin{itemize}
\item $\State{1} \transfuncNFA{v_1} \State{i} \transfuncNFA{v_B} \State{k} \transfuncNFA{v_C} \State{j} \transfuncNFA{v_3} F$,
\item $v' = v_1 v_B v_C v_3$ is a subword-marked word from $\lang(M)$,
\item $\getWord{v'} = \doc$ (this follows from $\getWord{v_B v_C} = \deriv{A}$). 
\end{itemize}
From these facts, it directly follows that $(B, i, k)$ satisfies property $(\dagger)$ (with $v_B$ playing the role of $v_2$) and that $(C, k, j)$ satisfies property $(\dagger)$ (with $v_C$ playing the role of $v_2$).\hfill \qed (\emph{Claim $1$})\smallskip\\
We can now use Claim $1$ in order to prove the upper bound on $|\markerTransMat_B[i, k]| \cdot |\markerTransMat_C[k, j]|$ claimed above: 
\smallskip\\
\noindent \emph{Claim $2$}: Let $A \in N$ and let $i, j \in [\states]$. For every $k \in \intStatesTransMat_A[i, j]$, we have $|\markerTransMat_B[i, k]| \cdot |\markerTransMat_C[k, j]| \leq |\llbracket M \rrbracket(\doc)|$.\smallskip\\
\noindent \emph{Proof of Claim $2$}: Let $A \to BC$ be the rule for $A$ and
let $k \in \intStatesTransMat_A[i, j]$ be chosen arbitrarily. By definition, $\markerTransMat_A[i, j] = \bigcup_{k \in \intStatesTransMat_A[i, j]} \mainNode{A}{i}{j}{k}$, where 
\begin{align*}
\mainNode{A}{i}{j}{k} &= \markerTransMat_B[i, k] \msprod{\card{\deriv{B}}} \markerTransMat_C[k, j]\\
&= \{\Lambda_B \cup \rightshift{\Lambda_C}{\card{\deriv{B}}} \mid \Lambda_B \in \markerTransMat_B[i, k],  \Lambda_C \in \markerTransMat_C[k, j]\}\,.
\end{align*}
We make the following observations:
\begin{itemize}
\item $|\markerTransMat_B[i, k]| \cdot |\markerTransMat_C[k, j]| = |\mainNode{A}{i}{j}{k}|$: First note that 
\begin{equation*}
|\markerTransMat_B[i, k]| \cdot |\markerTransMat_C[k, j]| \geq |\mainNode{A}{i}{j}{k}| 
\end{equation*}
holds by definition, and now assume that 
\begin{equation*}
|\markerTransMat_B[i, k]| \cdot |\markerTransMat_C[k, j]| > |\mainNode{A}{i}{j}{k}|\,,
\end{equation*}
which means that there are $\Lambda_B, \Lambda'_B \in \markerTransMat_B[i, k]$ and $\Lambda_C, \Lambda'_C \in \markerTransMat_C[k, j]$ with 
\begin{equation*}
\Lambda_B \msprod{\card{\deriv{B}}} \Lambda_C = \Lambda'_B \msprod{\card{\deriv{B}}} \Lambda'_C\,, 
\end{equation*}
but $\Lambda_B \neq \Lambda'_B$ or $\Lambda_C \neq \Lambda'_C$. By Lemma~\ref{simpleDisjointnessLemma}, this is not possible and therefore $|\markerTransMat_B[i, k]| \cdot |\markerTransMat_C[k, j]| = |\mainNode{A}{i}{j}{k}|$.
\item $|\mainNode{A}{i}{j}{k}| \leq |\markerTransMat_A[i, j]|$: This follows from $\mainNode{A}{i}{j}{k} \subseteq \markerTransMat_A[i, j]$ (see Lemma~\ref{mainInductiveLemma}).
\item $|\markerTransMat_A[i, j]| \leq |\llbracket M \rrbracket(\doc)|$: Since $(A, i, j)$ satisfies property $(\dagger)$ (see Claim~$1$), there is some subword-marked word $v \in \lang(M)$ with $\getWord{v} = \doc$ and $v = v_1 v_2 v_3$ with $\getWord{v_2} = \deriv{A}$, $\State{1} \transfuncNFA{v_1} \State{i} \transfuncNFA{v_2} \State{j} \transfuncNFA{v_3} F$. Consequently, for every $\Lambda \in \markerTransMat_A[i, j]$ there is the subword-marked word $u_{\Lambda} = v_1 \insertmarkers{\deriv{A}}{\Lambda} v_3 \in \lang(M)$ that represents the element $\markedpositions{u_{\Lambda}}$ from $\llbracket M \rrbracket(\doc)$. The mapping $(\Lambda \mapsto \markedpositions{u_{\Lambda}})_{\Lambda \in \markerTransMat_A[i, j]}$ is an injective mapping (from $\markerTransMat_A[i, j]$ to $\llbracket M \rrbracket(\doc)$). Thus, $|\markerTransMat_A[i, j]| \leq |\llbracket M \rrbracket(\doc)|$.
\end{itemize}
Consequently, we have 
\begin{equation*}
|\markerTransMat_B[i, k]| \cdot |\markerTransMat_C[k, j]| = |\mainNode{A}{i}{j}{k}| \leq |\markerTransMat_A[i, j]| \leq |\llbracket M \rrbracket(\doc)|\,,
\end{equation*}
which concludes the proof of the claim.\hfill \qed (\emph{Claim $2$})\smallskip\\
In summary, sorted lists of all sets $\mainNode{A}{i}{j}{k}$ with $k \in \intStatesTransMat_A[i, j]$ can be computed in total time 
\begin{align*}
\bigO(\sum_{k \in \intStatesTransMat_A[i, j]} (|\varset| \cdot |\markerTransMat_B[i, k]| \cdot |\markerTransMat_C[k, j]|)) = \\
\bigO(\states \cdot |\varset| \cdot |\llbracket M \rrbracket(\doc)|) = \bigO(\states \cdot \size{\llbracket M \rrbracket(\doc)})\,.
\end{align*}
Moreover, with these sorted lists, we can now compute $\markerTransMat_A[i, j]$ by computing $\bigcup_{k \in \intStatesTransMat_A[i, j]} \mainNode{A}{i}{j}{k}$ in time 
\begin{align*}
\bigO(|\intStatesTransMat_A[i, j]| \cdot |\varset| \cdot \sum_{k \in \intStatesTransMat_A[i, j]} |\mainNode{A}{i}{j}{k}|) &= \bigO(\states^2 \cdot |\varset| \cdot |\llbracket M \rrbracket(\doc)|)\\
&= \bigO(\states^2 \cdot \size{\llbracket M \rrbracket(\doc)})\,.
\end{align*}
Since in Step~(\ref{CompAlgoStepOne}), we have computed sorted lists for all $\markerTransMat_{T_x}[i, j]$ for every $x \in \Sigma$ and $i, j \in [\states]$, we can assume in the recursive calls of $\recProc{A}{i}{j}$ that we always have the already computed sets of marker sets as sorted lists. \par
Since there are at most $|N| \cdot \states^2$ sets $\markerTransMat_A[i, j]$ to be computed (note that due to the matrices $\compMatrix{A}$ we compute each entry $\markerTransMat_A[i, j]$ at most once), the total running-time of Step~(\ref{CompAlgoStepTwo}) is $\bigO(|N| \cdot \states^4 \cdot \; \size{\llbracket M \rrbracket(\doc)})$. Thus, the total running time of the algorithm is 
\begin{align*}
&\bigO((\Sort{|M|} \cdot \states^2 +\; |N|\states^3)) + (|N| \cdot \states^4 \cdot \; \size{\llbracket M \rrbracket(\doc)})) = \\
&\bigO(\Sort{|M|} \cdot \states^2 + \;|N| \cdot \states^4 \cdot \; \size{\llbracket M \rrbracket(\doc)})\,.
\end{align*}
This completes the proof of Theorem~\ref{computeSetTheorem}. 
\end{proof}

\section{Details omitted in Section~\ref{sec:enumeration}}

\subsection*{Alternative Characterisation of (M,A)-Trees}

We give an alternative characterisation of $(M, A)$-trees that is also helpful for the following proofs. More precisely, in order to characterise $(M, A)$-trees, we give a (non-deterministic) recursive construction procedure $\constAlgo(A, i, k, j)$ (presented in Algorithm~\ref{constAlgo}) that, for any $A \in N$, $i, j \in [\states]$ and $k \in \intStatesTransMat_{A}[i, j] \cup \{\emptyMarker\}$, constructs a tree with a root labelled by $\treeNode{A}{i}{k}{j}$, $\treeNodeEmpty{A}{i}{j}$ or $\treeNodeTerminal{A}{i}{j}$. In Algorithm~\ref{constAlgo}, and also in the remainder of this section, we use the following convenient notation. For trees $\mathcal{T}_1$, $\mathcal{T}_2$ and $s_1, s_2 \in \mathbb{N}$, and a single node $P$ (or node label $P$), we denote by $P((\mathcal{T}_1, s_1), (\mathcal{T}_2, s_2))$ the tree with root $P$ that has the root of $\mathcal{T}_1$ as left child (with an arc labelled by $s_1$) and the root of $\mathcal{T}_2$ as right child (with an arc labelled by $s_2$).\par
The algorithm requires the data-structures $\reachTransMat_A$ and $\intStatesTransMat_A$, which we assume to be at our disposal (since this algorithm serves the purpose of defining $(M, A)$-trees, we are not concerned with complexity issues at this point). Moreover, we assume that, for every $A \in N$ and for every $i, j \in [\states]$, we have the set $\mathcal{I}_A[i, j]$ at our disposal, which is defined as follows. If $A = T_x$ or $\reachTransMat_{A}[i, j] = \emptyMarker$ then $\mathcal{I}_A[i, j] = \{\baseCaseMarker\}$, and $\mathcal{I}_A[i, j] = \intStatesTransMat_A[i, j]$ otherwise. This means that $\mathcal{I}_A[i, j] = \{\baseCaseMarker\}$ denotes that the triple $A, i, j$ describes a \emph{base} case of the recursion, i.\,e., $\reachTransMat_{A}[i, j] = \emptyMarker$ or $A$ is a leaf non-terminal.

\begin{algorithm}
\SetAlgoNoEnd
\LinesNumbered
\SetSideCommentRight
\SetFillComment
\SetKwInOut{Input}{Input}
\SetKwInOut{Output}{Output}
\Input{Non-terminal $A \in N$, $i, j \in [\states]$, $k \in \intStatesTransMat_A[i, j] \cup \{\baseCaseMarker\}$.}
\Output{A tree with a root $\treeNode{A}{i}{k}{j}$, $\treeNodeEmpty{A}{i}{j}$ or $\treeNodeTerminal{A}{i}{j}$}

\If{$k = \baseCaseMarker$}
{
	\If{$\reachTransMat_{A}[i, j] = \emptyMarker$}
	{
	\Return single node with label $\treeNodeEmpty{A}{i}{j}$\;\label{produceEmptyTreeLine}
	}
	\Else
	{
	\Return single node with label $\treeNodeTerminal{A}{i}{j}$\;\label{produceTerminalTreeLine}
	}
}
\ElseIf{$A$ is inner non-terminal with $A \to BC$}
{
	let $k_B \in \mathcal{I}_B[i, k]$ be chosen non-deterministically\;
	let $k_C \in \mathcal{I}_C[k, j]$ be chosen non-deterministically\;
	compute $\mathcal{T}_B = \constAlgo(B, i, k_B, k)$ \;
	compute $\mathcal{T}_C = \constAlgo(C, k, k_C, j)$ \;
	\Return $\treeNode{A}{i}{k}{j}((\mathcal{T}_B, 0), (\mathcal{T}_C, \card{\deriv{B}}))$\;
}  		
\caption{$\constAlgo(A, i, k, j)$}\label{constAlgo}
\end{algorithm}

Next, we make some observations about the algorithm $\constAlgo$. If $k \in \intStatesTransMat_A[i, j] \cup \{\baseCaseMarker\}$, then either $\constAlgo(A, i, k, j)$ terminates without further recursive calls, or there are two recursive calls $\constAlgo(B, i, k_B, k)$ and $\constAlgo(C, k, k_C, j)$ that also satisfy $k_B \in \intStatesTransMat_B[i, k] \cup \{\baseCaseMarker\}$ and $k_C \in \intStatesTransMat_C[k, j] \cup \{\baseCaseMarker\}$. 
Thus, $\constAlgo(A, i, k, j)$ is well-defined if $k \in \intStatesTransMat_A[i, j] \cup \{\baseCaseMarker\}$. \par
If we carry out $\constAlgo(A, i, k, j)$ with 
$k \in \mathcal{I}_A[i, j]$, then we construct a tree in which \emph{all} inner nodes labelled $\treeNode{A'}{i'}{k'}{j'}$ must satisfy that $\reachTransMat_{A'}[i', j'] = \nonEmptyMarker$, \emph{all} leaves labelled $\treeNodeEmpty{A'}{i'}{j'}$ 
must satisfy that $\reachTransMat_A'[i', j'] = \emptyMarker$, and \emph{all} leaves labelled $\treeNodeTerminal{A}{i'}{j'}$ must satisfy that $\reachTransMat_{A}[i', j'] = \nonEmptyMarker$ and that $A$ is a leaf non-terminal. We further note that in each call of $\constAlgo$, the only non-deterministic elements are the possible choices of $k_B \in \mathcal{I}_B[i, k]$ and $k_C \in \mathcal{I}_C[k, j]$.

\begin{observation}
For every $A \in N$, $(M, A)$-trees are exactly the trees that can be constructed by $\constAlgo(A, i, k, j)$ for some $i, j \in [\states]$ and $k \in \mathcal{I}_A[i, j]$.
\end{observation}

\subsection*{Proof of Lemma~\ref{treeSizeLemma}}

\begin{proof}
Let $v_1, v_2, \ldots, v_{\ell}$ be the terminal-leaves of $\mathcal{T}$. We claim that for any node $u$ of $\mathcal{T}$, if the subtree rooted by $u$ contains $\ell' \geq 1$ terminal-leaves, then there is some partial marker set $\Lambda \in \yield{u}$ with $|\Lambda| \geq \ell'$. This implies that $\ell \leq \max\{|\Lambda| \mid \Lambda \in \yield{\mathcal{T}}\}$, and since partial marker sets can contain at most $2|\varset|$ elements, we obtain that $\ell \leq 2|\varset|$, i.\,e., $\mathcal{T}$ has at most $2|\varset|$ terminal-leaves. We next prove this claim by induction.\par
As the basis of the induction, we first prove this statement for the case that $u$ is a terminal-leaf $v_i$ with $i \in [\ell]$. Since $v_i$ is labelled $\treeNodeTerminal{T_x}{i'}{j'}$ for some $x \in \Sigma$ and $i', j' \in [\states]$ with $\reachTransMat_{T_x}[i', j'] = \nonEmptyMarker$, there is some $\Lambda_i \in \yield{v_i}$ with $\Lambda_i \neq \emptyset$, i.\,e., $|\Lambda_i| \geq 1$.\par
 Now let $u$ be an arbitrary inner node such that the subtree rooted by $u$ contains terminal-leaves $v_{i_1}, v_{i_2}, \ldots, v_{i_{\ell'}}$. Moreover, assume that $u$ has a left child $u_l$ and a right child $u_r$, and that the subtrees rooted with $u_l$ and $u_r$ contain the terminal-leaves $v_{i_1}, \ldots, v_{i_{\ell''}}$ and $v_{i_{\ell''}}, \ldots, v_{i_{\ell'}}$, respectively. By induction, we conclude that there is a partial marker set $\Lambda_{u_l} \in \yield{u_l}$ with $|\Lambda_{u_l}| \geq \ell''$ and there is a partial marker set $\Lambda_{u_r} \in \yield{u_r}$ with $|\Lambda_{u_r}| \geq \ell' - \ell''$. Consequently, $\Lambda_u = \Lambda_{u_l} \msprod{\card{\deriv{B}}} \Lambda_{u_r} \in \yield{u}$, where $B$ is the non-terminal of $u_l$, i.\,e., $u_l$ is labelled $\treeNode{B}{i'}{k'}{j'}$, $\treeNodeTerminal{B}{i'}{j'}$ or $\treeNodeEmpty{B}{i'}{j'}$ for some $i', j', k' \in [\states]$. This means that $|\Lambda_u| = |\Lambda_{u_l}| + |\Lambda_{u_r}| \geq \ell'' + (\ell' - \ell'') = \ell'$.\par
By the definition of $(M, A)$-trees, each empty-leaf $v$ is a child of an inner node $u$ that lies on a path from some terminal-leaf to the root (otherwise, there would be a node $v$ with two empty-leaves as children, but this would mean that $v$ must already be an empty-leaf). Obviously, each inner node of such a path has at most one adjacent empty-leaf. Consequently, $|\mathcal{T}| \leq 2 |\mathcal{T}'|$, where $\mathcal{T}'$ is obtained from $\mathcal{T}$ by erasing all empty-leaves. We can also note that the leaves of $\mathcal{T}'$ are exactly the terminal-leaves of $\mathcal{T}$, since it is impossible that two empty-leaves are siblings (by definition, this would mean that the parent node must already be an empty-leaf). \par
Since the depth of $\mathcal{T}$ is at most $\depth{A}$, the tree $\mathcal{T}'$ has depth $\depth{A}$ and at most $2|\varset|$ leaves. This means that $|\mathcal{T}'| \leq 2|\varset| \cdot \depth{A}$ and therefore $|\mathcal{T}| \leq 2|\mathcal{T}'| \leq 4|\varset| \cdot \depth{A}$.
\end{proof}

\subsection*{Proof of Lemma~\ref{singleTreeEnumLemma}}

\begin{proof}
Let $L_1, L_2, \ldots, L_\ell$ be the terminal-leaves from $\mathcal{T}$ ordered from left to right, and assume that, for every $r \in [\ell]$, $L_r$ is labelled by $\treeNodeTerminal{T_{x_r}}{i_r}{j_r}$.
 For each $r \in [\ell]$, we compute the \emph{total shift} $s_r$, which is the sum of all arc-labels on the path from the root to $L_r$. Computing all these total shifts can be done by one top-down traversal of $\mathcal{T}$ and one addition of constant numbers of size $\card{\doc}$ at each node. Then, we construct an array $\mathbf{A}$ of size $\ell$ such that $\mathbf{A}[r]$ for every $r \in [\ell]$ stores a pointer to the first element of the list representing $\markerTransMat_{T_{x_r}}[i_r, j_r]$. All this can be done in time $\bigO(|\mathcal{T}|)$, and it concludes the preprocessing. 
Since, by Lemma~\ref{treeSizeLemma}, $|\mathcal{T}| \leq 4|\varset| \cdot \depth{A}$, the preprocessing can be done in time $\bigO(\depth{A}|\varset|)$. \par
In the enumeration phase, by $\ell$ nested loops, we iterate through all sequences $(p_{1}, p_{2}, \ldots, p_{\ell})$ of $\ell$ pointers to elements of the lists representing the sets $\markerTransMat_{T_{x_r}}[i_r, j_r]$, $r \in [\ell]$, and for each such sequence, we produce the partial marker set 
\begin{equation*}
\rightshift{\Lambda_{p_1}}{s_1} \cup \rightshift{\Lambda_{p_2}}{s_2} \cup \ldots \cup \rightshift{\Lambda_{p_\ell}}{s_\ell}\,, 
\end{equation*}
where, for every $r \in [\ell]$, $\Lambda_{p_r}$ is the element of $\markerTransMat_{T_{x_r}}[i_r, j_r]$ corresponding to pointer $p_r$. \par
It can be easily verified that, according to Definition~\ref{yieldDefinition}, the thus enumerated partial marker sets are exactly the partial marker sets from $\yield{\mathcal{T}}$. Moreover, due to the shifts applied to the partial marker-sets of the sets $\markerTransMat_{T_{x_r}}[i_r, j_r]$ with $r \in [\ell]$, there are no duplicates in this enumeration.\par
The maximum delay of this procedure is bounded by the depth of the nested loops, i.\,e., $\ell$. Since, by Lemma~\ref{treeSizeLemma}, $\ell \leq 2|\varset|$, the delay is $\bigO(|\varset|)$.
\end{proof}

\subsection*{Proof of Lemma~\ref{mainCorrectnessLemmaEnumPartOne}}

\begin{proof}
Let $A \to BC$ be the rule of $A$. By Definition~\ref{kSetsDefinition} we have 
\begin{align*}
\mainNode{A}{i}{j}{k} &\:= \:\markerTransMat_B[i, k] \msprod{\card{\deriv{B}}} \markerTransMat_C[k, j]\\
&\:= \:\{\Lambda_B \msprod{\card{\deriv{B}}} \Lambda_C \mid \Lambda_B \in \markerTransMat_B[i, k],  \Lambda_C \in \markerTransMat_C[k, j]\}\,.
\end{align*}
By Lemma~\ref{mainInductiveLemma} we have 
\begin{equation*}
\markerTransMat_A[i, j] = \bigcup_{k \in \intStatesTransMat_A[i, j]} \mainNode{A}{i}{j}{k}\,.
\end{equation*}
By definition of $(M, A)$-trees and by the definition of the function $\yield{\cdot}$, the set $\yield{\allTrees{A}{i}{k}{j}}$ contains exactly the elements $\Lambda_B \msprod{\card{\deriv{B}}} \Lambda_C$, where $\Lambda_B$ is in
\begin{itemize}
\item $\bigcup_{k_B \in \intStatesTransMat_{B}[i, k]} \yield{\allTrees{B}{i}{k_B}{k}}$, if $B$ is an inner non-terminal with $\reachTransMat_B[i, k] = \nonEmptyMarker$,
\item $\yield{\treeNodeTerminal{T_x}{i}{k}} = \markerTransMat_{T_x}[i, k]$, if $B = T_x$ with $\reachTransMat_{T_x}[i, k] = \nonEmptyMarker$,
\item $\yield{\treeNodeEmpty{B}{i}{k}} = \markerTransMat_{B}[i, k] = \{\emptyset\}$, if $\reachTransMat_B[i, k] = \emptyMarker$,
\end{itemize}
and, analogously, $\Lambda_C$ is in
\begin{itemize}
\item $\bigcup_{k_C \in \intStatesTransMat_{C}[k, j]} \yield{\allTrees{C}{k}{k_C}{j}}$, if $C$ is an inner non-terminal with $\reachTransMat_C[k, j] = \nonEmptyMarker$,
\item $\yield{\treeNodeTerminal{T_x}{k}{j}} = \markerTransMat_{T_x}[k, j]$, if $C = T_x$ with $\reachTransMat_{T_x}[k, j] = \nonEmptyMarker$,
\item $\yield{\treeNodeEmpty{C}{k}{j}} = \markerTransMat_{C}[k, j] = \{\emptyset\}$, if $\reachTransMat_C[k, j] = \emptyMarker$.
\end{itemize}
Therefore, if $B = T_x$ with $\reachTransMat_{T_x}[i, k] = \nonEmptyMarker$, or if $\reachTransMat_B[i, k] = \emptyMarker$, then $\Lambda_B$ is from $\markerTransMat_{B}[i, k]$. Moreover, if $B$ is an inner non-terminal with $\reachTransMat_B[i, k] = \nonEmptyMarker$,
then we conclude by induction that $\Lambda$ is from 
\begin{align*}
&\bigcup_{k_B \in \intStatesTransMat_{B}[i, k]} \yield{\allTrees{B}{i}{k_B}{k}} = \\
&\bigcup_{k_B \in \intStatesTransMat_{B}[i, k]} \mainNode{B}{i}{k}{k_B} \: = \: \markerTransMat_B[i, k]\,. 
\end{align*}
Note that the last equality is due to Lemma~\ref{mainInductiveLemma}.\par
Analogously, we get that $\Lambda_C$ is from $\markerTransMat_C[k, j]$. Consequently, the set $\yield{\allTrees{A}{i}{k}{j}}$ equals
\begin{equation*}
\{\Lambda_B \msprod{\card{\deriv{B}}} \Lambda_C \mid \Lambda_B \in \markerTransMat_{B}[i, k], \Lambda_C \in \markerTransMat_{C}[k, j]\}
\end{equation*}
and therefore $\yield{\allTrees{A}{i}{k}{j}} = \mainNode{A}{i}{j}{k}$.
\end{proof}

\subsection*{Proof of Lemma~\ref{mainInductiveLemmaEnumVersion}}

\begin{proof}
Let $M$ be a $\DFA$ and let $A' \to BC$ be the rule of the inner non-terminal $A'$. 
\begin{itemize}
\item $\markerTransMat_A[i, j] \cap \markerTransMat_A[i, j'] = \emptyset$: For contradiction, assume that there is some $\Lambda_A \in \markerTransMat_A[i, j] \cap \markerTransMat_A[i, j']$. Let $w = \insertmarkers{\deriv{A}}{\Lambda_A}$. By definition, this means that $\State{i} \transfuncNFA{w} \State{j}$ and $\State{i} \transfuncNFA{w} \State{j'}$ with $j \neq j'$, which is a contradiction to the assumption that $M$ is deterministic.
\item $\mainNode{A'}{i}{j}{k} \cap \mainNode{A'}{i}{j}{k'} = \emptyset$: For contradiction, assume that there is some $\Lambda \in \mainNode{A'}{i}{j}{k} \cap \mainNode{A'}{i}{j}{k'}$. By definition, this means that $\Lambda = \Lambda_B \msprod{\card{\deriv{B}}} \Lambda_C$ for some $\Lambda_B \in \markerTransMat_B[i, k]$ and $ \Lambda_C \in \markerTransMat_C[k, j]$, and $\Lambda = \Lambda'_B \msprod{\card{\deriv{B}}} \Lambda'_C$ for some $\Lambda'_B \in \markerTransMat_B[i, k']$ and $ \Lambda'_C \in \markerTransMat_C[k', j]$. However, since every $(\sigma, p) \in \Lambda_B \cup \Lambda'_B$ satisfies $p \leq \card{\deriv{B}}$, and every $(\sigma, p) \in \rightshift{\Lambda_C \cup \Lambda'_C}{\card{\deriv{B}}}$ satisfies $p > \card{\deriv{B}}$, we conclude that $\Lambda_B \msprod{\card{\deriv{B}}} \Lambda_C = \Lambda'_B \msprod{\card{\deriv{B}}} \Lambda'_C$ is only possible if $\Lambda_B = \Lambda'_B$ and $\Lambda_C = \Lambda'_C$. However, this means that $\Lambda_B \in \markerTransMat_B[i, k] \cap \markerTransMat_B[i, k']$, which is a contradiction to the lemma's first statement.
\end{itemize}
\end{proof}

\subsection*{Proof of Lemma~\ref{mainCorrectnessLemmaEnumPartTwo}}

Before presenting the proof of Lemma~\ref{mainCorrectnessLemmaEnumPartTwo}, we first discuss in more detail what it means if two $(M, A)$-trees are non-equal. To this end, let $\mathcal{T}_1$ and $\mathcal{T}_2$ be $(M, A)$-trees. If $\mathcal{T}_1 \neq \mathcal{T}_2$, but the roots are nevertheless corresponding, then there must be corresponding nodes $P_1$ and $P_2$ (possibly the roots) with some common label $\treeNode{A'}{i}{k}{j}$, such that their left children $L_1$ and $L_2$ are not corresponding, or their right children $R_1$ and $R_2$ are not corresponding. Let us first assume that neither $L_1$ nor $L_2$ is a leaf. By definition of $(M, A)$-trees, if $A' \to BC$ is the rule of $A'$, then $L_1$ and $L_2$ are labelled by $\treeNode{B}{i}{k_{B, 1}}{k}$ and $\treeNode{B}{i}{k_{B, 2}}{k}$, respectively. Thus, if they are not corresponding, then they only differ in $k_{B, 1} \neq k_{B, 2}$. Analgously, if neither $R_1$ nor $R_2$ is a leaf, then they are labelled by $\treeNode{C}{k}{k_{C, 1}}{j}$ and $\treeNode{C}{k}{k_{C, 2}}{j}$, respectively, and $k_{C, 1} \neq k_{C, 2}$ in case they do not correspond. On the other hand, if $\reachTransMat_B[i, k] = \emptyMarker$, then both $L_1$ and $L_2$ are leaves labelled with $\treeNodeEmpty{B}{i}{k}$ and therefore they correspond. If $B = T_x$ with $\reachTransMat_{T_x}[i, k] = \nonEmptyMarker$, then both $L_1$ and $L_2$ are leaves labelled with $\treeNodeTerminal{T_x}{i}{k}$ and they correspond as well. The situation is analogous with respect to $R_1$ and $R_2$.

\begin{proof}
For contradiction, assume that $\yield{\mathcal{T}_1} \cap \yield{\mathcal{T}_2} \neq \emptyset$.\par
By definition $\mathcal{T}_1 \in \allTrees{A}{i}{k_1}{j_1}$ and $\mathcal{T}_2 \in \allTrees{A}{i}{k_2}{j_2}$, and, by Lemma~\ref{mainCorrectnessLemmaEnumPartOne}, $\yield{\mathcal{T}_1} \subseteq \mainNode{A}{i}{j_1}{k_1} \subseteq \markerTransMat_A[i, j_1]$ and $\yield{\mathcal{T}_2} \subseteq \mainNode{A}{i}{j_2}{k_2} \subseteq \markerTransMat_A[i, j_2]$. 
It then follows from the first statement of Lemma~\ref{mainInductiveLemmaEnumVersion} that $j_1 = j_2$. From the second statement of Lemma~\ref{mainInductiveLemmaEnumVersion} we hence obtain that $k_1 = k_2$. In the following, we set $j = j_1 = j_2$ and $k = k_1 = k_2$. Hence, $\mathcal{T}_1$ and $\mathcal{T}_2$ have corresponding roots labelled by $\treeNode{A}{i}{k}{j}$.\par
Let $\widehat{\mathcal{T}}$ be the tree of the nodes of $\mathcal{T}_1$ and $\mathcal{T}_2$ that are corresponding (since the roots correspond, $\widehat{\mathcal{T}}$ is non-empty); we denote by $R$ the root of $\widehat{\mathcal{T}}$. \par
Let $P$ be some node of $\widehat{\mathcal{T}}$ (thus, a node of both $\mathcal{T}_1$ and $\mathcal{T}_2$) labelled by $\treeNode{A'}{i'}{k'}{j'}$ such that $A'$ is an inner non-terminal with a rule $A' \to BC$ (note that the root satisfies these properties). This node $P$ has a left child $L_1$ and a right child $R_1$ in $\mathcal{T}_1$, and a left child $L_2$ and a right child $R_2$ in $\mathcal{T}_2$. \par
We first consider the case that neither of these children are leaves in $\mathcal{T}_1$ or $\mathcal{T}_2$, respectively (the case that we have leaves among those nodes will be discussed later on). These children $L_1, L_2, R_1, R_2$ may or may not be in $\widehat{\mathcal{T}}$. Furthermore, since $P$ is labelled by $\treeNode{A'}{i'}{k'}{j'}$ and since there is a rule $A' \to BC$, we can assume that $L_1$ and $R_1$ are labelled with $\treeNode{B}{i'}{k_{B, 1}}{k'}$ and $\treeNode{C}{k'}{k_{C, 1}}{j'}$, respectively, and that 
$L_2$ and $R_2$ are labelled $\treeNode{B}{i'}{k_{B, 2}}{k'}$ and $\treeNode{C}{k'}{k_{C, 2}}{j'}$ in $\mathcal{T}_2$, respectively. 
If $\yieldPara{\mathcal{T}_1}{P} \cap \yieldPara{\mathcal{T}_2}{P} \neq \emptyset$, then, by definition of the yield, there are 
\begin{itemize}
\item $\Lambda_{B, 1} \in \yieldPara{\mathcal{T}_1}{L_1}$,
\item $\Lambda_{C, 1} \in \yieldPara{\mathcal{T}_1}{R_1}$,
\item $\Lambda_{B, 2} \in \yieldPara{\mathcal{T}_2}{L_2}$,
\item $\Lambda_{C, 2} \in \yieldPara{\mathcal{T}_2}{R_2}$
\end{itemize}
with $\Lambda_{B, 1} \msprod{\card{\deriv{B}}} \Lambda_{C, 1} = \Lambda_{B, 2} \msprod{\card{\deriv{B}}} \Lambda_{C, 2}$. Since, due to Lemma~\ref{mainCorrectnessLemmaEnumPartOne}, we have
\begin{itemize}
\item $\yieldPara{\mathcal{T}_1}{L_1} \subseteq \mainNode{B}{i'}{k'}{k_{B, 1}} \subseteq \markerTransMat_B[i', k']$,
\item $\yieldPara{\mathcal{T}_1}{R_1} \subseteq \mainNode{C}{k'}{j'}{k_{C, 1}} \subseteq \markerTransMat_C[k', j']$,
\item $\yieldPara{\mathcal{T}_2}{L_2} \subseteq \mainNode{B}{i'}{k'}{k_{B, 2}} \subseteq \markerTransMat_B[i', k']$,
\item $\yieldPara{\mathcal{T}_2}{R_2}  \subseteq \mainNode{C}{k'}{j'}{k_{C, 2}} \subseteq \markerTransMat_C[k', j']$,
\end{itemize}
we can conclude with Lemma~\ref{simpleDisjointnessLemma} that $\Lambda_{B,1} = \Lambda_{B,2}$ and $\Lambda_{C,1} = \Lambda_{C,2}$. Consequently, 
\begin{align*}
&\yieldPara{\mathcal{T}_1}{P} \cap \yieldPara{\mathcal{T}_2}{P} \neq \emptyset \Longrightarrow\\
&\yieldPara{\mathcal{T}_1}{L_1} \cap \yieldPara{\mathcal{T}_2}{L_2} \neq \emptyset \wedge
\yieldPara{\mathcal{T}_1}{R_1} \cap \yieldPara{\mathcal{T}_2}{R_2} \neq \emptyset\,.
\end{align*}
This holds, if neither of $L_1, L_2, R_1, R_2$ are leaves. \par
Let us now turn to the case where $L_1$ or $L_2$ is a leaf in $\mathcal{T}_1$ or $\mathcal{T}_2$, respectively. Then, since their parent nodes are corresponding, they must both be corresponding leaves. Thus, they have the same label and therefore also the same yields with respect to $\mathcal{T}_1$ and $\mathcal{T}_2$, and since yields of nodes are always non-empty, we conclude that $\yieldPara{\mathcal{T}_1}{L_1} \cap \yieldPara{\mathcal{T}_2}{L_2} \neq \emptyset$. An analogous observation can be made in the case that $R_1$ or $R_2$ is a leaf in $\mathcal{T}_1$ or $\mathcal{T}_2$. Consequently, the implication displayed above holds for all nodes $P$ of $\widehat{\mathcal{T}}$. \par
Since, by assumption, $\mathcal{T}_1 \neq \mathcal{T}_2$ and $\yieldPara{\mathcal{T}_1}{R} \cap \yieldPara{\mathcal{T}_2}{R} \neq \emptyset$ (where $R$ is the root of $\widehat{\mathcal{T}}$), we can now also conclude (by inductively using the implication proved above) that there is a node $P$ of $\widehat{\mathcal{T}}$ with $\yieldPara{\mathcal{T}_1}{P} \cap \yieldPara{\mathcal{T}_2}{P} \neq \emptyset$, such that $P$'s left children are not corresponding or $P$'s right children are not corresponding. Let us assume that this is the case with respect to the left children $L_1$ and $L_2$ of $P$ (the argument for the right children is analogous). As shown above, we know that $\yieldPara{\mathcal{T}_1}{L_1} \cap \yieldPara{\mathcal{T}_2}{L_2} \neq \emptyset$.\par
Now assume that $P$ is labelled by $\treeNode{A'}{i'}{k'}{j'}$ such that $A'$ is an inner non-terminal with a rule $A' \to BC$. Consequently, $L_1$ is labelled by $\treeNode{B}{i'}{k_{B, 1}}{k'}$, $L_2$ is labelled by $\treeNode{B}{i'}{k_{B, 2}}{k'}$ and, since $L_1$ and $L_2$ do not correspond, we have $k_{B, 1} \neq k_{B, 2}$. \par
Again by Lemma~\ref{mainCorrectnessLemmaEnumPartOne}, it follows that
\begin{align*}
\yieldPara{\mathcal{T}_1}{L_1} \subseteq \mainNode{B}{i'}{k'}{k_{B, 1}} \text{ and } \yieldPara{\mathcal{T}_2}{L_2} \subseteq \mainNode{B}{i'}{k'}{k_{B, 2}}\,.
\end{align*}
Therefore, we can conclude that $\mainNode{B}{i'}{k'}{k_{B, 1}} \cap \mainNode{B}{i'}{k'}{k_{B, 2}} \neq \emptyset$, which is a contradiction to Lemma~\ref{mainInductiveLemmaEnumVersion}.
\end{proof}

\subsection*{Proof of Lemma~\ref{enumAllLemma}}

The following can directly be concluded from the definition of $(M, A)$-trees and the definition of algorithm $\enumAll$.

\begin{lemma}\label{enumBaseCasesLemma}
Let $A \in N$ and let $i, j \in [\states]$ with $\reachTransMat_{A}[i, j] \neq \undefin$.
\begin{enumerate}
\item If $\reachTransMat_{A}[i, j] = \emptyMarker$, then the algorithm $\enumAll(A, i, \baseCaseMarker, j)$ enumerates the set $\allTrees{A}{i}{\baseCaseMarker}{j}$ with constant preprocessing and constant delay. 
\item If $A = T_x$, then the algorithm $\enumAll(T_x, i, \baseCaseMarker, j)$ enumerates the set $\allTrees{T_x}{i}{\baseCaseMarker}{j}$ with constant preprocessing and constant delay.
\end{enumerate}
\end{lemma}
\noindent
This serves as the induction base for the proof of Lemma~\ref{enumAllLemma}:

\begin{proof}
We proceed by induction and prove the following stronger statement:
\begin{enumerate}[$(*)$]
\item[$(*)$] 
 There is a constant $c$ such that for all inputs 
 $A \in N$, $i, j \in [\states]$, $k \in \mathcal{I}_A[i, j] \cup \{\baseCaseMarker\}$,
 such that $k=\baseCaseMarker$ or $\reachTransMat_{A}[i, j] = \nonEmptyMarker$,
 the algorithm $\enumAll(A, i, k, j)$ enumerates (without duplicates) the set $\allTrees{A}{i}{k}{j}$ such that it
 takes time at most
 \begin{itemize}
  \item $c\cdot \Max{A,i,k,j}$ before the first output is created
  \item $2c\cdot \Max{A,i,k,j}$ between any two consecutive output trees
  \item $c\cdot\Max{A,i,k,j}$ between outputting the last tree and the end-of-enumeration message $\eol$.
 \end{itemize}
\end{enumerate}
The case where $k=\baseCaseMarker$ serves as the induction base; and for this case $(*)$ is provided by
Lemma~\ref{enumBaseCasesLemma}. 
For the induction step consider an input $(A,i,k,j)$ where $A$ is a non-terminal with rule 
$A \to BC$ and $k\in \intStatesTransMat_A[i, j]$.
Our induction hypothesis is as follows: 
For every $k_B \in \mathcal{I}_B[i,k]$ and every $k_C\in\mathcal{I}_C[k,j]$, a call of $\enumAll(B, i, k_B, k$) and of $\enumAll(C,k,k_C,j)$ 
enumerates (without duplicates) the sets $\allTrees{B}{i}{k_B}{k}$ and $\allTrees{C}{k}{k_C}{j}$, respectively, and moreover, 
satisfies the time bounds stated in $(*)$ (where 
$\Max{A,i,k,j}$ has to be replaced with $\Max{B,i,k_B,k}$ and with $\Max{C,k,k_C,j}$, respectively). 

For simplicity, we shall denote the loops of 
Lines~\ref{enumStatesLine}, \ref{enumBLine} and \ref{enumCLine} by \emph{states-loop}, \emph{$B$-loop} 
and \emph{$C$-loop}, respectively, and we denote Line~\ref{recursiveOutputLine} by \emph{output-line}.  
We shall also denote $\mathcal{I}_B[i, k]$ and $\mathcal{I}_C[k, j]$ simply by $\mathcal{I}_B$ and $\mathcal{I}_C$, respectively.

We first observe that $\enumAll(A, i, k, j)$ does in fact enumerate the set $\allTrees{A}{i}{k}{j}$. Since $k \neq \baseCaseMarker$, for every $(k_B, k_C) \in (\mathcal{I}_B \times \mathcal{I}_C)$, for every $\mathcal{T}_B \in \allTrees{B}{i}{k_B}{k}$ and every $\mathcal{T}_C \in \allTrees{C}{k}{k_C}{j}$, the algorithm will output as next element of the output sequence the tree with a root labelled by $\treeNode{A}{i}{k}{j}$, and with the roots of $\mathcal{T}_B$ and $\mathcal{T}_C$ as left and right child, respectively. After having done this for all $(k_B, k_C) \in (\mathcal{I}_B \times \mathcal{I}_C)$, it will produce $\eol$ and terminate. By definition of $(M, A)$-trees, all $(M, A)$-trees with a root labelled by $\treeNode{A}{i}{k}{j}$ can be constructed in this way, and therefore all elements of $\allTrees{A}{i}{k}{j}$ are constructed at some point in the output-line. Furthermore, every tree that is constructed in the output-line does indeed belong to $\allTrees{A}{i}{k}{j}$.

We next show that it is impossible to create duplicates in the output-line. To this end, assume that we reach the output-line once with $k_B, k_C, \mathcal{T}_B, \mathcal{T}_C$, and once with $k'_B, k'_C, \mathcal{T}'_B, \mathcal{T}'_C$. Moreover, let $\mathcal{T} = \treeNode{A}{i}{k}{j}(\mathcal{T}_B, \mathcal{T}_C)$ and $\mathcal{T}' = \treeNode{A}{i}{k}{j}(\mathcal{T}'_B, \mathcal{T}'_C)$ be the trees produced in these two calls of the output-line. Obviously, $\mathcal{T} = \mathcal{T}'$ is only possible if $\mathcal{T}_B = \mathcal{T}'_B$ and $\mathcal{T}_C = \mathcal{T}'_C$. Since we can assume by induction that the recursive calls do not produce any duplicates, this cannot happen in the same iteration of the states-loop. Consequently, $k_B \neq k'_B$ or $k_C \neq k'_C$. However, if $k_B \neq k'_B$, then the roots of $\mathcal{T}_B$ and $\mathcal{T}'_B$ are different, and if $k_C \neq k'_C$, then the roots of $\mathcal{T}_C$ and $\mathcal{T}'_C$ are different. Thus, $\mathcal{T}_B \neq \mathcal{T}'_B$ or $\mathcal{T}_C \neq \mathcal{T}'_C$, which means that $\mathcal{T} \neq \mathcal{T}'$.

It remains to show that the runtime guarantees stated in $(*)$ are satisfied.

A call of $\enumAll(A,i,k,j)$ starts by checking that $k \neq \baseCaseMarker$, and then it starts a cycle of the states-loop, the $B$-loop and the $C$-loop and reaches the output-line, where it produces the first output. 
The induction hypothesis tells us that it takes time at most $c\cdot \Max{B,i,k_B,k}$ until the first 
$\mathcal{T}_B$ is produced in the $B$-loop and it takes time at most $c\cdot \Max{C,k,k_C,j}$ until the first $\mathcal{T}_C$ is produced in the $C$-loop. Afterwards, we only have to create one new root node and set two pointers to construct the first output tree $\treeNode{A}{i}{k}{j}(\mathcal{T}_B, \mathcal{T}_C)$.
All this is done within time 
\[
c\cdot \Max{B,i,k_B,k} + c\cdot \Max{C,k,k_C,j} + c 
\ \ \leq \ \ 
c\cdot\Max{A,i,k,j}.
\]
Now assume that we have just produced an output in the output-line. Moreover, let $(k_B, k_C)$ be the current pair from $\mathcal{I}_B \times \mathcal{I}_C$. There are several possibilities of which lines are executed before the algorithm reaches another line that produces an output, i.\,e., the output-line or Line~\ref{theEndLine}. All these possibilities obviously depend on whether or not we are currently in the last iteration of the states-loop, the $B$-loop and the $C$-loop. Moreover, to estimate the required time, we should count all starts of a new iteration (since these require a new output from some recursive call), all necessary checks that there is no further iteration (since this requires nevertheless the request of $\eol$ from a recursive call), and the start of a new cycle of a loop (since this requires a new recursive call). In particular, we can consider these operations with respect to the states-loop to require only constant time, since we have the sets $\mathcal{I}_B$ and $\mathcal{I}_C$ at our disposal. There are four cases to consider:
\begin{enumerate}[(1)]
\item\label{item:CLoopMiddle} We are not in the last iteration of the $C$-loop.
\item\label{item:CLoopEnd} We are in the last iteration of the $C$-loop (but not the $B$-loop).
\item\label{item:CandBLoopEnd} We are in the last iterations of the $C$-loop and the $B$-loop (but not the states-loop).
\item\label{item:AllLoopsEnd} We are in the last iterations of the $C$-loop, the $B$-loop and the states-loop.
\end{enumerate}

\smallskip

\noindent
In case~\eqref{item:CLoopMiddle} we request the next tree from $\enumAll(C, k, k_C, j)$,
this tree is (by the induction hypothesis) provided within time at most $2c\cdot\Max{C,k,k_C,j}$, 
and then  the output tree $\treeNode{A}{i}{k}{j}(\mathcal{T}_B, \mathcal{T}_C)$ 
can be constructed immediately. The time spent for all this is at most $2c\cdot\Max{A,i,k,j}$.

\medskip

\noindent
In case~\eqref{item:CLoopEnd} we
request the next element $\eol$ from $\enumAll(C, k, k_C, j)$, then request the next tree
$\mathcal{T}_B$ from $\enumAll(B,i,k_B,k)$, then request the first tree $\mathcal{T}_C$ from
$\enumAll(C,k,k_C,j)$, and then take constant time to produce the next output tree
$\treeNode{A}{i}{k}{j}(\mathcal{T}_B, \mathcal{T}_C)$.
By the induction hypothesis, the requests are answered within time $c\cdot\Max{C, k, k_C, j}$, 
time $2c\cdot\Max{B,i,k_B,k}$, and time $c\cdot\Max{C,k,k_C,j}$, respectively.
Note that
\[
\begin{array}{rcl}
\Max{A,i,k,j} & \geq & \Max{B, i, k_B, k} + \Max{C,k,k_C,j}+1.
\end{array}
\]
Hence, the total time spent for producing the next output tree is at most
$2c\cdot\Max{A,i,k,j}$.

\medskip

\noindent
In case~\eqref{item:CandBLoopEnd}
the algorithm requests the next element $\eol$ from $\enumAll(C, k, k_C, j)$, 
the next element $\eol$ from $\enumAll(B, i, k_B, k)$, 
and then requests the first element $\mathcal{T}_B$ of $\enumAll(B, i, k'_B, k)$ and the first element $\mathcal{T}_C$ of $\enumAll(C, k, k'_C, j)$ 
(where $(k'_B, k'_C)$ is the pair of the next iteration of the states-loop). 
By the induction hypothesis, each of these requests is answered within time
$c\cdot\Max{C, k, k_C, j}$, time $c\cdot\Max{B, i, k_B, k}$, time 
$c\cdot\Max{B, i, k'_B, k}$, and time $c\cdot\Max{C, k, k'_C, j}$.
And afterwards it takes constant time until it outputs the tree
$\treeNode{A}{i}{k}{j}(\mathcal{T}_B, \mathcal{T}_C)$.
Note that 
\[
\begin{array}{rcl}
\Max{A,i,k,j} & \geq & \Max{B, i, k_B, k} + \Max{C,k,k_C,j}+1 \quad\text{and}
\\
\Max{A,i,k,j} & \geq & \Max{B, i, k'_B, k} + \Max{C,k,k'_C,j}+1.
\end{array}
\]
Thus, the algorithm produces the next output tree within time at most 
$2c\cdot\Max{A,i,k,j}$.

\medskip

\noindent
In case~\eqref{item:AllLoopsEnd} we request the next element $\eol$ from $\enumAll(C, k, k_C, j)$, then request the next element $\eol$ from $\enumAll(B, i, k_B, k)$, and then notice that the states-loop has no further iteration. We therefore terminate in Line~\ref{theEndLine} by outputting $\eol$. 
By the induction hypothesis, the requests are answered within time 
$c\cdot\Max{C,k,k_C,j}$ and time $c\cdot\Max{B,i,k_B,k}$, respectively. Since
\[
\begin{array}{rcl}
\Max{A,i,k,j} & \geq & \Max{B, i, k_B, k} + \Max{C,k,k_C,j}+1,
\end{array}
\]
the total time spent until outputting $\eol$ is at most $c\cdot\Max{A,i,k,j}$.

In summary, we have shown that $(*)$ is satisfied when calling $\enumAll(A,i,k,j)$. This completes the
induction step and completes the proof of Lemma~\ref{enumAllLemma}.
\end{proof}

\subsection*{Proof of Lemma~\ref{enumSetTheorem}}

\begin{proof}
We describe the preprocessing phase and the enumeration phase separately.\medskip\\
\textbf{Preprocessing}: 
First, we compute all the matrices $\reachTransMat_A$ for every $A \in N$, $\intStatesTransMat_{A'}$ for every inner non-terminal $A' \in N$, and\,$\markerTransMat_{T_x}$ for every $x \in \Sigma$.
According to Lemma~\ref{inductionBaseLemma}, this can be done in time $\bigO(|M| + (\size{\mathcal{S}}\states^3))$.
After this, we also compute, for every $A \in N$ and for every $i, j \in [\states]$, the sets $\mathcal{I}_A[i, j]$ as required by Algorithm~\ref{enumAllAlgo}, i.\,e., if $A = T_x$ or $\reachTransMat_{A}[i, j] = \emptyMarker$ then $\mathcal{I}_A[i, j] = \{\baseCaseMarker\}$, and $\mathcal{I}_A[i, j] = \intStatesTransMat_A[i, j]$ otherwise. Computing all these sets can obviously be done in time $\bigO(\size{\mathcal{S}}\states^2)$. Then we compute the set $F' = \{j \in F \mid \reachTransMat_{S_0}[1, j] \neq \undefin\}$, which can be done in time $\bigO(\states)$. This concludes the preprocessing. We note that the preprocessing time is $\bigO(|M| + (\size{\mathcal{S}} \cdot \states^3))$. \medskip\\
\textbf{Enumeration}: We first formulate an enumeration procedure that receives an $(M, A)$-tree $\mathcal{T}$ as input.\smallskip\\
$\enumSingleTree(\mathcal{T})$:
\begin{enumerate}
\item We transform $\mathcal{T}$ into the corresponding $(M, A)$-tree $\mathcal{T}'$ with leaf-pointers. 
\item We enumerate $\yield{\mathcal{T}'}$ according to Lemma~\ref{singleTreeEnumLemma}.
\end{enumerate}
\noindent \emph{Claim $1$}: The procedure $\enumSingleTree(\mathcal{T})$ enumerates $\yield{\mathcal{T}'}$ with preprocessing time $\bigO(\depth{A}|\varset|)$ and delay $\bigO(|\varset|)$. \smallskip\\
\noindent \emph{Proof of Claim $1$}: The first step can be done in time $\bigO(|\mathcal{T}|)$ by simply adding to the terminal-leaves of $\mathcal{T}$ the pointers to the corresponding sets $\markerTransMat_{T_x}[i', j']$ (which have been computed in the preprocessing). By Lemma~\ref{treeSizeLemma}, $|\mathcal{T}'| \leq 4 |\varset| \cdot \depth{A}$. By Lemma~\ref{singleTreeEnumLemma}, the second step can be done with preprocessing $\bigO(\depth{A}|\varset|)$ and delay $\bigO(|\varset|)$. \hfill \qed (\emph{Claim $1$})\smallskip\\
Next, for any $j \in F'$ and $k \in \intStatesTransMat_{S_0}[1, j]$, we formulate a second enumeration procedure.\smallskip\\
$\enumSingleRoot(j, k)$:
\begin{enumerate}
\item By calling $\enumAll(S_0, 1, k, j)$, we produce a sequence 
\begin{equation*}
(\mathcal{T}_1, \mathcal{T}_2, \ldots, \mathcal{T}_{n_{j, k}})
\end{equation*}
of $(M, S_0)$-trees followed by $\eol$.
\item In this enumeration, as soon as we have received $\mathcal{T}_\ell$ for some $\ell \in [n_{j, k}]$, we carry out $\enumSingleTree(\mathcal{T}_{\ell})$ and produce the output sequence of $\enumSingleTree(\mathcal{T}_{\ell})$ as output elements. 
\end{enumerate}
\noindent \emph{Claim $2$}: The procedure $\enumSingleRoot(j, k)$ enumerates $\mainNode{S_0}{1}{j}{k}$ with preprocessing time and delay $\bigO(\depth{S_0}|\varset|)$. \smallskip\\
\noindent \emph{Proof of Claim $2$}: According to Lemma~\ref{enumAllLemma}, $(\mathcal{T}_1, \mathcal{T}_2, \ldots, \mathcal{T}_{n_{j, k}})$ is an enumeration of the set $\allTrees{S_0}{1}{k}{j}$ without duplicates, i.\,e., a sequence of exactly the $(M, S_0)$-trees with root $\treeNode{S_0}{1}{k}{j}$. Furthermore, $\enumAll(S_0, 1, k, j)$ has preprocessing time and delay $\Max{S_0,1,k,j}$
which, by Lemma~\ref{treeSizeLemma}, means that the preprocessing time and delay is $\bigO(\depth{S_0}|\varset|)$. We also note that, for every $\ell \in [n_{j, k}]$, since $\reachTransMat_{S_0}[1, j] \neq \undefin$ and $k \in \intStatesTransMat_{S_0}[1, j]$, we know that $\yield{\mathcal{T}_{\ell}} \neq \emptyset$. Since by Claim~$1$ every call of $\enumSingleTree(\mathcal{T}_{\ell})$ uses preprocessing time $\bigO(\depth{S_0}|\varset|)$ and delay $\bigO(|\varset|)$, we can conclude that both the preprocessing time and the delay of $\enumSingleRoot(j, k)$ is $\bigO(\depth{S_0}|\varset|)$. \par
Since $(\mathcal{T}_1, \mathcal{T}_2, \ldots, \mathcal{T}_{n_{j, k}})$ is an enumeration of $\allTrees{S_0}{1}{k}{j}$, we know that $\enumSingleRoot(j, k)$ enumerates $\yield{\allTrees{S_0}{1}{k}{j}}$ which, by Lemma~\ref{mainCorrectnessLemmaEnumPartOne} is equal to 
$\mainNode{S_0}{1}{j}{k}$. It remains to observe that the enumeration is without duplicates. Indeed, for all $\ell, \ell' \in [n_{j, k}]$ with $\ell \neq \ell'$, we have $\yield{\mathcal{T}_{\ell}} \cap \yield{\mathcal{T}_{\ell'}} = \emptyset$ by Lemma~\ref{mainCorrectnessLemmaEnumPartTwo}, and, furthermore, the enumeration of each $\yield{\mathcal{T}_{\ell}}$ and the enumeration of $\allTrees{S_0}{1}{k}{j}$ have no duplicates due to the correctness of enumeration procedure $\enumSingleTree(\mathcal{T})$ and enumeration procedure $\enumAll(S_0, 1, k, j)$, respectively.\hfill \qed (\emph{Claim $2$})\smallskip\\
The complete enumeration phase is now as follows. For every $j \in F'$ and for every $k \in \intStatesTransMat_{S_0}[1, j]$, we perform $\enumSingleRoot(j, k)$ and produce the elements of its output sequence as output elements. From Claims~$1$~and~$2$, we obtain that the delay of this enumeration phase is $\bigO(\depth{S_0}|\varset|)$. \par
Since each $\enumSingleRoot(j, k)$ enumerates $\mainNode{S_0}{1}{j}{k}$, Lemma~\ref{mainInductiveLemma} implies that this enumeration procedure enumerates 
\begin{align*}
\bigcup_{j \in F'} \bigcup_{k \in \intStatesTransMat_{S_0}[1, j]} \mainNode{S_0}{1}{j}{k} \: = \: \bigcup_{j \in F'} \markerTransMat_{S_0}[1, j] \: = \: \bigcup_{j \in F} \markerTransMat_{S_0}[1, j]\,.
\end{align*}
Since, by Lemma~\ref{markerTransMatSolutionSetComputationLemma}, $\llbracket M \rrbracket(\doc) = \bigcup_{\State{j} \in F} \markerTransMat_{S_0}[1, j]$, we can conclude that the enumeration procedure enumerates $\llbracket M \rrbracket(\doc)$.\par
Note that this enumeration does not produce any duplicates: As already observed above, for each $j \in F'$ and $k \in \intStatesTransMat_{S_0}[1, j]$, the set $\mainNode{S_0}{1}{j}{k}$ is enumerated without any duplicates. And by Lemma~\ref{mainInductiveLemmaEnumVersion}, 
$\mainNode{S_0}{1}{j}{k} \cap \mainNode{S_0}{1}{j}{k'} = \emptyset$ for all distinct $k, k' \in \intStatesTransMat_{S_0}[1, j]$.\par
Finally, we consider states $j, j' \in F'$ with $j \neq j'$. We know that $\bigcup_{k \in \intStatesTransMat_{S_0}[1, j]} \mainNode{S_0}{1}{j}{k} = \markerTransMat_{S_0}[1, j]$ and $\bigcup_{k \in \intStatesTransMat_{S_0}[1, j']} \mainNode{S_0}{1}{j'}{k} = \markerTransMat_{S_0}[1, j']$ and
by Lemma~\ref{mainInductiveLemmaEnumVersion}, $\markerTransMat_{S_0}[1, j] \cap \markerTransMat_{S_0}[1, j'] = \emptyset$. Thus,
\begin{equation*}
\left(\bigcup_{k \in \intStatesTransMat_{S_0}[1, j]} \mainNode{S_0}{1}{j}{k}\right) \cap \left(\bigcup_{k \in \intStatesTransMat_{S_0}[1, j']} \mainNode{S_0}{1}{j'}{k}\right) = \emptyset\,.
\end{equation*}

Therefore, the enumeration is without duplicates. This completes the proof of Theorem~\ref{enumSetTheorem}.
\end{proof}

\end{document}